\documentclass[11pt,headings=big,numbers=noenddot,DIV=14,a4paper]{article}%

\pdfoutput=1

\usepackage[margin=1 in]{geometry}
\usepackage{comment}
\usepackage{graphicx}  
\usepackage{dcolumn}   
\usepackage{bm,relsize}        
\usepackage{amsfonts,amsmath,amssymb,amsthm,mathtools}
\usepackage{slashed}
\usepackage{placeins}
\usepackage{lipsum, color}
\usepackage[usenames,dvipsnames,svgnames,table]{xcolor}
\usepackage[linktoc=page,bookmarks=false,colorlinks=true,linkbordercolor=RoyalBlue,citebordercolor=ForestGreen,urlbordercolor=CornflowerBlue]{hyperref}
\hypersetup{
    colorlinks=true,
    linkcolor=ForestGreen,
    filecolor=ForestGreen,      
    urlcolor=ForestGreen,
    citecolor=ForestGreen,
}
\usepackage[sort&compress,numbers,merge]{natbib}
\usepackage{afterpage}
\usepackage{floatpag}
\usepackage{tikz-feynman}
\usepackage{mathtools}

\def\beq{\begin{equation}}
\def\eeq{\end{equation}}
\def\bea{\begin{eqnarray}}
\def\eea{\end{eqnarray}}

\def\xFO			{x_{\mathsmaller{\rm FO}}}

\def\MDM		{m_{\mathsmaller{\rm DM}}}
\def\MPl		{M_{\mathsmaller{\rm Pl}}}
\def\gSM		{g_{\mathsmaller{\rm SM}}}
\def\gDM		{g_{\mathsmaller{\rm DM}}}

\newcommand{\dilaton}{\chi}
\newcommand{\dilatonvev}{f}
\newcommand{\DM}{\mathsmaller{\rm DM}}

\newcommand{\Tnuc}{T_{\rm nuc}}
\newcommand{\Tstart}{T_{\rm start}}
\newcommand{\Treh}{T_{\mathsmaller{\rm RH}}}

\newcommand{\msigma}{m_{\sigma}}
\newcommand{\fconf}{f}
\newcommand{\stoptocwriting}{%
  \addtocontents{toc}{\protect\setcounter{tocdepth}{-5}}}
\newcommand{\resumetocwriting}{%
  \addtocontents{toc}{\protect\setcounter{tocdepth}{\arabic{tocdepth}}}}

\def\gwp		{\gamma_{\mathsmaller{\text{wp}}}}

\def\gcw		{\gamma_{\mathsmaller{\text{cw}}}}
\def\gcp		{\gamma_{\mathsmaller{\text{cp}}}}
\def\ECM		{E_{\mathsmaller{\text{CM}}}}
\def\TC			{{\mathsmaller{\text{TC}}}}
\def\YDM		{Y_{\mathsmaller{\rm DM}}}

\begin{document}

\begin{flushright}
\footnotesize
ULB-TH/21-17 \\
DESY 21-171
\end{flushright}
\color{black}

\begin{center}

{\LARGE \bf
Supercool Composite Dark Matter Beyond 100~TeV
\large
}

\medskip
\bigskip\color{black}\vspace{0.5cm}

{
{\large Iason Baldes},$^a$
{\large Yann Gouttenoire},$^{b,c}$
{\large Filippo Sala}$,^d$
{\large G\'eraldine Servant}$^{b,e}$
}
\\[7mm]

{\it \small $^a$ Service de Physique Th\'eorique, Universit\'e Libre de Bruxelles, \\ Boulevard du Triomphe, CP225, B-1050 Brussels, Belgium}\\
{\it \small $^b$ Deutsches Elektronen-Synchrotron DESY, Notkestr. 85, 22607 Hamburg, Germany}\\
{\it \small $^c$ School of Physics and Astronomy, Tel-Aviv University, Tel-Aviv 69978, Israel}\\
{\it \small $^d$ Laboratoire de Physique Th\'eorique et Hautes \'Energies, Sorbonne Universit\'e, CNRS, Paris, France}\\
\it \small $^e$ II. Institute of Theoretical Physics, Universit\"{a}t Hamburg, D-22761 Hamburg, Germany \\
\end{center}

\bigskip

\centerline{\bf Abstract}
\begin{quote}
\color{black}

Dark matter could be a composite state of a confining sector with an approximate scale symmetry. We consider the case where the associated pseudo-Goldstone boson, the dilaton, mediates its interactions with the Standard Model. When the confining phase transition in the early universe is supercooled, its dynamics allows for dark matter masses up to $10^6$~TeV. We derive the precise parameter space compatible with all experimental constraints, finding that this scenario can be tested partly by telescopes and entirely by gravitational waves.

\medskip

\end{quote}

\clearpage
\noindent\makebox[\linewidth]{\rule{\textwidth}{1pt}} 
\tableofcontents
\noindent\makebox[\linewidth]{\rule{\textwidth}{1pt}}

\section{Introduction}

In the coming years we will be flooded by data offering new windows on particle physics at high energy scales. 
Telescopes such as CTA~\cite{CTAConsortium:2018tzg} and LHAASO~\cite{Bai:2019khm} will collect, for the first time, $\gamma$ rays at and above 100 TeV, and KM3NeT~\cite{KM3NeT} will explore a similar energy range in upward going neutrinos from the Galactic Centre. Underground experiments for  Dark Matter (DM) direct detection, such LZ~\cite{Akerib:2018lyp} and DARWIN~\cite{Aalbers:2016jon}, will test new parameter regions of heavy DM mass.
Further in the future, interferometers such as LISA~\cite{amaroseoane2017laser}, DECIGO~\cite{Kawamura:2020pcg}, and the Einstein Telescope~\cite{Maggiore:2019uih} will be sensitive to gravitational waves (GWs) generated at (currently unexplored) high temperatures in the early universe, thus offering yet another beacon on physics at high energies.
These prospects are particularly exciting as they could unveil some of the current mysteries of Nature at a fundamental level. It appears unavoidable that some physics beyond the Standard Model (BSM) lives at high energy scales, given the success of the Standard Model (SM) in explaining physics below a TeV.

In comparison with the information extracted from colliders, these upcoming experiments will offer indirect information. 
However, they are highly complementary with each other.
It is therefore important to cross-correlate the signals that any given BSM sector could yield in each of these experiments.
It is likely that this would be the only way to tell apart different models, which can easily appear indistinguishable in a single dataset. 
In this paper we take a step in this direction. We determine how telescopes, underground labs and GW interferometers interplay in testing a concrete class of heavy DM models, that are motivated by a wide picture of solutions to other problems of the~SM.

\medskip

We will consider Dark Matter as a composite state of a new confining sector, which acquires mass precisely at the associated confining phase transition in the early universe.
From the DM point of view, new strongly coupled sectors are appealing because they are associated with global symmetries of high quality below the confinement scale. These symmetries imply the existence of stable or very long-lived composite states, much like the proton in QCD, and therefore render confining sectors a natural starting point in building DM models, see e.g.~\cite{Antipin:2015xia}.
This property also makes confining sectors useful for BSM more generally, for example they can solve the quality problem of the QCD axion~\cite{Rubakov:1997vp,Redi:2016esr}. Such sectors have also attracted enormous attention because they dynamically generate a mass scale via dimensional transmutation, when their characteristic coupling runs from perturbative values in the UV to large values in the~IR. This adds to their appeal for DM model building, as they provide an explanation for the origin of the DM mass. Independently of DM, strongly-coupled sectors are ubiquitous in extensions of the SM because dimensional transmutation enables one to obtain large and stable hierarchies among mass scales. This has, for example, been used to address the hierarchy problem in composite~\cite{Contino:2010rs,Panico:2015jxa},
Twin Higgs~\cite{Chacko:2005pe,Geller:2014kta,Barbieri:2015lqa,Low:2015nqa}  and relaxion models~\cite{Graham:2015cka,Batell:2017kho} and to explain the separation between the supersymmetry breaking and the Planck scale~\cite{Witten:1981nf,Shadmi:1999jy}. 
Needless to say, a new confining sector could address several of the SM issues mentioned above at the same time.

\medskip

Here we will focus on strongly-coupled sectors where an approximate scale symmetry is spontaneously broken at confinement, so that a light pseudo-Goldstone boson exists in the spectrum, the dilaton~\cite{Carruthers:1971uy}.
When the explicit breaking of conformal invariance is sufficiently small, the dilaton can be significantly lighter than the confinement scale set by the dilaton vacuum expectation value (vev).
As the dilaton vev sets all scales in the strong sector, if the dilaton is light, other dynamical fields can be integrated out and 
the confinement phase transition can be described  in terms of the dilaton vev only.
In these models the confining phase transition (PT) in the early universe is supercooled, namely bubble nucleation is so delayed that the radiation energy density becomes smaller than the vacuum energy, so that the universe undergoes a stage of inflation until the PT ends~\cite{Kolb:1979bt}.
Supercooled PTs have been predicted in the 5D holographic warped description ~\cite{Creminelli:2001th,Randall:2006py,Nardini:2007me,Hassanain:2007js, Konstandin:2010cd, Konstandin:2011dr, vonHarling:2017yew,Fujikura:2019oyi,Bunk:2017fic, Dillon:2017ctw, Megias:2018sxv,Megias:2020vek, Agashe:2020lfz}, which is dual to nearly-conformal 4D theories~\cite{Arunasalam:2017ajm,Bruggisser:2018mus, Bruggisser:2018mrt, Baratella:2018pxi, Agashe:2019lhy, DelleRose:2019pgi, vonHarling:2019gme,Bloch:2019bvc,Baldes:2020kam}, and have attracted a lot of interest recently. We call attention to \cite{Gouttenoire:2022gwi} for a review on supercooled PTs. Unique dynamical features of any such PT were pointed out and modelled in~\cite{Baldes:2020kam}: fundamental quanta of the confining sector, upon being swallowed by the expanding bubbles, connect to their walls via strings which then fragment, producing high-energy populations of hadrons. These dynamics have several important effects, for example the relic abundance of composite states, which acquire mass at the PT, is affected by many orders of magnitudes with respect to the (wrong) treatment where these effects are ignored.

\begin{figure}[t]
\centering
\hspace{-1cm}
\raisebox{0cm}{\makebox{\includegraphics[height=0.5\textwidth, scale=1]{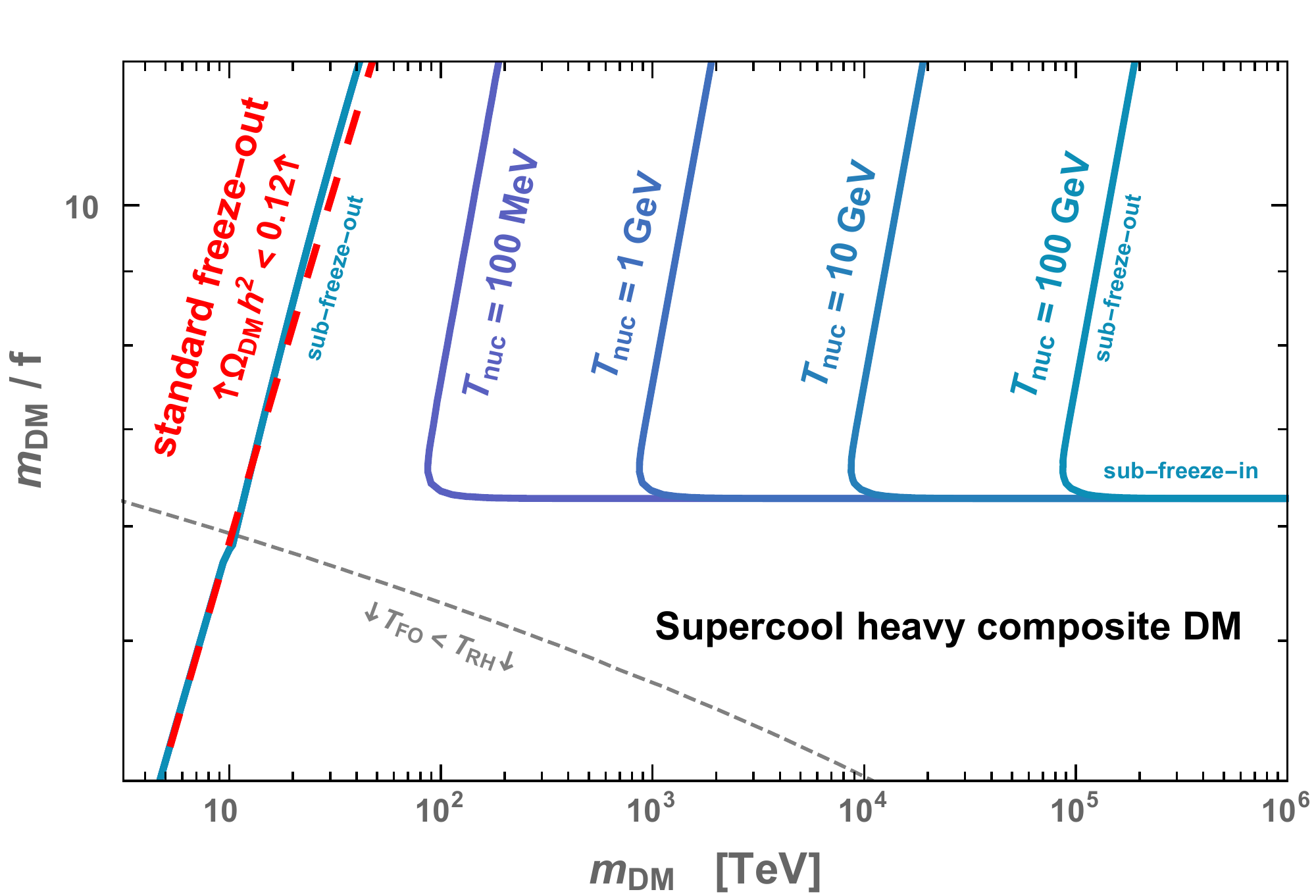}}}
\caption{\it \small The abundance of scalar composite DM is reproduced along the red dashed line in the case of standard freeze-out, and along the blue lines in presence of a supercooled phase-transition, for different values of the nucleation temperature $\Tnuc$. See Eq.~\eqref{eq:sub-thermal-bottom-right} and Eq.~\eqref{eq:sub-thermal-top-left} for a description of the sub-freeze-out and sub-freeze-in regimes.
See Fig.~\ref{fig:DM_SC_abund} for analogous figures with experimental limits and sensitivities superimposed. This Figure has been drawn for $Z=5$ (defined in Sec.~\ref{sec:dilaton_Teq0}) and BR~$=10^{-6}$ (defined in Eq.~(\ref{eq:DIS})). 
If the reheating temperature is smaller than the freeze-out one $T_{\rm RH} \lesssim T_{\rm FO}$, then the phase transition results in a slightly different DM abundance than the standard case even for vanishing supercooling (large $\Tnuc$), cf. Eq.~\eqref{eq:sub-thermal-bottom-right}, so that the red and the leftmost blue lines do not overlap. }
\label{fig:sketch_intro} 
\end{figure}

In the present work we build upon the results of~\cite{Baldes:2020kam} and apply them to a concrete model of composite Dark Matter which interacts with the SM through the dilaton portal, thus generalizing, extending and updating former studies in Ref.~\cite{Bai:2009ms,Blum:2014jca,Efrati:2014aea,Kim:2016jbz,Fuks:2020tam}. We offer a preview of our result for the DM abundance in Figure~\ref{fig:sketch_intro}, where one can appreciate that a wide parameter space opens up for DM heavier than what predicted by standard freeze-out.

This paper is organised as follows.
In Sec.~\ref{sec:nearly-conformal-sector} we introduce a light dilaton, which controls the physics of the phase transition, and calculate the associated nucleation temperature.
In Sec.~\ref{sec:DM in Composite Higgs} we assume that the dilaton mediates the interactions between DM and the SM, which allows us to make precise statements regarding the DM abundance and, in Sec.~\ref{sec:SupercoolDM}, to unify its treatment with that of the PT. It also allows us to explore the complementarity among the myriad experimental signatures of this scenario, which were not addressed in~\cite{Baldes:2020kam}: we explore DM signals in Secs~\ref{sec:IndirectDetection},~\ref{sec:DirectDetection} and GW ones in Sec.~\ref{sec:SGWB  prediction}.
We conclude in Sec.~\ref{sec:outlook}, and complement this paper with a series of technical results in the various Appendices.

\section{Supercooling from a nearly-conformal sector}
\label{sec:nearly-conformal-sector}

\paragraph{Spontaneous breaking of scale invariance.}
We assume that, at high-energies, the theory is described by a strongly-interacting conformal field theory (CFT) plus a small explicit breaking of scale invariance, which we parameterise with the slightly relevant operator $\epsilon \, \mathcal{O}_\epsilon\,,$ with scaling dimension $d = 4 + \gamma_\epsilon \lesssim 4$ and $\epsilon \ll 1$ in the UV.
We assume that $\epsilon$ grows at lower energies, until the CFT is  spontaneously broken and a confinement scale $f$ is generated.
Similarly to QCD, we expect this to give rise to a tower of resonances of mass of order $f$.

\paragraph{EFT described by the dilaton.}
We further assume that the dilaton $\chi$, the pseudo-Goldstone boson associated to the spontaneous breaking of scale invariance, is lighter than these resonances. It is then possible to integrate them out and describe the phase transition in terms of the dilaton field only.
A model building effort is required to obtain a parametrically light dilaton in strongly-coupled theories~\cite{Contino:2010unpub,Appelquist:2010gy,Bellazzini:2013fga,Coradeschi:2013gda,Chacko:2013dra,Megias:2014iwa,Megias:2016jcw}. In this paper we will not make progress in this direction, and we will limit ourselves to dilaton masses $m_\sigma \gtrsim f/5$.

\paragraph{Connection to the lattice.}
On the lattice side, evidence for a light dilaton with a mass $m_{\sigma} \simeq 0.5 \, m_\rho$ where $\rho$ is the lightest vector meson\footnote{If the scalar $\sigma$ is indeed the dilaton, we expect $m_{\sigma}/m_\rho$ to go to zero in the large volume limit \cite{Gorbenko:2018ncu}.} have been shown in $SU(3)$ gauge theory with $N_f = 8$ massless Dirac fermions in the fundamental representation \cite{Appelquist:2018yqe} or with $N_f = 2$ massless Dirac fermions in the sextet (two index symmetric) representation \cite{Fodor:2019vmw,Fodor:2020niv}. In that context, see the studies \cite{Appelquist:2017wcg,Appelquist:2017vyy,Appelquist:2020bqj} where free parameters of some dilaton EFT are adjusted on the results of the lattice simulations.

\subsection{The dilaton potential and interactions}
\label{sec:dilaton_Teq0}

\paragraph{pNGB of scale invariance.}
The dilaton is defined by its transformation rule, $\chi'(x) \to \lambda \chi' (x/\lambda)$, under dilatations $x_\mu \to x_\mu/\lambda$ .
The prime index is to account for the option that the kinetic term of $\chi'$ is non-canonically normalised, a possibility which turns out to have phenomenological relevance, and which we therefore treat in some detail.
A non-canonically normalised kinetic term is for example a well known prediction of 5D duals of our 4D CFT picture (see e.g.~\cite{Creminelli:2001th,Randall:2006py,Rattazzi:2003ea,Nardini:2007me,Bellazzini:2012vz}). We review such duality in App.~\ref{app:5D} and stick to a purely 4D description in the rest of the exposition.

Scale invariance allows one to write a potential term of the form $\chi'^4$, which is either unbounded from below or implies $\langle \chi'\rangle = 0$, with no pNGB in the spectrum. The slightly relevant coupling $\epsilon$, which comes from a small explicit breaking of the scale invariance, generates an additional potential for the dilaton, $\epsilon(\chi') \chi'^{4+\gamma_\epsilon}$, that in turn can lead to spontaneous symmetry breaking $\langle \chi'\rangle \neq 0$.
We then parametrize the dilaton field as
	\beq
	\chi'(x) = f e^{\frac{\sigma'(x)}{f}}
	\eeq
where $\sigma'(x)$ transforms non-linearly under dilatations, $\sigma'(x) \to \sigma'(\lambda x) + f \log\lambda$.
Assuming $\mathcal{O}_\epsilon$ stays close-to-marginal till confinement\footnote{
This does not happen in QCD, where $|\beta_{\alpha_s}| \sim O(1)$ at confinement, and where indeed there is no evidence for a light dilaton.
See~\cite{Bellazzini:2013fga,Coradeschi:2013gda} for models that achieve $|\beta_\epsilon| \ll 1$ (or $|\gamma_\epsilon| \ll 1$, equivalent for our discussion).}, $|\gamma_\epsilon| \ll 1$, one has $\partial \epsilon/\partial \log \mu \simeq \gamma_\epsilon \,\epsilon$, where $\mu$ is the renormalization scale. As anticipated, $\epsilon$ grows in the IR\footnote{
Subleading terms in the evolution of $\epsilon$ can be parametrized as $\partial \epsilon/\partial \log \mu \simeq \gamma_\epsilon + c_\epsilon\,\epsilon^2/g_\chi^2$, where $c_\epsilon$ is a small parameter that e.g. in CHMs receive contributions from loops of partially-elementary SM fields.
The new term tames the growth of $\epsilon$ in the IR, causing it to flow to the IR fixed point $\epsilon_\text{IR} = -\gamma_\epsilon g_\chi^2/c_\epsilon$.
Subleading terms do not play a crucial role in our discussion and we do not explore them further here, the interested reader can find more details in~\cite{Bruggisser:2018mrt}.} until scale invariance gets spontaneously broken.
We parameterise the dilaton potential as
	\beq
	V(\chi') = c_\chi g_\chi^2 \chi'^4 \Big[1 - \frac{1}{1 + \gamma_\epsilon/4} \Big(\frac{\chi'}{f}\Big)^{\gamma_\epsilon} \Big]\,,
	\eeq
where $c_\chi$ is an order one parameter and $g_\chi \gtrsim 1$ is a coupling, about which we give more details in App.~\ref{app:5D}. We omit for simplicity a constant term ensuring the smallness of the cosmological constant. This potential implies $\langle \chi'\rangle = f$.

\paragraph{Dilaton EFT.}
The field $\chi'$ is responsible for the generation of the masses $m_\psi$ of the composite resonances $\psi$ (and via the Higgs also of the SM fields, in CHMs). We conveniently employ  $\chi'$ as a compensator for any non-marginal operator in the Lagrangian~\cite{Goldberger:2008zz}. All in all, accounting for the possibility of a non-canonical normalisation $Z \neq 1$ of the kinetic term ($Z^2 = 24$ in the 5D dual reviewed in App.~\ref{app:5D}), the Lagrangian for $\chi'$ reads
\beq
\mathcal{L}_\chi'
= \frac{Z^2}{2} (\partial_\mu \chi')^2
- V(\chi')
- y_\psi \bar{\psi} \psi \chi',
\label{eq:chiprimelag}
\eeq
where we have introduced the coupling $y_\psi$ such that
\beq
m_\psi = y_\psi f .
\eeq

\paragraph{Canonically normalized dilaton.}
\label{par:Znormalization}
We define the canonically normalised compensator field $\chi$ and the dilaton field $\sigma$ as
\begin{equation}
\label{eq:def_sigma}
\chi \equiv Z \chi',  \qquad \textrm{and} \qquad \sigma \equiv  \chi - Z f\,.
\end{equation}
This leads to the potential
\beq
V(\chi) = \frac{c_\chi}{Z^4} g_\chi^2 \chi^4 \Big[1 - \frac{1}{1+\gamma_\epsilon/4} \Big(\frac{\chi}{Z f}\Big)^{\gamma_\epsilon} \Big]\,,
\label{eq:dilaton_zero_T_pot}
\eeq
with a minimum at $\langle\chi \rangle  = Z\, f$. The dilaton interactions read
\beq
y_\psi  \bar{\psi} \psi \frac{\chi}{Z} = m_\psi \bar{\psi} \psi  \left(1+\frac{\sigma}{Z f}\right),
\eeq
from which it is manifest that the interactions of $\sigma$ are suppressed by a factor $Z$ with respect to the naive expectation $m_\psi/f$.\footnote{Had we reabsorbed $Z$ in the definition of $f$, as $f' = Z\,f$, then $Z$ would have appeared in the definition of the masses in terms of $f'$, and the physics would have of course not changed.
}
Since the value of $Z$ sizeably impacts the phenomenology presented in this paper, in the main text we will show results for different values of this parameter.

\paragraph{Dilaton mass.}
Remembering $\gamma_\epsilon < 0$, we obtain for the dilaton mass 
\beq
m_\sigma^2
= \frac{\partial^2}{\partial\sigma^2} V\big(\chi(\sigma)\big)\big|_{\sigma = 0}
= - 4\,\gamma_\epsilon\, c_\chi g_\chi^2 \, \frac{f^2}{Z^2}.
\label{eq:msigma}
\eeq
The smallness of $m_\sigma$ is then tied to the smallness of the anomalous dimension $\gamma_\epsilon$ at the confinement scale.
Note that in the limit where $|\gamma_{\epsilon}| \ll 1$, the dilaton potential at zero-temperature in Eq.~\eqref{eq:dilaton_zero_T_pot} reduces to the Coleman-Weinberg potential~\cite{Coleman:1973jx}
\begin{equation}
V(\dilaton)   \simeq  \frac{m_{\sigma}^2}{ Z^2 f^2}\frac{\chi^4}{4}  \, \log\left( \frac{\chi}{ e^{1/4} Z f} \right).
\label{eq:dilaton_potential_taylor_cw}
\end{equation}

\paragraph{Vacuum energy.}
Finally, the vacuum energy in the unbroken phase reads
\beq
\Delta V  = V(\chi = 0) -  V(\chi = Z f)
\simeq \left( \frac{ Z m_\sigma f }{ 4 } \right)^2.
\label{eq:vacuum_energy}
\eeq
This defines the quantity $c_{\rm vac}$ in \cite{Baldes:2020kam}.
If the Higgs is also a condensate of the same strong sector, as in CHM, then one has  $\Delta V= (Z m_\sigma f/4)^2 + (m_h v_{\rm H})^2/8$, where $m_h = 125$~GeV is the physical Higgs mass and $v_{\rm H} \simeq 246$~GeV its vev.
Given the experimental limits on $f$ and the dilaton masses we consider, the Higgs contribution is always negligible with respect to the one of Eq.~(\ref{eq:vacuum_energy}).

\paragraph{Consequences of non-canonical kinetic term.}
To summarise, a non-canonical kinetic term for the dilaton has the effect of
\begin{itemize}
\item suppressing its physical mass,
\item suppressing its couplings,
\item enhancing the difference in potential energy between the false and the true vacuum.
\end{itemize}

\subsection{Finite-temperature corrections}
\label{sec:Finite-temperature-corrections}

\paragraph{Deconfined phase.}
In the previous section, we discussed the potential of the dilaton in the confined phase which we expect to be the thermodynamically most favorable phase when the temperature is notably below $g_{\dilaton} \dilatonvev$.\footnote{The appearance of $g_{\dilaton}$ here can be understood by restoring Planck's constant: $[g_{\dilaton}]=\hbar^{-1/2}$ and $[\dilatonvev] = \hbar^{1/2} E$, such that $g_{\dilaton} \dilatonvev $ has the dimension of an energy scale.} However, when the temperature is notably above $g_{\dilaton} \dilatonvev$, we expect the deconfined phase to be thermodynamically most favorable. The EFT contains only the dilaton and the SM, so it is inappropriate to use it to describe the departure from the confined phase to the deconfined phase at $T \sim \mathcal{O}(g_{\dilaton} \dilatonvev)$, since we expect the excitation of heavier bound states as well as their deconfined constituents to play an important role at this temperature and above. Instead of looking for a finer description of the strong sector, which would be very model dependent, we assume that we can approximate the deconfined phase by an $\mathcal{N}=4$ $SU(N)$ super Yang-Mills which in the large $N$ limit is dual to an AdS-Schwarzschild space-time~\cite{Witten:1998qj, Witten:1998zw}. This is known to have free energy~\cite{Creminelli:2001th}\footnote{Note that the free-energy of an interacting gluon gas is lower than the free-energy of a non-interacting gluon gas, $F= -\frac{\pi^2}{90} 2N^2 T^4$, 
by a factor of $\sim 5$, in conformity with the intuition that switching on the strong interaction would stabilize the phase. Note that we also neglect the number of degrees of freedom $\gSM$ of the SM and of the techni-quarks $g_{\mathsmaller{\rm q}}$ compared to the effective number of the techni-gluons $ \frac{45}{4} N^2 $, namely, $-\frac{\pi^2}{90} (\gSM  + g_{\mathsmaller{\rm q}} + \frac{45}{4} N^2 ) T^4 \simeq -\frac{\pi^2}{8} N^2 T^4$.}
	\begin{equation}
	\label{eq:free_energy_CFT}
	F_{\rm dec} = -\frac{\pi^2}{8} N^2 T^4.
	\end{equation}
	
\paragraph{Confined phase at finite-temperature.}
Without a precise UV description, we are unable to determine the potential for $0<\dilaton<T/g_{\dilaton}$. Nevertheless, we follow \cite{Randall:2006py, Bruggisser:2018mrt} and proceed by assuming that the dilaton still approximately exists in this regime, and that its potential can be estimated. With regard to the thermal corrections, we assume that they arise from the $N^2$ bosons of the CFT in the deconfined phase and model them as  \cite{Randall:2006py, Bruggisser:2018mrt}
 	\begin{equation}
 	\label{eq:finiteTempPot}
	V_{\mathsmaller{\rm T}}( \dilaton,T) = \sum_{\rm\mathsmaller{CFT}~ bosons} \frac{n T^4}{2 \pi^2} J_B\left( \frac{m_{\rm \mathsmaller{CFT}}^2}{T^2} \right),
	\end{equation}
where each bosonic state comes with a degeneracy factor $n$, and mass $m_{\mathsmaller{\rm CFT}} = g_{\dilaton}  \dilaton/Z$. 
The thermal function $J_{B}$ is given by
	\begin{equation}
	J_{B}(x) = \int_0^\infty dk \; k^2 \, \log[1-e^{-\sqrt{k^2+x}}].
	\end{equation}
In order to recover the free energy Eq.~\eqref{eq:free_energy_CFT} in the deconfined phase, we set
	\begin{equation}
	\sum_{\mathsmaller{\rm CFT}~\rm bosons} n = \frac{45 N^2}{4}.
	\label{eq:nbr_states_Ncft}
	\end{equation}
\paragraph{Effective potential.}
Combining the above, the finite-temperature effective potential for the dilaton is
	\begin{equation}
	\label{eq:tot_pot_dil}
	V(\dilaton,T) = V(\dilaton) +  V_{\mathsmaller{\rm T}}(\dilaton,T),
	\end{equation}
with $V(\dilaton)$ and $V_{\mathsmaller{\rm T}}(\dilaton,T)$ defined in Eq.~\eqref{eq:dilaton_zero_T_pot} and Eq.~\eqref{eq:finiteTempPot}.
Note that we have neglected the thermal corrections induced by the light degrees of freedom in the confined phase, as these will also be light in the deconfined phase, and irrelevant for our purposes.

\subsection{Computation of the nucleation temperature}
\label{sec:nucleation-temperature}

\begin{figure}[t]
\centering
\hspace{-1cm}
\raisebox{0cm}{\makebox{\includegraphics[height=0.5\textwidth, scale=1]{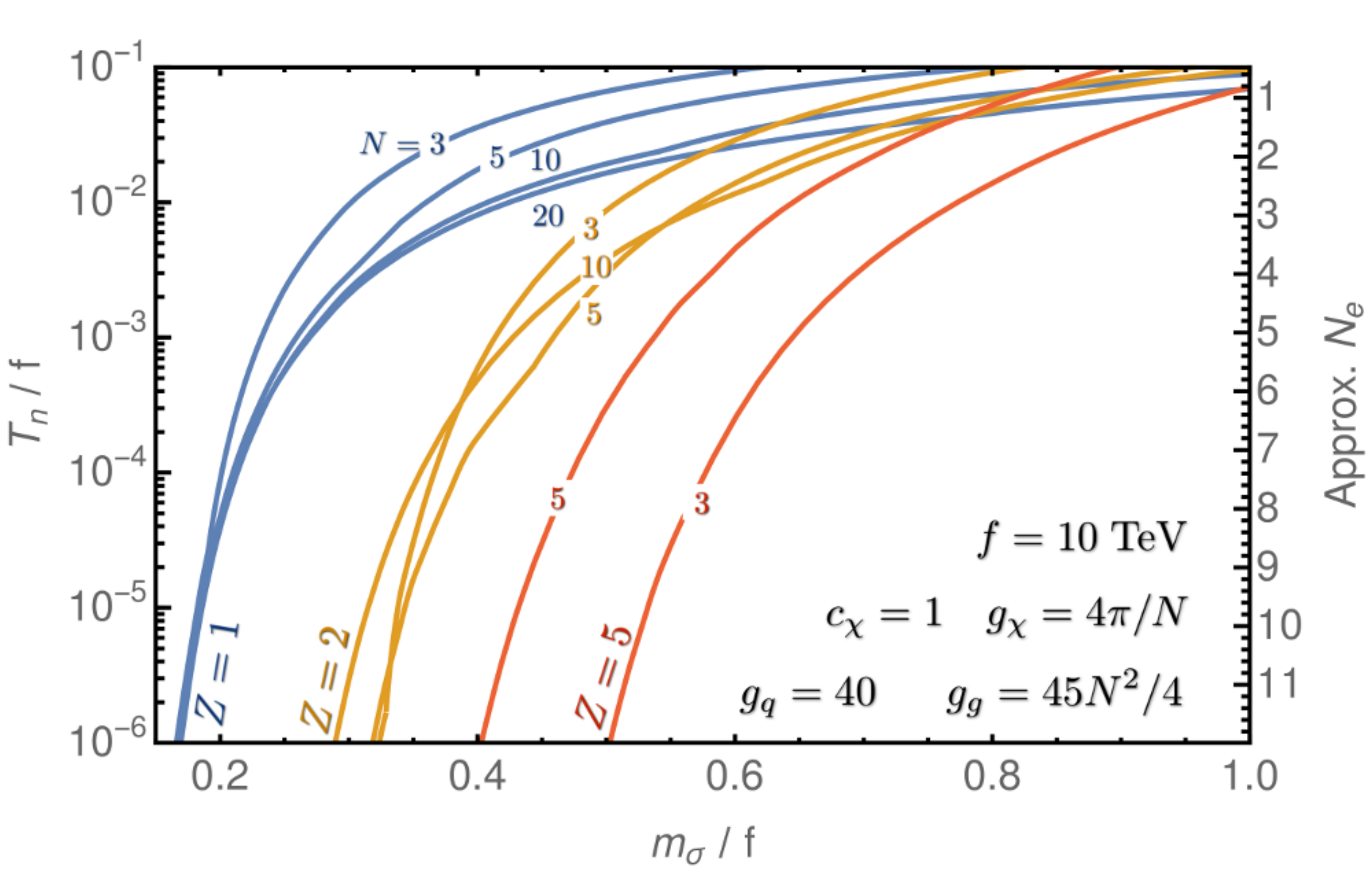}}}
\caption{\it \small Nucleation temperature, solution of Eq.~\eqref{eq:nucTempInstApprox}, after numerically solving the bounce action, as a function of the dilaton mass for different values of the  kinetic rescaling factor, $Z$, and strong sector group $SU(N)$ rank. We see that strong supercooling is associated with a light dilaton. We set $f=10$ TeV,  $c_\chi = 1$, and a dilaton-as-glueball inspired  $g_{\chi}=4\pi/N$. With this scaling of $g_{\chi}$, increasing $N$ leads to an increase in $\gamma_{\epsilon}$, which can counteract the thermal effects of the additional CFT bosons and lead to a weaker transition. For the same reason, large choices of $Z$, $N$, and $m_{\sigma}$ can also invalidate the perturbative analysis, which is why we restrict ourselves to smaller $N$ for large $Z$.
On the right axis we show approximate values for the number of e-foldings $N_e$, where we use Eq.~\eqref{eq:Ne_def} with $m_\sigma = 0.5~f$, $N=5$ and $Z=2$ for definiteness (so the difference between the approximate $N_e$ and the actual $N_e$ is $<1$ for the entire plot).}
\label{fig:IB_supercooling_nuc_temp} 
\end{figure}

\paragraph{Tunneling rate.}
The phase transition, being first order, completes through bubble nucleation via quantum or thermal tunneling. The decay rate of the false vacuum is given by~\cite{Coleman:1977py,Callan:1977pt,Linde:1980tt,Linde:1981zj,Coleman:1985rnk}
\begin{equation}
\Gamma(T) \simeq {\rm max} \left[   T^4\left( \frac{S_3}{2\pi T} \right)^{3/2} {\rm Exp} \left(-S_3/T\right) , \quad \frac{1}{R_c^{4}} \left( \frac{S_4}{2\pi} \right)^2 {\rm Exp} \left(-S_4 \right)  \right],
\label{eq:tunneling_rate}
\end{equation}
where $R_c$ is the bubble radius, and $S_3$ and $S_4$ are the three and four dimensional Euclidean actions of the $O(3)$ and $O(4)$ symmetric tunneling solutions respectively. The former is thermally-induced while the later is purely quantum. The $S_3$ and $S_4$ bounce actions read
\begin{equation}
S_{3}=4\pi\int dr~r^2~\left(\frac{1}{2}\phi^{'}(r)^2+V\left(\phi(r)\right)\right), \qquad S_{4}=2\pi^2\int dr~r^3~\left(\frac{1}{2}\phi^{'}(r)^2+V\left(\phi(r)\right)\right)
\end{equation}
where $V(\phi)$ is the finite temperature effective potential and $\phi(r)$ is the field configuration which interpolates between the two asymptotic vacua. Extremization of the action leads to the Euclidean equation of motion
\begin{equation}
\phi''(r) + \frac{d-1}{r}\phi'(r) = \frac{dV}{d\phi},
\label{eq:bounce_eom}
\end{equation}
with boundary conditions
\begin{equation}
\phi'(0)=0, \qquad \textrm{and} \qquad  \lim_{r \to \infty} \phi(r) = 0.
\label{eq:BC_bounce}
\end{equation}
The phase transition first becomes energetically allowed at the critical temperature,
\begin{equation}
\label{eq:crit_temp_PT}
T_c =  \left(  \frac{Z \,  \msigma \dilatonvev}{\sqrt{2} \pi N}  \right)^{1/2},
\end{equation}
when the two minima are degenerate.
\paragraph{Nucleation.}
We consider the phase transition to take place when the number of bubble nucleations per Hubble volume and per Hubble time is of $\sim \mathcal{O}(1)$\footnote{We refer to App.~\ref{app:Bubble nucleation and percolation temperature} for justification, following a comparison of $T_{\rm nuc}$ to the temperature at which bubbles percolate.}
	\begin{equation}
	\label{eq:nucTempInstApprox}
	\Gamma	(T_{\rm nuc}) \simeq H(T_{\rm nuc})^4.
	\end{equation}
This occurs at an associated $S_{\mathsmaller{\rm crit}}$, typically $\sim 100$ for the present scenario. We determine $T_{\rm nuc}$ in two different ways.
	\begin{itemize}
	\item \underline{Numerical approach.} First of all, we solve the bounce action numerically using an ``undershoot-overshoot'' method. Nucleation temperatures computed numerically, showing the extreme supercooling achievable with light dilaton masses in this model, are shown in Fig.~\ref{fig:IB_supercooling_nuc_temp}.
	\item \underline{Analytical approach.} To aid our understanding, we also derive an analytical expression of the nucleation temperature. Restricting ourselves to the $S_{4}$ action for simplicity, we find
	\begin{equation}
	\label{eq:analyticalTnuc}
	\left(\frac{T_{\rm nuc}}{T_c} \right)^4 \propto \text{Exp}\left( -c_{*}\, \frac{8\, \pi^2}{ S_{\mathsmaller{\rm crit}} } \frac{Z^2 f^2}{ \msigma^2} \right),
	\end{equation}
where $c_{*} \simeq 2$. We refer the reader to App.~\ref{app:O(3)_vs_O4} for the derivation, together with further details on the $O(3)$ bounce and the minimal nucleation temperature, below which the universe remains forever stuck in the false vacuum.
\end{itemize}
From Fig.~\ref{fig:IB_supercooling_nuc_temp}, or Eq.~(\ref{eq:analyticalTnuc}), we see that for small dilaton mass, or equivalently for small anomalous coupling $|\gamma_{\epsilon}|$, the nucleation temperature is exponentially suppressed, thus leading to a large amount of supercooling. The origin of this behaviour can be traced back to the shallowness of the zero-temperature potential at small dilaton field values.

\paragraph{Catalysis through QCD.}
Finally, we note that for extreme supercooling, QCD effects may enter the effective potential~\cite{Witten:1980ez,Iso:2017uuu,vonHarling:2017yew,Baratella:2018pxi,Bloch:2019bvc,Fujikura:2019oyi} and interestingly can trigger the new confinement phase transition. Therefore, there could be some non-trivial connection between the nucleation temperature and QCD effects. The inclusion of these corrections induces a certain degree of model dependence, due to the altered running of the QCD gauge coupling in the presence of the CFT states~\cite{vonHarling:2017yew,Baratella:2018pxi} leading to two QCD phase transitions in the cosmological history and a modified QCD confinement scale at early times. The first confinement of QCD may then occur at a value far below its value today $T_{\rm QCD} \ll 100$ MeV. Note that even once QCD confines, leading to a reduction in the barrier separating the phases, the tunnelling rate can remain suppressed so that supercooling continues to some much lower temperature~\cite{Iso:2017uuu,vonHarling:2017yew}. We will not attempt to capture these effects here, but we keep in mind the possibility that our determination of $T_{\rm nuc}$ may be altered, and also that there may be an underlying reason for $T_{\rm nuc}$ to be related to the QCD scale (that has interesting phenomenological implications, such as enabling cold baryogenesis \cite{Konstandin:2011ds} and the use of strong CP violation for baryogenesis \cite{Servant:2014bla}).

\section{Dilaton-mediated composite DM}
\label{sec:DM in Composite Higgs}

\subsection{DM candidates in Composite Higgs}
\label{sec:DM-Candidates}
We consider DM, which we denote $\eta$, as being a resonance of the composite sector communicating to the SM through the dilaton portal~\cite{Bai:2009ms,Blum:2014jca,Efrati:2014aea,Kim:2016jbz,Fuks:2020tam}. 
Numerous other proposals of DM as a resonance of a composite sector exist in the literature, and we refer the interested reader to the review \cite{Kribs:2016cew}. In this paper, we focus on the case where DM interacts with the SM through the dilaton only.
This is a consistent assumption if we do not consider processes at energy scales larger than the confining one. Therefore it is fully justified for the direct detection and dilaton collider phenomenology that we will study later. Coming to the annihilation cross sections, the relevant energies are of the order of the DM mass, and we comment on the validity of our assumption in Sec.~\ref{sec:unitary-bound}.


We define here all the five models that we will study throughout our paper.
To keep the material easy to read, when presenting summary plots of the phenomenological results we will restrict ourselves to two of the models in the main text: scalar DM, and pNGB DM with shift symmetry broken by the top quark. The phenomenologies of the other three models (fermion DM, vector DM, and pNGB DM with shift symmetry broken by bottom quark) are presented in App.~\ref{app:more_pheno_plots}.

\begin{itemize}

\item \underline{{\bf Scalar DM.}}
The coupling of scalar DM with the dilaton $\dilaton$ can be found treating as usual $\dilaton' = \chi/Z$ as a conformal compensator, and reads
	\begin{equation}
	-\frac{\dilaton^2}{Z^2 \dilatonvev^2}  \frac{1}{2} \MDM^2 \, \eta^2\,.
	\end{equation}
This gives the correct mass term after the confining phase transition.
Then, the coupling between the DM and the excitation of the dilaton $\sigma$, defined in Eq.~(\ref{eq:def_sigma}), is
	 \begin{equation}
	\mathcal{L}_\DM \supset - \left(2\frac{\sigma}{Z\dilatonvev}+\frac{\sigma^2}{Z^2\dilatonvev^2}\right)\frac{1}{2}\MDM^2 \, \eta^2. \label{eq:scalar_DM_dilaton}
	\end{equation}

\item \underline{{\bf Fermion DM.}} Analogously, for the (Dirac) fermionic case we have
	\begin{align}
	\mathcal{L}_\DM \supset -  \frac{\sigma}{Z\dilatonvev} \MDM \,\bar{\eta}\eta. \label{eq:fermion_DM_dilaton}
	\end{align}

\item \underline{{\bf Vector DM.}} 
Similarly, for the vector case we have
	\begin{equation}
	\mathcal{L}_\DM \supset  \left(2\frac{\sigma}{Z\dilatonvev}+\frac{\sigma^2}{Z^2\dilatonvev^2}\right)\frac{1}{2}m_{\DM}^2 \, \eta_\mu\eta^\mu.\label{eq:vector_DM_dilaton}
	\end{equation}

\item{ \underline{{\bf pNGB DM.}}} Finally, another possibility for the DM is to be a pNGB of a spontaneously broken global symmetry $\mathcal{G}$. The main difference with the previous cases is the additional Higgs portal.

\textbf{Dilaton-mediated channel.}
In presence of a dilaton, the DM shifts under a $\mathcal{G}$ transformation $\omega$ as $\eta \to \eta + \omega \chi' = \eta + \omega \chi/Z$.
The genuine dynamical degree of freedom, with kinetic term invariant under $\mathcal{G}$, is then $\eta / \dilaton' = Z\eta/\chi$ \cite{Bruggisser:2018mrt}.
By defining $V_\eta$ as the potential generated by the sources of small explicit breaking of $\mathcal{G}$, which we do not specify here, we then find that the operators invariant under both scale and $\mathcal{G}$ transformations are (also see~\cite{Chacko:2012sy,Kim:2016jbz})
	\begin{equation}
	\label{eq:L_DMpNGB_dilaton}
	\mathcal{L}_{\eta\dilaton}
	= \frac{\dilaton^2}{2\,Z^2} \left( \partial \frac{Z\eta}{\dilaton}  \right)^2 + \frac{\chi^4}{(Zf)^4} V_\eta\left(\frac{Zf\,\eta}{\chi}\right).
	\end{equation}
The couplings between pNGB DM and the dilaton, at first order in the excitation $\sigma$ defined in Eq.~(\ref{eq:def_sigma}), then read
	\beq
	\mathcal{L}_\DM \supset
	\frac{\sigma}{Z\,f} \Big((\partial \eta)^2 - 2 \MDM^2 \eta^2 \Big) + \frac{\sigma^2  \eta^2}{2(Z\,f)^2} \Big(m_{\sigma}^2 -\MDM^2 \Big),\, \label{eq:derivative_coupling_dilaton_pNGB}
	\eeq
where we have used integration by parts and the equations of motion for both $\eta$ and $\sigma$.
Note that the derivative origin of the dilaton coupling to pNGB DM results into different coefficients of the dilaton linear and quadratic interactions with DM, with respect to the case of DM as a scalar resonance Eq.~\eqref{eq:scalar_DM_dilaton}.

\textbf{Composite Higgs \& DM.}
The couplings of pNGB DM with SM particles are model dependent. 
While the possibility of DM being a pNGB is interesting irrespective of whether the Higgs boson is also a pNGB or not, here, for concreteness, we will consider the case where the Higgs is also a pNGB. We further assume that the Higgs and DM arise from the spontaneous breaking of the same compact group $\mathcal{G} \to \mathcal{H}$.
Minimal composite Higgs models based on the coset $\mathcal{G} / \mathcal{H} = SO(5)/SO(4)$ contain just enough broken generators to accommodate the Higgs complex doublet. However, larger cosets can offer additional room for having DM among one of its broken generators. The next-to-minimal coset $SO(6)/SO(5)$ with its five pNGBs naturally contains an additional scalar singlet \cite{Gripaios:2009pe}.
However, the requisite of DM stability imposes an additional $\mathcal{Z}_2$ or $U(1)$, so that the minimal cosets containing DM and at the same time justifying its stability are $O(6)/O(5)$~\cite{Frigerio:2012uc, Marzocca:2014msa} or $SO(7)/SO(6)$~\cite{Balkin:2017aep,Balkin:2018tma,DaRold:2019ccj}.\footnote{Many other examples have been studied in the literature, e.g $SO(6)/SO(5)$ \cite{Fonseca:2015gva,Bruggisser:2016ixa,Chala:2018qdf,Ramos:2019qqa}, $SO(7) \rightarrow G2$ in \cite{Ballesteros:2017xeg}, $SU(4)^2/SU(4)$ in \cite{Ma:2017vzm}, $SU(4)/Sp(4)$, $SU(5)/SO(5)$ and $SU(4)^2/SU(4)$ in \cite{Cacciapaglia:2018avr}, $SU(3) / SU(2)\times U(1)$ and $SU(2)^2 \times U(1) / SU(2) \times U(1)$ in \cite{Carmona:2015haa}, $SO(6)/SO(5)$, $SO(7)/SO(6)$, $SO(7)/G2$, $SO(5)\times U(1)/SO(4)$ and $SO(6)/SO(4)$ in \cite{Chala:2018qdf}, little Higgs $SU(5)/SO(5)$ in \cite{Balkin:2017yns}, $SU(4)\times SU(2) \times U(1) / Sp(4) \times U(1)\times \mathcal{Z}_2$ in \cite{Alanne:2018xli}, $SU(6)/SO(6)$ in \cite{Cacciapaglia:2019ixa}, two-Higgs doublet models $SO(6)/SO(4)\times SO(2)$ in \cite{Mrazek:2011iu, Fonseca:2015gva,DeCurtis:2018iqd,DeCurtis:2018zvh} or  $SO(7)/SO(5)\times SO(2)$ in \cite{Davoli:2019tpx}. We refer to \cite{Bellazzini:2014yua} for a review on existing composite Higgs models.}
Note further that we have made the simplifying assumption that the $f$ of the composite Higgs and the one of the scale breaking coincide, while they really do only up to the ratio of two O(1) couplings, see e.g.~\cite{Bruggisser:2018mus}.

\textbf{Higgs-mediated channel and quark contact interactions.}
An intrinsic property of pNGB DM is its derivative coupling with the Higgs, which is entirely fixed by the choice of the coset. A second feature is the DM-Higgs marginal mixing $\lambda_{h\eta}$, which depends on the incomplete representation of the global symmetry $\mathcal{G}$ in which the third generation of quarks is embedded~\cite{Balkin:2018tma}. We focus on the case where the symmetry breaking pattern is $\mathcal{G} / \mathcal{H} =O(6)/O(5)$ so that \cite{Frigerio:2012uc} 
		\begin{multline}
	\mathcal{L}_{\rm DM}  \supset \frac{1}{2}(\partial_\mu \eta)^2 - \frac{1}{2} \mu_{\eta}^2 \eta^2 -  \lambda_{h\eta} \left|H\right|^2 \eta^2 \\+ \frac{1}{2(Zf)^2}\left(\partial_\mu \left|H\right|^2 + \frac{1}{2}\partial_\mu \eta^2  \right)^2 + \frac{\eta^2}{(Zf)^2}\left( c_t\, y_t \, \bar{q}_{\rm L} \tilde{H} t_{\rm R} + c_b \, y_b\, \bar{q}_{\rm L} H t_{\rm R}   + {\rm h.c.}\right),
	\label{eq:pNGBDM_Higgs}
	\end{multline}
where $v_{\rm H} \equiv \sqrt{2} \left<H\right> = 246~$GeV, $\MDM^2 = \mu_\eta^2 + \lambda_{h\eta}v_{\rm H}^2$, and we have neglected terms suppressed by $v_{\rm H}^2/f^2$. The corresponding Feynman rules are given in table 2 of \cite{Frigerio:2012uc}.

Two options now present themselves:

\begin{itemize}
\item \textbf{DM shift symmetry broken by top quark.}
The largest breaking of the shift symmetry $U(1)_{h}$ protecting the Higgs mass is given by the top contribution. Having the top in some representation that explicitly breaks $\mathcal{G}$ is necessary to give mass to the Higgs, otherwise these models would not be viable. It is model dependent whether the top also breaks the shift symmetry $U(1)_\eta$ protecting the mass of DM. If it does, then one expects the DM mass and the DM-Higgs quartic coupling to be, respectively, (see e.g.~\cite{Marzocca:2014msa}) 
	\begin{equation}
	 m_{\DM} \lesssim \frac{m_h}{v_{\rm H}}  f \simeq 0.5\, f
	\end{equation}
	if the contribution is not tuned to be small, and
	\begin{equation}
	 \lambda_{h\eta} \simeq \lambda_h/ 2 \simeq m_h^2/4v_{\rm H}^2 \simeq  0.065. \label{eq:top_breaking_lambda}
	\end{equation}
For definiteness, we fix  $c_b=c_t = 1/2$. 

\item \textbf{DM shift symmetry broken by bottom quark.}
The DM-Higgs mixing can be lowered and the direct detection constraints weakened if the DM shift symmetry $U(1)_\eta$ is only broken at the bottom level~\cite{Balkin:2018tma}
	\begin{equation}
	\lambda_{h\eta} \sim y_b^2 \sim 10^{-4}.\label{eq:bottom_breaking_lambda}
	\end{equation}
In that case we fix $c_t =0$ \cite{Frigerio:2012uc}, while keeping $c_b =1/2$.
\end{itemize}

\end{itemize}

\subsection{Dilaton-mediated interactions between DM and SM}
\label{sec:DM-SM-interact}

\paragraph{Connection to the hierarchy problem.}
In order to describe the dilaton interactions with the SM, we assume that the sector that breaks scale invariance is not secluded from the electroweak sector. In other words, the Higgs originates from the same sector that breaks scale invariance, as we have already assumed when writing Eq.~(\ref{eq:pNGBDM_Higgs}) for the Higgs coupling of pNGB~DM.

Therefore, our description automatically covers the usual composite Higgs models, and is thus of interest if one cares about them as a solution of the hierarchy problem of the Fermi scale~\cite{Contino:2010rs,Panico:2015jxa}.
In that case, the interesting values of $f$ --- which are still allowed by experiments~\cite{Panico:2015jxa} --- lie as close as possible to a TeV. (Or just above a TeV, if the composite sector solves the big hierarchy problem but the little hierarchy is taken care of otherwise, e.g.~by a composite twin Higgs mechanism~\cite{Geller:2014kta,Barbieri:2015lqa,Low:2015nqa}.)
With this in mind, we will display our phenomenological results for the standard thermal freezeout DM in a zoomed-in region of parameter space, where $f$ is not too far from the TeV scale. In turn, this will allow us to connect with past studies of dilaton-mediated DM~\cite{Bai:2009ms,Blum:2014jca,Efrati:2014aea,Kim:2016jbz,Fuks:2020tam}, improving them with our more refined treatment (which e.g. includes the dilaton wave-function renormalization) and updating them by the use of more recent experimental results. 

Nevertheless, our description is not necessarily tied to a natural solution of the hierarchy problem, even within our assumption that the Higgs sector communicates with the one that breaks scale invariance. Large values of $f$ could for example be motivated by other reasons, such as solutions of the strong CP problem. As we will see below, our supercooling mechanism for DM will allow for much larger values of $f$ than the standard freeze-out, so we will extend our plots accordingly when applicable.

The hierarchy problem would then need to be addressed by some other mechanism, so that values of $f$ much larger than a TeV could be considered. An example of such mechanism is cosmological relaxation of the electroweak scale~\cite{Graham:2015cka}. All known successful relaxion proposals require an inflationary stage or additional assumptions about reheating. It was shown in~\cite{Fonseca:2019lmc} that the duration of the inflationary stage can be rather short (also see~\cite{Fonseca:2019ypl,Morgante:2021bks}), and that new physics at scales $f$ up to $\sim 100$~TeV can be made technically natural, via relaxation, with an inflation period of about 10 e-folds.
As it will turn out, the amount of supercooling we will need, in order to reproduce the measured DM abundance, corresponds to $\mathcal{O}$(10) e-folds. It is interesting and non-trivial that this same amount of inflation generated by the supercooled PT could be used for the relaxation mechanism mentioned above, if the Higgs is external to the CFT of the DM sector.\footnote{In the concrete example studied in this paper, the Higgs and its potential do not exist prior to the dilaton gaining a vev, and hence cannot be relaxed using the supercooled PT.} This makes composite heavy DM technically natural up to PeV masses (because in our setup $m_\DM \sim 10 f$), or even heavier if its communication with the Higgs sector is somewhat suppressed.
We leave the exploration to such UV physics to future work.

\medskip
\paragraph{Dilaton portal to SM.}
After electroweak symmetry breaking the dilaton $\dilaton$ couples to the SM fields as~\cite{Csaki:2007ns, Goldberger:2008zz, Chacko:2012sy, Bellazzini:2012vz, Blum:2014jca, Peskin:1995ev}
	\begin{align}
\mathcal{L} \; \supset & \; \; \frac{1}{2} \partial_{\mu} \sigma \partial^{\mu}\sigma - \frac{1}{2} m_{\sigma}^2\sigma^2 -  \frac{5}{6}\frac{m_{\sigma}^2}{Z f}\sigma^3 - \frac{11}{24} \frac{m_{\sigma}^2}{Z^2 f^2}\sigma^4 + \cdots   \notag\\ & \; - \left( \frac{\sigma}{Z \, f} \right) \, \sum_{q} (1+\gamma_q) m_{q}\bar{q}q + \left( \frac{2\sigma}{Z \, f}+\frac{\sigma^2}{Z^2 \, f^2} \right) \left[ m_{W}^2W^{+\mu}W_{\mu}^{-} + \frac{1}{2}m_{Z}^2 Z_{\mu}Z^{\mu} - \frac{1}{2} m_{h}^2 h^2  \right]  \notag \\
& \; + \frac{\alpha_{\rm EM}}{8\pi \, Z \, f}\,c_{\rm EM} \, \sigma \, F_{\mu\nu}F^{\mu\nu} +  \frac{\alpha_{\rm s}}{8\pi \, Z \, f}\, c_{\rm G} \, \sigma \, G_{a\mu\nu}G^{a\mu\nu} \notag\\
& \; + \frac{\sigma}{Zf}\partial_\mu h\partial^\mu h - \frac{\sigma}{Zf} m_h^2 h^2  + \frac{m_{\sigma}^2 - m_h^2}{2(Z\,f)^2}\sigma^2  h^2,
	\label{eq:eff_lag_dilaton}
	\end{align}
where the rescaling of the couplings with $Z$ comes from the assumption of a non-canonical kinetic term for the dilaton, Eq.~\eqref{eq:chiprimelag}. Here, $\gamma_q$ is the anomalous dimension of the fermionic operators, which we assume in what follows to be $\gamma_q\simeq 0$.

\paragraph{Dilaton self-couplings.}
The dilaton self-coupling terms, scaling as $\sigma^3$ and $\sigma^4$, are obtained after Taylor-expanding the dilaton potential in Eq.~\eqref{eq:dilaton_zero_T_pot} in the small $|\gamma_\epsilon|$ limit. 

\paragraph{Couplings to Higgs.}
In order to respect the Goldstone equivalence theorem \cite{Chanowitz:1985hj,Peskin:1995ev}, we have added the dilaton coupling to the Higgs kinetic term in the last line of Eq.~\eqref{eq:eff_lag_dilaton}. This term and the other ones on the same line can be derived analogously to how we derived the dilaton couplings to pNGB DM, see the discussion around Eq.~(\ref{eq:L_DMpNGB_dilaton}), and indeed arise naturally in composite Higgs scenarios~\cite{Chacko:2012sy,Kim:2016jbz,Bruggisser:2018mrt}.

\paragraph{Couplings to gauge bosons.}
Couplings to transverse gauge bosons are induced by trace anomalies and triangle diagrams containing heavy charged fields. We make explicit the coupling to gluons, which plays an important role for direct detection,
	\begin{equation}
	c_G = b_{\rm IR}^{(3)} - b_{\rm UV}^{(3)} + \frac{1}{2}F_{1/2}(x_t).
	\label{eq:def_cG}
	\end{equation}
Here, the function $F_{1/2}$ is the triangle diagram induced by a fermion~\cite{Csaki:2007ns, 	Bellazzini:2012vz, Blum:2014jca}
	\begin{align}
	F_{1/2}(x) = 2x[1+(1-x)f(x)], 
	 \qquad 
	f(x) = \left\{
                \begin{array}{ll}
                 [\sin^{-1}(1/\sqrt{x})]^2 \qquad \qquad \qquad x\geq 1, \\
                 -\frac{1}{4}\left[ \log\left(\frac{1+\sqrt{x-1}}{1-\sqrt{x-1}}\right) - i\pi \right]^2 \quad \; x<1,
                \end{array} 
              \right. 
	\end{align}
where $x_t = 4m_t^2/m_\sigma^2$. The constant $b_{\rm UV}^{(3)}$ is the QCD beta function above the confining scale $\mu \gtrsim f$, so it contains the contributions from both elementary states of the SM and the CFT of the techni sector.  On the other hand, $b_{\rm IR}^{(3)}$ is the QCD beta function below the confining scale $\mu \lesssim f$, so it receives contributions from elementary states of the SM and light composite states, $m_* \lesssim f$, of the techni sector \cite{Chacko:2012sy}. Hence, the term $b_{\rm IR}^{(3)} - b_{\rm UV}^{(3)} $ can be interpreted as the contribution to the QCD beta function due to loops of composite states of the techni sector lighter than $f$ \cite{Bellazzini:2012vz}. In the SM, the QCD $\beta$-function coefficient reads\footnote{$\mu \frac{d}{d\mu} \frac{1}{g_3^2} =  \frac{1}{8\pi^2} \left( \frac{11}{3} N_c - \frac{2}{3}N_f \right)$.} \cite{Caswell:1974gg}
	\begin{equation}
	b_{\rm SM}^{(3)} = \frac{11}{3} N_c - \frac{2}{3}N_f,
	\end{equation}
where $N_c$ and $N_f$ are the number of colours and fermions respectively. 
An appealing feature of composite Higgs models is the possibility to explain the hierarchy in the Yukawa matrix from the mixing between elementary quarks and composite resonances charged under QCD: a small degree of compositeness leading to a small Yukawa coupling~\cite{Kaplan:1991dc, Contino:2004vy, Contino:2006nn}. 
To be conservative, we consider that only the right-handed top and the longitudinal modes of the electroweak vector bosons receives a strong compositeness fraction, and that the CFT contribution to the QCD beta function is minimal \cite{Blum:2014jca}
	\begin{align}
	b_{\rm IR}^{(3)} - b_{\rm UV}^{(3)} = -\frac{1}{3}. \label{eq:assumption_cG}
	\end{align}

\subsection{DM annihilations and unitarity bound}
\label{sec:unitary-bound}

\paragraph{DM annihilation cross-sections.}
The main DM annihilation channels are the annihilation into WW, ZZ as well into dilaton pair $\sigma\sigma$. We compute all the DM to SM annihilation channels and report them in App.~\ref{app:ann_cross-section}. In the limit of large $\MDM$ these are given by 
	\begin{align}
	\label{eq:cross-section-DM-WW}
	&\sigma v_{\mathsmaller{\rm rel}}|_{\mathsmaller{ \eta \eta \rightarrow WW, ZZ}} \simeq c_{W}\, \frac{ 2\pi\alpha}{v_{\mathsmaller{\rm rel}}} \,  \frac{\MDM^2}{ (Zf)^4}, \\ 
	\label{eq:cross-section-DM-ss}
	&\sigma v_{\mathsmaller{\rm rel}}|_{\mathsmaller{ \eta \eta \rightarrow \sigma\sigma}} \simeq  c_{\sigma} \, \frac{2\pi \alpha}{v_{\mathsmaller{\rm rel}}} \, \frac{\MDM^2}{ (Zf)^4} ,
	\end{align}
where 
	\begin{equation} 
	\label{eq:DMannfactors}	
	\bigg(c_W, \, c_\sigma \bigg) = \left(\frac{3}{16 \pi} , \, \frac{1}{16 \pi} \right),
	 \quad	 \left(\frac{1}{16 \pi}, \, \frac{1}{48 \pi} \right), \quad 
	  \left( \frac{3v_{\mathsmaller{\rm rel}}^2}{128 \pi} \left[1+\frac{ \alpha^{2} }{ v_{\mathsmaller{\rm rel}}^2} \right], \, \frac{3\,v_{\mathsmaller{\rm rel}}^2}{128 \pi } \left[1+\frac{ \alpha^{2} }{ v_{\mathsmaller{\rm rel}}^2} \right] \right),
	\end{equation}
for scalar, vector and fermion DM respectively, where $\alpha$ is an effective coupling defined later. In our plots, we instead use the full expression for the DM annihilation cross-section given in App.~\ref{app:ann_cross-section}. In the case of pNGB DM, due to the high number of terms we do not write any analytical formula for the annihilation cross-sections, but instead we only display plots of $\sigma v$ in Fig.~\eqref{fig:sigmav_pNGB}.
Note that the scalar and vector DM models therefore have very similar indirect detection phenomenology, while the fermion one has velocity-suppressed annihilations and hence weaker signals. 

\paragraph{Sommerfeld enhancement.}
\label{par:sommerfeld_enhancement}
The effective coupling $\alpha$ governs the strength of the dilaton-mediated attractive Yukawa potential between the incoming DM wave functions,
	\begin{equation}
	V_Y(r) = -\alpha\, \frac{ e^{-m_{\dilaton} r} }{ r }, 
	\end{equation}
responsible for the Sommerfeld enhancement.  The non-relativistic potential between the DM wave-functions can be computed as the Fourier transform
	\begin{equation}
	V_Y(r) = -\frac{1}{4i\MDM^2} \int \frac{d\vec{k}^3}{(2\pi)^3} \mathcal{W}(\vec{k}) \, e^{-i\vec{k}\vec{r}}
	\end{equation}
of the four-point function amplitude $\mathcal{W}(\vec{k})$ with one-boson mediator exchange~\cite{Petraki:2015hla}, where $\vec{k}$ is the momentum of the exchanged boson.

As shown in App.~\ref{app:NR_potential}, for fermion, scalar and vector resonance DM, we obtain the dilaton-mediated effective coupling
	\begin{equation}
	\alpha = \frac{1}{4\pi}\frac{ \MDM^2 }{ (Z f)^2 }.
	\label{eq:alpha_def}
	\end{equation}
In contrast, in the pNGB DM scenario, the Yukawa potential is both dilaton and Higgs-mediated 
	\begin{equation}
	V_Y(r) = -\alpha_{\sigma}\, \frac{ e^{-m_{\dilaton} r} }{ r } -  \alpha_{h}\, \frac{ e^{-m_{h} r} }{ r }  ,
	\end{equation}
	with 
	\begin{equation}
	\alpha_{\sigma} = \frac{1}{4\pi}\frac{ \MDM^2 }{ (Z f)^2 }, \qquad \quad\alpha_{h} = \frac{1}{16\pi}\frac{\lambda_{h\eta}^2 v_{\rm H}^2 }{  \MDM^2 }.
	\label{eq:alpha_def_pNGB}
	\end{equation}
See App.~\ref{app:NR_potential} for the derivation of Eq.~\eqref{eq:alpha_def_pNGB} and for a discussion on the contribution of the derivative coupling $\eta \partial \eta \partial h $ to the potential.
In the following, we assume that $\alpha_{\sigma} \gg \alpha_{h}$ is always realized such that we can neglect Sommerfeld effects coming from the Higgs channel.

Then, the enhancement of the annihilation cross-section can be obtained from the standard analytical formula for the Sommerfeld enhancement factor, derived from the Hulth\'en potential
	\begin{equation}
	 V_{H} = - \frac{\alpha\, \overline{m}_{\sigma}\,e^{-\overline{m}_{\sigma}r}}{(1-e^{-\overline{m}_{\sigma}r})}, \qquad  \mathrm{where} \qquad  \overline{m}_{\sigma}= \frac{\pi^2}{6} \,m_{\sigma},
	\end{equation}
and which we report here for arbitrary $l$-wave process~\cite{Cassel:2009wt}
\begin{equation}
S_{H}^{(l)} = \left| \frac{\Gamma(a^-)\Gamma(a^+)}{\Gamma(1+l+2i\omega))} \frac{1}{l!}   \right|^2, \label{eq:Sommerfeld_factor_Cassel}
\end{equation}
with $a^{\pm} \equiv 1+ l + i \omega(1 \pm \sqrt{1-x/\omega})$, $x \equiv 2\alpha/v_{\mathsmaller{\rm rel}}$ and $\omega \equiv \MDM v_{\mathsmaller{\rm rel}} / (2 \overline{m}_{\sigma})$.
Explicitly, for s-wave annihilation one finds~\cite{Slatyer:2009vg}
	\begin{equation}
	S_{H}^{(l=0)} = \frac{2\pi\alpha}{v_{\mathsmaller{\rm rel}}} \frac{\sinh (\pi \MDM v_{\mathsmaller{\rm rel}}/\overline{m}_{\sigma})}{\cosh{(\pi\MDM v_{\mathsmaller{\rm rel}} /\overline{m}_{\sigma})} - 		\cosh{ \pi \sqrt{\MDM^2 v_{\mathsmaller{\rm rel}}^2/\overline{m}_{\sigma}^2  -  4\MDM \alpha/\overline{m}_{\sigma}}}}.
	\end{equation}
In Eq.~\eqref{eq:cross-section-DM-WW} and \eqref{eq:cross-section-DM-ss}, we have reported the Sommerfeld enhancement factor in the limit where the range of the dilaton-mediated force $m_{\sigma}^{-1}$ is larger than the de Broglie wavelenght of the incoming wave-packets $(\MDM v_{\mathsmaller{\rm rel}}/2)^{-1}$, itself larger than the typical DM bound-state
\footnote{
We safely neglect the effective contribution, to DM annihilations, of dilaton-mediated formation and decay of bound states, because selection rules imply it is suppressed parametrically by $\alpha^2$ with respect to that of Sommerfeld-enhanced annihilations~\cite{Petraki:2015hla, Petraki:2016cnz}. Dilaton self-couplings are small enough to not affect this suppression~\cite{Oncala:2018bvl}.
  Furthmore, the pNGB coupling to the Higgs doublet in the symmetric EW phase is not of the form $H\eta^{2}$, which could otherwise lead to potentially significant bound state formation via charged scalar emission~\cite{Oncala:2019yvj}.
} 
size $(\MDM\, \alpha/2)^{-1}$
	\begin{equation}
	S_{H}^{(l=0)} \overset{ \MDM v_{\mathsmaller{\rm rel}}/2 \, \gg \, m_{\sigma}}{\xrightarrow{\hspace*{2.5cm}}} \frac{2\pi\alpha}{v_{\mathsmaller{\rm rel}}}
	 \frac{1}{1-e^{2\pi \alpha/	v_{\mathsmaller{\rm rel}}}}  \overset{\alpha \, \gg \, v_{\mathsmaller{\rm rel}} } {\xrightarrow{\hspace*{1.5cm}}}
	  \frac{2\pi\alpha}{v_{\mathsmaller{\rm rel}}} .
	\end{equation}
The full expressions can be easily recovered by not taking the above limits. For the p-wave ($l=1$) annihilation of fermionic DM, we have also included the important higher-order term $1+ \alpha^{2} / v_{\mathsmaller{\rm rel}}^2$ in Eq.~(\ref{eq:DMannfactors}), which lifts the velocity suppression when the Sommerfeld enhancement is large~\cite{Cassel:2009wt, Baldes:2017gzw}.

\paragraph{Unitarity bound.}
The unitarity of the S-matrix, $SS^{\dagger} =1$, leads to an upper bound on each coefficient of the decomposition of the inelastic cross-section into partial waves $J$~\cite{Griest:1989wd,vonHarling:2014kha,Baldes:2017gzw,Smirnov:2019ngs,Gouttenoire:2022gwi}
\begin{equation}
\label{eq:uni_bound}
\sigma_{\mathsmaller{\rm inel} }^{(J)} v_{\mathsmaller{\rm rel}} \leq \sigma_{\mathsmaller{\rm uni} }^{(J)} v_{\mathsmaller{\rm rel}} = \frac{4\pi(2J+1)}{\MDM^2 v_{\mathsmaller{\rm rel}}}.
\end{equation}
Saturating the unitarity bound corresponds to having a maximal probability for the inelastic process, and a vanishing probability for the elastic scattering. The scaling of the DM annihilation cross-section in Eq.~\eqref{eq:cross-section-DM-WW} and Eq.~\eqref{eq:cross-section-DM-ss} implies that unitarity is violated for DM mass larger than $\MDM \gtrsim 2.6\, Z \,f$ for scalar DM, $\MDM \gtrsim 3.1\, Z \,f$ for vector DM, and $ \MDM \gtrsim 3.9 \, Z f$ for fermion DM. This apparent violation indicates that the perturbative methods used to calculate the annihilation cross section have broken down~\cite{Baldes:2017gzw}. We recover the unitarity by using for value of the DM annihilation cross-section, the minimum between the perturbative cross-section in Eq.~\eqref{eq:cross-section-DM-WW} or Eq.~\eqref{eq:cross-section-DM-ss}, and the cross-section which saturates the unitarity bound in Eq.~\eqref{eq:uni_bound}.

\paragraph{Possible breakdown of the EFT.}
The energies involved in DM annihilations are of the order of the DM mass, so that one may worry about the validity of the EFT we used to perform our computations.  
Our predictions are solid if DM, the dilaton and the SM constitute the lightest states, and they are lighter than the EFT cutoff $\sim 4\pi f$.
This is realised in the case of pNGB DM, is not inconceivable in the models of heavier DM, and is the reason why in all Figures we shade in gray the parameter space where $\MDM > 4\pi f$.
If other composite states are light, however, the annihilation cross section could be modified with respect to those that we have computed.
%
In addition, our perturbative calculations become less accurate for values of the couplings, $\MDM/(Z \,f)$, larger than 1, i.e. close to the region of unitarity saturation discussed in the previous paragraph.
For example, either of the two comments above could imply that processes with more than two final states are relevant.
	All this would have a small impact for DM heavier than $O(100)$~TeV, because in that regime the cross section is anyway close to the unitarity limit.
	Note that the intrinsic uncertainties associated to the coupling apply also to models of elementary DM, because their annihilation cross section is also affected by non-perturbative uncertainties in that regime.

\begin{figure}[t!]
\centering
\vspace{-1.5cm}
\raisebox{0cm}{\makebox{\includegraphics[height=0.5\textwidth, scale=1]{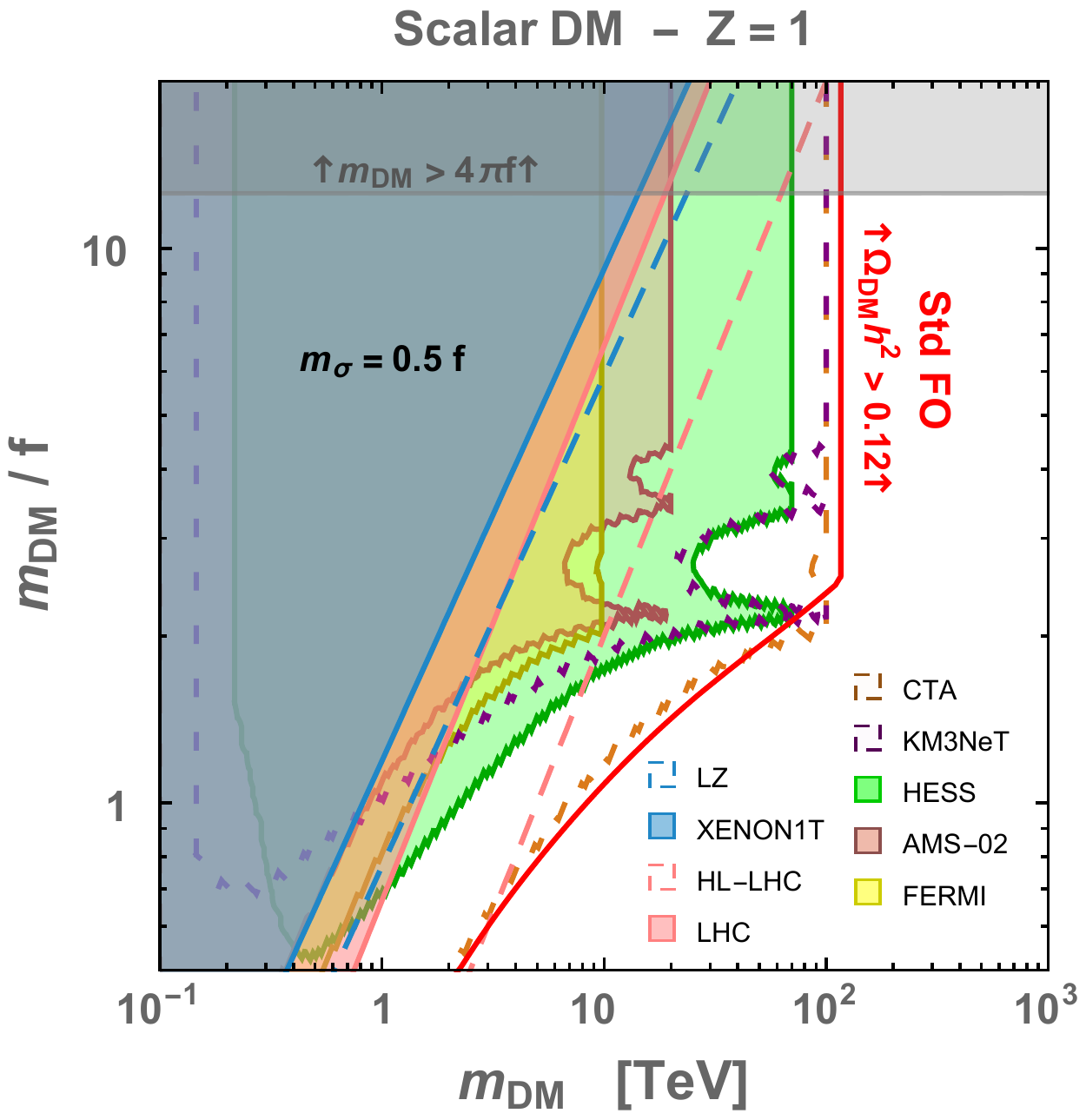}}}
\raisebox{0cm}{\makebox{\includegraphics[height=0.5\textwidth, scale=1]{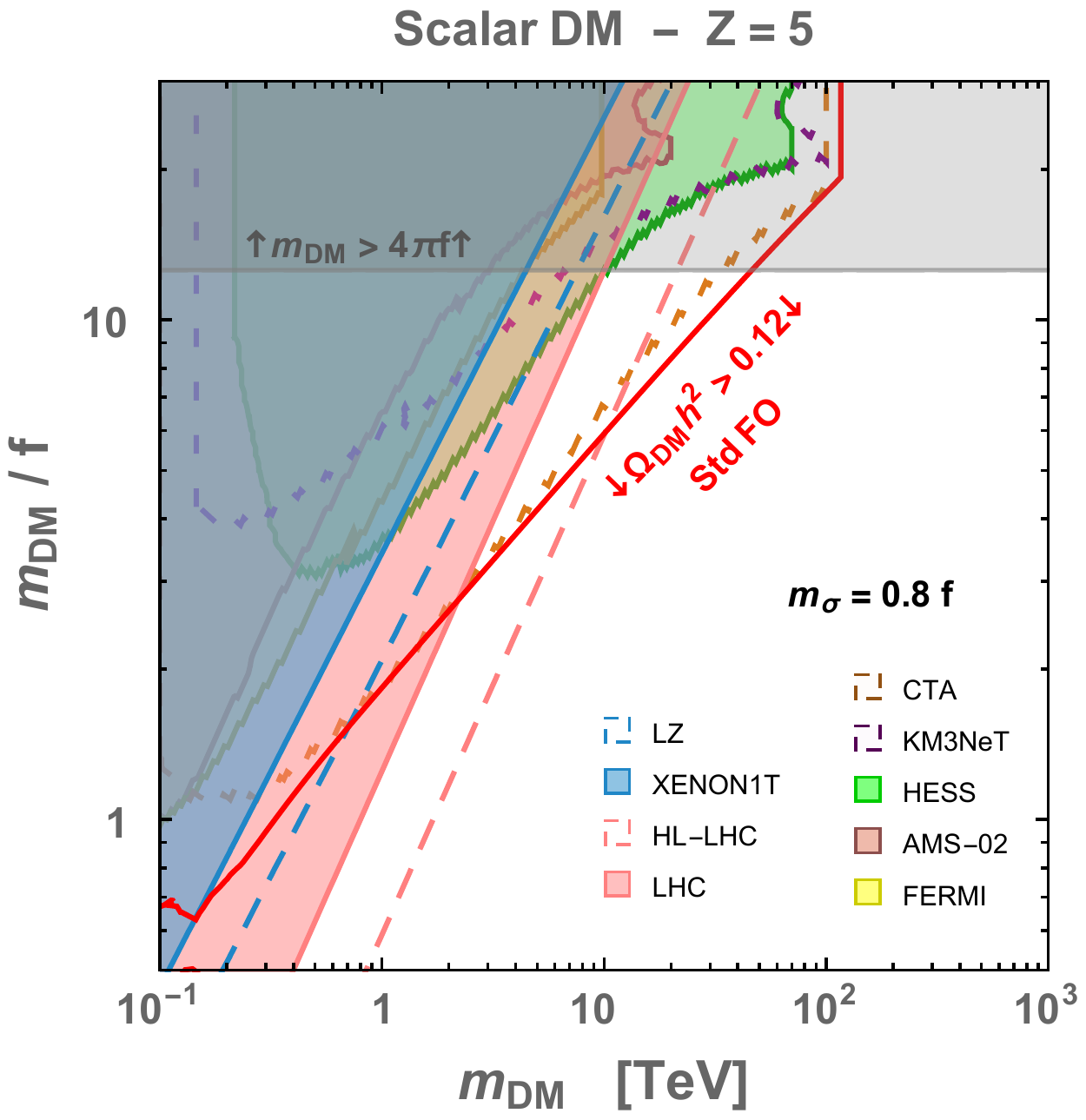}}}
\caption{\it \small \textbf{Dilaton-mediated scalar DM in the standard freeze-out case}. On the left (right) we set the dilaton normalization strength to $Z=1$ ($Z=5$), cf. Sec.~\ref{par:Znormalization}, and the dilaton mass to $m_\sigma = 0.5 f$ ($m_\sigma = 0.8 f$) to avoid significant supercooling, see Fig.~\ref{fig:IB_supercooling_nuc_temp}.
The red lines show the points leading to the correct DM abundance assuming the standard freeze-out paradigm, cf. Sec.~\ref{sec:DM_abundance_freeze-out}.
As $\MDM$ increases, the required coupling strength, controlled by $\MDM/f$, also increases to keep the annihilation cross section at the thermal freeze-out value. We impose the DM annihilation cross-section to be smaller than the unitarity bound in Eq.~\eqref{eq:uni_bound}, leading to a maximal DM mass of order $\sim 100 \; \mathrm{TeV}$.
In yellow, red plum and green we show the current indirect-detection constraints from FERMI, AMS-02 and HESS, cf. Sec~\ref{sec:IndirectDetection}. We add the prospects from KM3NeT and CTA in dashed purple and dashed orange. The resonances in the ID regions are due to the non-perturbative Sommerfeld effect, cf. Sec.~\ref{par:sommerfeld_enhancement}. 
In blue, we show direct-detection constraints from the current XENON-1T and the future LZ, cf. Sec.~\ref{sec:DirectDetection}. All (in)direct-detection constraints in this and subsequent plots assume our DM matches the observed relic abundance. 
In gray we also show a sketch of the current constraints from LHC and of the prospects at HL-LHC, cf. Sec.~\ref{sec:colliders}. The fermion and vector cases are shown in Fig.~\ref{fig:stdFO_pheno_app} of App.~\ref{app:more_pheno_plots}.
}
\label{fig:stdFO_pheno_msigma_p25a} 
$\quad$ \\
\raisebox{0cm}{\makebox{\includegraphics[height=0.5\textwidth, scale=1]{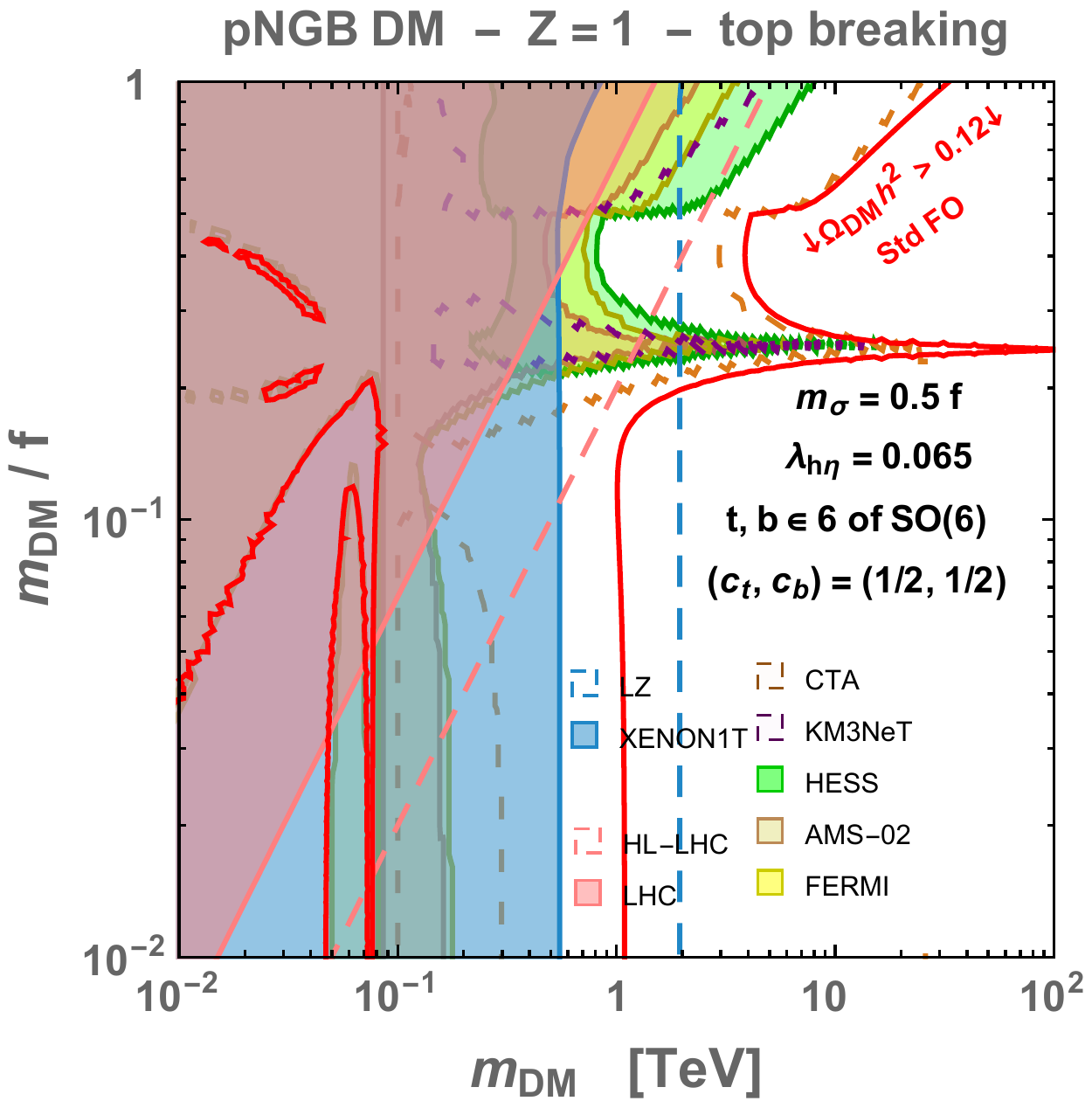}}}
\raisebox{0cm}{\makebox{\includegraphics[height=0.5\textwidth, scale=1]{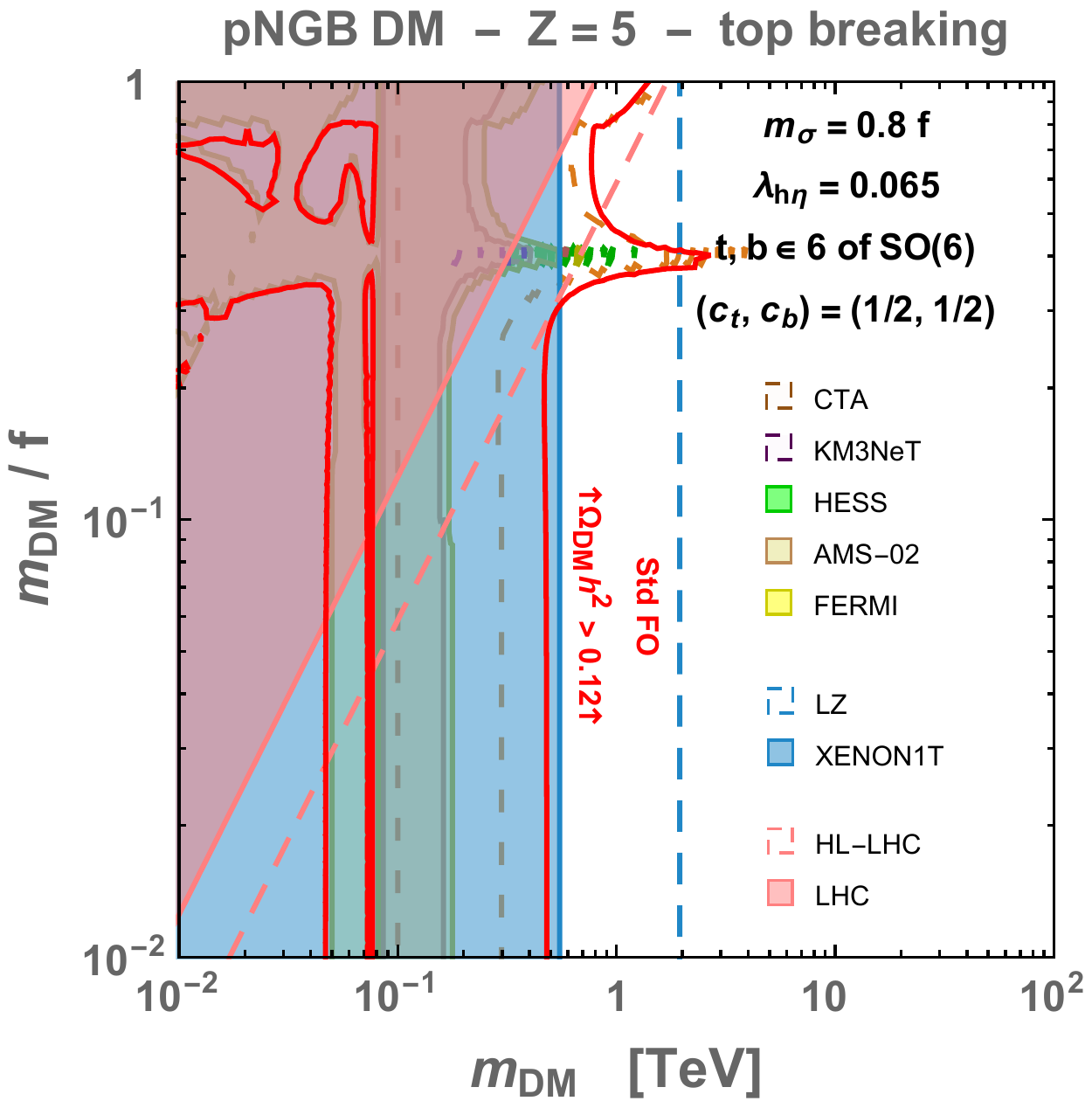}}}
\caption{\it \small \textbf{Dilaton-mediated O(6)/O(5) pNGB DM in the standard freeze-out case}. Lines and shadings as in Fig.~\ref{fig:stdFO_pheno_msigma_p25a}. The constants $c_b$ and $c_t$ are defined in Eq.~(\ref{eq:pNGBDM_Higgs}).
In addition to the dilaton channel, pNGB DM is  coupled to the SM through the composite Higgs interactions,  cf. Sec.~\ref{sec:DM-Candidates}, which lead to a non-trivial behaviour of the red thermal relic abundance contours. We can see a horizontal resonance around $\MDM \simeq m_{\sigma}/2$, a vertical resonance around $\MDM \simeq m_{\rm h}/2$ and a broad funnel in the $[80~\rm GeV,~1~\rm TeV]$ or $[80~\rm GeV,~500~\rm GeV]$ region due to efficient annihilation into $WW$, $ZZ$, $hh$ and $tt$. In Fig.~\ref{fig:stdFO_pNGB_pheno_app}, we also show the pNGB DM scenario where the shift symmetry is broken by the bottom quark instead of the top quark, such that the Higgs-DM mixing is reduced to $\lambda_{\rm h\eta} \simeq 0.065 \to 10^{-4}$, cf. Sec.~\ref{sec:DM-Candidates}. }
\label{fig:stdFO_pheno_msigma_p25} 
\thisfloatpagestyle{empty}
\end{figure}

\subsection{Indirect detection}
\label{sec:IndirectDetection}

\paragraph{Criterion.}
We consider Indirect Detection (ID) constraints coming from gamma-ray measurements of the Galactic Centre from HESS, measurements of gamma-rays from dwarf spheroidal galaxies (dSphs) from FERMI, and measurements of the anti-proton flux from AMS-02.
We exclude regions of parameter space which satisfy
	\begin{equation}
	\label{eq:ID_constraint}
	\mathsmaller{ {\rm BR}(\sigma \rightarrow WW,ZZ,bb)} \,  \sigma v_{\mathsmaller{\rm rel}} |_{\rm{ \eta \eta \rightarrow \sigma \sigma}} 
		+ \sigma v_{\mathsmaller{\rm rel}} |_{\rm{ \eta \eta \rightarrow WW, ZZ,bb}} 
			~> ~ 2 \left<\sigma v_{\mathsmaller{\rm rel}} \right>_{\rm limit}^{\rm ID},
	\end{equation}
where $v_{\mathsmaller{\rm rel}} = \sqrt{2}\, v_0$, $v_0 = 220$~km/s for DM annihilations in the Milky Way and $v_0 = 10$~km/s for the dSphs, see e.g.~\cite{Cirelli:2016rnw}.
 We checked that considering a proper averaging over a truncated Maxwellian DM velocity distribution, e.g. \cite{Lewin:1995rx}, does not lead to a significant change in the constraints.
 We introduced a factor two on the right-hand side of Eq.~\eqref{eq:ID_constraint} because we are considering non-self-conjugate DM, while all the limits we consider are given by the experimental collaborations for self-conjugate DM. 
 The dilaton branching ratios are given in App.~\ref{app:dilaton_decay_width}. 

 By writing Eq.~\eqref{eq:ID_constraint} we mean that we take the strongest limit among $WW$, $ZZ$ and $bb$ and not their sum weighted in our model, because that would require an analysis of data that goes beyond the purposes of our paper (plus, for some experiments, these data are not publicly available). This is a conservative assumption, because it effectively reduces the flux of cosmic rays produced by DM annihilations. On the other hand, the energy spectra of cosmic-rays coming from DM annihilations into dilatons are not the same of those considered by the experimental collaborations in casting their limits on the `pure' channels $WW$, $ZZ$ and $bb$, because of the extra step in the cascade.
Whether our procedure is conservative or aggressive, in this respect, depends on the DM mass and on the specific properties of each telescope, like the energy in which it is more sensitive and its energy resolution.
   Again, performing a more refined analysis goes beyond the purposes of our paper and requires access to data.

\paragraph{HESS, FERMI, AMS-02.}
The limits from HESS~\cite{Abdallah:2016ygi} come from $254$ hours of observation, assuming a standard NFW DM profile and local DM density $\rho_{\odot} = 0.39$~GeV/cm$^3$.
We take them from~\cite{Profumo:2017obk}, where they are given for secluded models and up to a DM mass of 10 TeV.
Since the original HESS publication gave limits for DM mass up to 70 TeV, we extend the~\cite{Profumo:2017obk} limits up to $\MDM=70$~TeV following~\cite{Cirelli:2018iax}, to which we refer also for a discussion of caveats and details.
The limits from FERMI result from the combined analysis of $15$ dSphs after six years of data taking. They assume a NFW profile and extend up to at $\MDM \simeq 10$~TeV \cite{Ackermann:2015zua}.
For AMS-02 we use the limits derived in~\cite{Giesen:2015ufa}, which are provided up to $\MDM \simeq 20$~TeV. These were found by requiring the primary $\bar{p}$ flux from DM annihilation not deteriorate the astrophysical-only fit of the AMS-02 data~\cite{Aguilar:2016kjl}. For both FERMI and AMS-02 limits we again refer the reader to~\cite{Cirelli:2018iax} for more details.

\paragraph{CTA, KM3NeT.}
Finally, we also consider the expected sensitivities from the future experiments CTA  and KM3NeT, as determined in~\cite{Acharyya:2020sbj} (where we used the `GC survey' sensitivity for the case without a core in the Galactic Centre) and~\cite{Gozzini:2020dom} respectively, for the $WW$ channel.
These sensitivities reach DM masses of at most 100~TeV not because of expected intrinsic experimental limits, but because the collaborations keep the habit of not extending their plots beyond that value.
Models of heavier DM, like the ones presented in this paper and many others, not only exist but are gaining intrinsic attention and motivation.
We therefore encourage the experimental collaborations to look into their data for the products of annihilations of DM with masses heavier than 100~TeV, and analogously to provide sensitivities in that regime.

\paragraph{Limits and sensitivities from telescopes.}

The limits and sensitivities described above, together with those from direct detection and colliders we will describe next, are displayed in Figs.~\ref{fig:stdFO_pheno_msigma_p25a} and~\ref{fig:stdFO_pheno_msigma_p25}, respectively for the cases of scalar and pNGB DM with shift symmetry broken by the top quark, and for two different values of the wave-function renormalization $Z=1$ and $Z=5$.
App.~\ref{app:Standard dilaton-mediated DM} contains analogous results for the cases of vector and fermion DM, displayed in Fig.~\ref{fig:stdFO_pheno_app}, and for pNGB DM with symmetry broken by the bottom quark, displayed in Fig.~\ref{fig:stdFO_pNGB_pheno_app}.
Their shape displays the resonances from the Sommerfeld-enhanced annihilation discussed in Sec.~\ref{sec:unitary-bound}, and becomes independent of $m_\DM/f$ for large values of this parameter just because the limits are not provided for larger DM masses. This implies that more information on models of heavy DM is contained in existing data, but is not exploited.

HESS limits are the most constraining indirect-detection ones for all models and in all the parameter space considered in this paper. For values of $m_\DM/f$ larger than a few, they exclude values of the DM mass smaller than 70~TeV.
We stress that HESS could reach larger values of $m_\DM$ already in existing data, if data were analysed in this sense.
For smaller values of $m_\DM/f$ they exclude smaller values of $m_\DM$, down to $m_\DM \sim$~TeV for $m_\DM/f \sim 0.7$.
HESS limits were derived under the assumption of an NFW DM density profile, if instead the profile had a core towards the Galactic Center then they would become much weaker.
The most constraining limits would then become those from AMS-02 at large values of $m_\DM/f$ and $m_\DM$, and those from FERMI dSphs at smaller values of the above parameters (see~\cite{Lefranc:2016fgn} for a quantitative comparison of the interplay of limits and sensitivities for varying sizes of a putative DM core).

Limits and sensitivities from telescopes are stronger than collider and direct detection ones in the case where DM is a scalar or vector resonance of the strong sector, over the entire parameter space except for large $Z$ and small $m_\DM/f$.
Indirect detection is weaker than collider and direct detection in the case of fermion DM (except at large values of $m_\DM/f$ and small $Z$), because that is the only model for which the tree-level annihilation cross section is $p$-wave suppressed, see Eq.~(\ref{eq:DMannfactors}). The fact that it is comparatively weaker also in the case of pNGB DM is instead due to the fact that the interesting values of $m_\DM/f$ are smaller in that case. 

\paragraph{Telescopes and DM masses larger than 100 TeV.}
\label{par:telescopes_1PeV}
Finally concerning sensitivities, as already mentioned they do not extend to $m_\DM > 100$~TeV simply because collaborations do not report them there. In the case of DM without supercooling, this is close to the maximal DM mass allowed by the unitarity limit, so we do not extend them further.
However, in the case of supercooled DM that we will study from Section~\ref{sec:SupercoolDM} on, larger masses are allowed and motivate searches for heavier DM in telescope data.
To avoid drawing the conclusion that these models will only be tested by gravitational waves, which would be wrong, we guesstimate that future $\gamma$ and $\nu$ telescope could well probe them up to $m_\DM \sim$~PeV, and display this in Fig.~\ref{fig:summary}.
This guesstimate is based on our naive extrapolation of those sensitivities at DM masses larger than 100~TeV, where we observe that they cross the perturbative unitarity limit (which lies close to the prediction of our model) around a PeV.

\subsection{Direct detection}
\label{sec:DirectDetection}

\paragraph{The dilaton-nucleon interactions.}
The direct detection constraints come from the possibility for the DM to scatter off a nucleon via a dilaton exchange. 
The dilaton-quark and gluon interaction in Eq.~\eqref{eq:eff_lag_dilaton}, lead to an effective dilaton-nucleon coupling of the form,
	\begin{equation}
	\mathcal{L}_{\sigma n n } = y_n \, \sigma \,\bar{n}n,
	\end{equation}
where the effective coupling $y_n$ is given by
	\begin{equation}
	y_n \equiv -\sum_{q=u,d,s,c,b,t}  f_q^n \frac{m_n}{Z \dilatonvev} + R^n \frac{c_G}{8\pi \, Z\dilatonvev},
	\label{eq:y_n_def}
	\end{equation}
	with $c_{G}$ given by Eq.~\eqref{eq:def_cG} and Eq.~\eqref{eq:assumption_cG}, and
	\begin{equation}
	m_n f_q^n \,\equiv\, \left \langle n|~ m_q \bar{q} q ~|n\right \rangle, \qquad R^n \,\equiv \alpha_s \,\left \langle n|~  GG ~ |n\right \rangle.
	\end{equation}
The quark form factors $f_q^n$ are taken from~\cite{Ellis:2018dmb} 
	\begin{align}
	\begin{split}
	\label{eq:fnq}
	&f_u^n \simeq  0.015, \quad f_s^n \simeq 0.037, \quad f_b^n \simeq 0.072, \\
	&f_d^n \simeq 0.034, \quad f_c^n \simeq 0.078, \quad f_t^n \simeq 0.069.
	\end{split}
	\end{align}
The trace anomaly $R^n$ can be determined from the fact that, at first order in the heavy-quark expansion, the form factors of heavy quarks coincides with the opposite sign of their contribution to the trace anomaly~\cite{Shifman:1978zn}
	\begin{equation}
	\label{eq:form_anomaly}
	\langle n|~ \sum_{q=c,b,t} m_q \bar{q}q ~|n \rangle \,\simeq - \langle n|\frac{\alpha_s}{8 \pi}c_G^{\rm heavy} GG |n \rangle,
	\end{equation}
with $c_G^{\rm heavy} =  2 N_q^{\rm heavy}/3 = 2$.  From using Eq.~\eqref{eq:form_anomaly} together with the fact that the nucleon mass $m_n$ is the trace of the trace of the QCD energy-momentum tensor~\cite{Jungman:1995df}
	\begin{equation}
	\label{eq:mass_nucleon_trace}
	m_n = \,\left \langle n|~ \theta_{\mu}^{\mu} ~|n\right\rangle \,= \,\langle n|~ \left( \sum_{q=u,d,s,c,b,t} m_q \bar{q}q + \frac{\alpha_s}{8\pi}c_G \, GG \right)~|n\rangle,
	\end{equation}
 we find\footnote{Note that Eq.~\eqref{eq:form_anomaly} and Eq.~\eqref{eq:mass_nucleon_trace} also imply the well-known \cite{Jungman:1995df} consistency relation $ \sum_{q=c,b,t} f_q^n \simeq \dfrac{c_{G}^{\rm heavy}}{c_{G}^{\rm heavy}-c_G}\left(1- \sum_{q=u,d,s} f_q^n\right) =\dfrac{2}{9}\left(1- \sum_{q=u,d,s} f_q^n\right) $.}
	\begin{equation}
 	R^n \equiv \alpha_s \langle n|~ GG ~|n \rangle\, \simeq \frac{8\pi}{c_G - c_G^{\rm heavy}}m_n \left( 1- \sum_{q=u,d,s} f_q^n \right) \simeq - 2.4~\rm GeV ,
	\end{equation}
where we have used $c_G - c_G^{\rm heavy} =  - \frac{11}{3} N_c + \frac{2}{3}N_q^{\rm light} = -9$.

\paragraph{The dilaton-mediated scattering cross-sections.}
The elastic cross-section between DM and nucleon mediated by dilaton interaction are computed in App.~\ref{app:DD_cross-section}. In the limit $\MDM \gtrsim m_n$, these are approximated by 
\begin{equation}
\sigma({\rm DMn \to DM n}) \simeq \frac{y_n^2}{\pi} \frac{m_n^2}{m_{\sigma}^4} \left(\frac{\MDM}{Z\,f} \right)^2,
\label{eq:sigma_DD_dil}
\end{equation}
for scalar DM, the fermion and vector cases differ only by the order one factor in front of this expression, and are given in App.~\ref{app:DD_cross-section}.

\paragraph{The pNGB DM case.}
When DM $\eta$ is a $O(6)/O(5)$ pNGB, the $\eta$-$\eta$-$h$ vertex and the DM-quark contact interaction in Eq.~\eqref{eq:pNGBDM_Higgs} lead to the effective DM-quark lagrangian \cite{Frigerio:2012uc, Marzocca:2014msa}
\begin{equation}
	\mathcal{L}_{\eta q q } = \sum_{q=u,d,s,c,b,t} a_q \, m_q \,\eta^2\,\bar{q}q, \label{eq:eta_qq_pNGB}
\end{equation}
with
\begin{equation}
a_{q} =  \frac{\lambda_{h\eta }(1-2\xi)}{m_h^2} + c_q \frac{\xi}{(1-\xi)v_{\rm H}^2}, \qquad \xi \equiv v_{\rm H}/Zf.\label{eq:a_q}
\end{equation}
We make the conservative assumption that the contact interaction terms with the first two quark generations are zero $c_{q=u,d,s,c} =0$. Then we consider the two cases $(\lambda_{h\eta },\,c_b,\,c_t)=(0.065,\,1/2,\,1/2)$ and $(\lambda_{ h\eta},\,c_b,\,c_t)=(10^{-4},\,1/2,\,0)$, described along Eq.~\eqref{eq:top_breaking_lambda} and Eq.~\eqref{eq:bottom_breaking_lambda}, according to whether $U(1)_\eta$ is broken by the top quark or by the bottom quark only.
The pNGB-DM-nucleon elastic cross-section is dominated by the dilaton-mediated contribution in Eq.~\eqref{eq:sigma_DD_dil} in the bottom-breaking case, and by the Higgs-mediated contribution~\cite{Frigerio:2012uc, Marzocca:2014msa} resulting from Eq.~\eqref{eq:eta_qq_pNGB} in the top-breaking case. In the limit $\MDM \gg m_n$, it reads
\begin{equation}
\label{eq:DD_xsec_pNGB}
\sigma({\rm DMn \to DM n}) \simeq
\left\{
                \begin{array}{ll}
                 \dfrac{y_n^2}{\pi} \dfrac{m_n^2}{m_{\sigma}^4} \left(\dfrac{\MDM}{Z\,f} \right)^2 \quad \text{bottom-breaking}\,,\\ \\
                \dfrac{c_n^2}{\pi} \dfrac{m_n^2}{\MDM^2} \quad \text{top-breaking}\,,
                \end{array}
              \right.
              \end{equation}
with $y_n$ given by Eq.~\eqref{eq:y_n_def} and 
\begin{equation}
c_n =  \sum_{q=u,d,s,c,b,t} f^n_q\, a_q\,m_n, \label{eq:cn_def}
\end{equation}
with $ f^n_q$ and $a_q$ given by Eq.~\eqref{eq:fnq} and Eq.~\eqref{eq:a_q}, respectively.

\paragraph{XENON-1T, LZ.}
We compare the DM-nucleon cross-section to the upper bound on the spin-independent WIMP-nucleon cross-section resulting from $279$ live days of data collecting in XENON-1T~\cite{Aprile:2018dbl}, and the projection for $1000$ live days of data collecting in LZ~\cite{Akerib:2018lyp}. 

\paragraph{The exclusion constraints.}
Let us now assume our DM matches the observed cosmological DM energy density. For normalization of the dilaton kinetic term equal to unity $Z=1$, the scattering cross-section for scalar and vector DM is too weak to lead to any signal by XENON-1T or even LZ, see Fig.~\ref{fig:stdFO_pheno_msigma_p25a} and Fig.~\ref{fig:stdFO_pheno_app}. For $Z=5$, XENON-1T constrains DM mass below $100$ and $400$~GeV, respectively.
For fermion DM, XENON-1T leads to the constraints  $\MDM \gtrsim 400$~GeV and $\MDM \gtrsim 3$~TeV, for $Z=1$ and $Z=5$ respectively, cf. Fig.~\ref{fig:stdFO_pheno_app}. For pNGB DM receiving its mass from top loops, XENON-1T leads to $\MDM \gtrsim 1$~TeV and $\MDM \gtrsim 600$~GeV for $Z = 1$ and $Z=5$, respectively, see Fig.~\ref{fig:stdFO_pheno_msigma_p25}. For pNGB DM receiving its mass from bottom loops, XENON-1T constraints drop to $\MDM \gtrsim 80$~GeV for both $Z = 1$ and $Z=5$, cf. Fig.~\ref{fig:stdFO_pNGB_pheno_app}.

\subsection{Collider limits}
\label{sec:colliders}

\paragraph{Higgs couplings.} For concreteness and as we have already made this assumption in the case of pNGB DM, we consider the case that the Higgs boson is a pNGB from the same strong sector of our DM models.
Higgs couplings measurements~\cite{CMS:2020gsy,ATLAS:2020qdt} then imply
\beq
\text{LHC~limit:} \; f \gtrsim 0.8~{\rm TeV}\,,
\label{eq:HiggsLimits_LHC}
\eeq
where we have translated the combined 95\%CL limit on the $k$-parameters $k_F$ and $k_V$ from Fig.~11 of~\cite{ATLAS:2020qdt} onto a limit on $f$ by using $k_V = 1-v^2/(2f^2)$ and $k_F = 1-3v^2/(2f^2)$.
Note that, for simplicity and because that would depend on further model details, we have neglected the contribution to Higgs-coupling deviations that comes from a dilaton-Higgs mixing.
Coming to future sensitivities, we consider those on Higgs coupling measurements at the HL-LHC as derived in~\cite{Cepeda:2019klc}, and translated on the confinement scale $f$ as in the scenario `SILH1b' there, which corresponds to the Higgs as a pNGB and to the generic expectation from partial compositeness, and matches the $k$-parameters that we assumed to derive the limit in Eq.~(\ref{eq:HiggsLimits_LHC}). The sensitivity then reads
\beq
\text{HL-LHC~sensitivity:}  \; f \gtrsim 1.7~{\rm TeV}\,.
\label{eq:HiggsLimits_HLLHC}
\eeq
In composite Higgs models LHC searches for strong sector resonances, like top-partners, lead to exclusions and sensitivities slightly stricter than those in Eqs.~(\ref{eq:HiggsLimits_LHC}) and~(\ref{eq:HiggsLimits_HLLHC}), see e.g.~\cite{Matsedonskyi:2015dns}. This is however not the case in models where the Higgs is a pNGB but another mechanism is in place to allow for heavier resonances, like in Twin Higgs models UV-completed by a composite sector (TH+CHM), where Higgs coupling measurements drive the limits on $f$~\cite{Geller:2014kta,Barbieri:2015lqa,Low:2015nqa}.
Moreover, in both CHMs and TH+CHM, the limits from top partner searches depend on further model-dependent details like the representations of the fermion resonances.
To keep our discussion general, therefore, we do not display these limits in our parameter space.
Similarly, we do not consider limits from EW precision tests, as they depend on unknown physics at the scale $m_*$ and are therefore less robust than those from Higgs couplings.

\paragraph{Dilaton searches.} 
To the best of our knowledge, the most recent interpretations of LHC resonance searches in the parameter space of a dilaton have been performed in~\cite{Fuks:2020tam} for $m_\sigma > 300$~GeV, and in~\cite{Ahmed:2019csf} for $m_\sigma < 300$~GeV.
The first analysis excludes values of $f$ of at most a few TeV, but it is of little use for our purposes, because it is performed under the hypothesis that the dilaton coupling to gluons of Eq.~(\ref{eq:eff_lag_dilaton}) reads $c_G = 7$ (which for us would correspond to the assumption that the entire SM is composite), while the benchmark we have chosen is smaller by a factor of $\sim 20$, cf. Eqs.~(\ref{eq:def_cG}) and~(\ref{eq:assumption_cG}).
In addition, dilaton couplings relevant for its production at the LHC receive a contribution also from the Higgs-dilaton mixing, which is model-dependent (see e.g. the discussions in~\cite{Bruggisser:2018mrt} and~\cite{Ahmed:2019csf}).
Ref.~\cite{Ahmed:2019csf} has instead casted its analysis using our same benchmark for $c_G$ (their `holographic dilaton') and, for $m_\sigma \sim 300$~GeV, found limits on $f$ in the range of 1.5 to 2 TeV depending on the assumption on the dilaton-Higgs mixing.
We can then infer that limits on $f$ in the same ballpark apply for the benchmark value $m_\sigma/f = 0.25$ that we will consider later. For larger values of $m_\sigma/f$, relevant to the non-supercooled DM case, one can naively expect that limits will not become stricter. In light of the above discussion, since performing a detailed collider analysis goes beyond the purposes of this paper, here we adopt the indicative limit
\beq
\text{LHC~limit:} \; Z f \gtrsim 1.5~{\rm TeV}\,,
\label{eq:DilatonLimits_LHC}
\eeq
where we have added the scaling of the limits with $Z$, neglected in Refs.~\cite{Ahmed:2019csf,Fuks:2020tam}.
Analogously, we make the simplifying assumption that the HL-LHC sensitivity will be in the ballpark of the one determined in ~\cite{Ahmed:2019csf} for the `holographic dilaton' and at the largest $m_\sigma$ value considered there,
\beq
\text{HL-LHC~sensitivity:} \; Z f \gtrsim 5~{\rm TeV}\,.
\label{eq:DilatonLimits_LHC}
\eeq

\paragraph{Summary.} To summarise collider limits and sensitivities on our picture, they are driven by dilaton searches at small $Z$, and by Higgs coupling measurements at large $Z$. Their precise determination would deserve a more detailed study, which however goes beyond the purposes of this paper, because our main interest lies in regions of $f$ beyond the multi-TeV.
Unlike the cases of indirect and direct detection, the collider limits we consider depend only on Higgs and dilaton physics, and so are independent of which of the five DM models we consider.

\subsection{DM abundance fixed by freeze-out}
\label{sec:DM_abundance_freeze-out}

\paragraph{The Boltzmann equation.}
Let us review the standard freeze-out scenario without supercooling. In kinetic equilibrium, the DM abundance can be tracked using the well-known Boltzmann equation,
	\begin{align}
	\label{eq:Bmann}
	\frac{dY_{\mathsmaller{\rm DM}}}{dx} = -\frac{ \lambda_n}{x^{2+n}} \left( Y_{\mathsmaller{\rm DM}}^2 -  Y_{\mathsmaller{\rm DM}}^{\rm eq \, 2} \right),
	\end{align}
where $x\equiv\MDM/T$, $\lambda_n=M_{\rm pl} \MDM \, \sigma_n \, \sqrt{8\pi^2 \gSM /45},$ $\sigma_n \equiv \langle \sigma v_{\mathsmaller{\rm rel}}\rangle\, x^{n}$ with $n$ chosen so that $\sigma_{n}$ is $x$ independent, and $M_{\rm pl} \simeq 2.44 \times 10^{18} ~ \text{GeV}$. In the standard analysis one has $n=0$ for s-wave and $n=1$ for p-wave annihilations. Note the Sommerfeld enhancement factor changes the velocity dependence of the cross-section, with $n=-1/2$ for both s-wave and p-wave in the large coupling/low mediator mass limit. However, for the sake of simplicity in the present discussion, we neglect it for now but include it in our plots.

\paragraph{Frozen-out DM abundance.}
Assuming a sufficiently high enough reheat temperature, the DM interactions are rapid and DM is initially in thermal equilibrium with the radiation bath. When the temperature drops below $\MDM$, the DM continues to annihilate preferentially into light degrees of freedom, until the annihilation rate becomes lower than the expansion rate of the universe and the abundance freezes-out. This occurs at the freeze-out temperature, $T_{\mathsmaller{\rm FO}} \equiv \MDM/\xFO$, where~\cite{Kolb:1990vq}
	\begin{equation}
	\xFO = \log\left[0.192(n+1)\,M_{\rm pl}\,\MDM\,\sigma_n \,  \gDM/\gSM\right]-(n+1/2)\log[\xFO].
	\label{eq:x_FO}
	\end{equation}
where $\gDM$ is the number of degrees of freedom of DM. It is equal to $\gDM = 1$, $\frac{3}{4}\cdot 4$ and $3$ for scalar, fermion and vector, respectively.
The abundance today is then the frozen-out abundance redshifted up to now
	\begin{equation}
	\label{eq:Omega_DM_FO}
	\Omega_{\rm DM} h^2 =2_{\mathsmaller{\rm DM\neq \bar{DM}}}  \frac{s_0 \MDM}{3 M_{\rm pl}^2 H_{100}^2}  Y_{\rm DM}^{\rm \mathsmaller{FO}}, 
	\end{equation}
where $H_{100} = 100$~km/s/Mpc, $s_0=2891.2$~cm$^{-3}$~\cite{Tanabashi:2018oca} and
	\begin{equation}
	\YDM^{\rm \mathsmaller{FO}} = \frac{(n+1)\,x_{\mathsmaller{\rm FO}}^{n+1}}{\lambda_n}.
	\label{eq:Y_DM_FO}
	\end{equation}
The potential factor $2$ in Eq.~\eqref{eq:Omega_DM_FO} stands for DM as being counted as $\Omega_{\rm DM} + \Omega_{\rm \bar{DM}}$. It must be removed if DM is its own anti-particle.

\paragraph{Summary of improvements.}
In the parameter space discussed so far, the supercooled PT does not affect the final relic abundance set by the freeze-out mechanism discussed here. This is the production mechanism considered for dilaton portal DM in the past literature~\cite{Bai:2009ms,Blum:2014jca,Efrati:2014aea,Kim:2016jbz,Fuks:2020tam}. We have updated these results by combining our freeze-out calculation with collider, direct and indirect detection limits, in Figs.~\ref{fig:stdFO_pheno_msigma_p25a} and~\ref{fig:stdFO_pheno_msigma_p25}. (In plotting the direct and indirect detection constraints, we assume our DM saturates the observed cosmological DM abundance, independently of the relation between $\MDM$ and $f$ required by the mechanism setting the relic abundance.) We have also extended the analysis to dilaton normalization strength $Z \neq 1$, which had not previously been considered in this context. We now go on to see how the picture is changed once the novel effects pointed out in \cite{Baldes:2020kam}, arising during supercooled confinement, are taken into account.

\FloatBarrier

\section{Supercooled dilaton-mediated composite DM}
\label{sec:SupercoolDM}

A late period of thermal inflation such as a supercooled phase transition will alter the abundance of primordial particles. Determining the abundance of composite states following supercooling requires the inclusion of a number of novel effects recently explored in a sister publication~\cite{Baldes:2020kam}, which we recapitulate here for completeness.

\paragraph{Sketch of the mechanism.}
The initial vacuum dominated phase leads to a dilution of the fundamental constituents (from now named dark techniquanta or quarks for brevity), which is an effect already familiar for studies considering elementary relics. As the dark quarks enter the bubbles of strongly-coupled phase, however, we must account for the dynamics of confinement. When entering the bubbles, the dark quarks are separated by large distances compared with the confinement scale, $1/\fconf$. In~\cite{Baldes:2020kam} it was argued that it is energetically favorable for the chromo-electric flux to form a string like configuration toward the bubble wall. Thus minimizing its length inside the strongly-coupled phase. The fragmentation of the flux string via pair-creation effects leads to the formation of additional composite states. Kinematics demands that these are boosted in the original plasma frame. These states undergo deep inelastic scatterings in their return to kinetic equilibrium which further enhances the yield of the composite DM. Finally thermal effects, such as the $2 \leftrightarrow 2$ scatterings familiar from freeze-out, must be taken into account following reheating.   

We now provide quantitative estimates of the above effects in the context and notation of the current model. This allows us to estimate the yield of composite DM following the supercooling and provide predictions in terms of the dilaton mass. This will allow us to compare the parameter space of interest with current and future experimental searches (direct, indirect, collider, gravitational wave).

\subsection{Dilution of the fundamental quanta}

\paragraph{Initial quark abundance.}
We assume that DM is a composite state formed of dark techniquanta. These are massless before the phase transition, so that their abundance follows a thermal distribution for massless particles which, when normalised to the entropy density, reads
	\begin{equation}
	Y_{\TC}^{\rm eq} \equiv \frac{n_{\TC}}{s} =\frac{45 \zeta(3) \, g_{\TC}}{2\pi^4 g_{*\rm s}},
	\end{equation}
where $g_{*\rm s}$ is the effective total number of relativistic degrees of freedom, and $g_{\TC}$ is the effective number of degrees of freedom of dark techniquanta, which typically consists of some model dependent number of dark quarks, $g_{q}$, and dark gluons, e.g.~$g_{g} = 2(N^{2}-1)$ for a dark $SU(N)$.

\paragraph{Supercooling stage.}
The supercooling stage starts when the free energy in Eq.~\eqref{eq:tot_pot_dil} becomes vacuum dominated,
	\begin{equation}
	\label{eq:Tstart_def}
	\frac{\pi^2}{30} \left( \gSM + g_{\mathsmaller{\rm q}} +45 N^2/4 \right) \Tstart^4 \simeq Z^2 \frac{\msigma^2 \dilatonvev^2}{16} \quad \implies \quad
	 \Tstart \overset{N\gg1}{\simeq }  \left( \frac{Z \, \msigma \dilatonvev}{\sqrt{6} \pi N}  \right)^{1/2}.
	\end{equation} 
In this model $\Tstart$ is generally slightly below $T_{c}$, defined in Eq.~\eqref{eq:crit_temp_PT}. 
Supercooling ends at the nucleation temperature, $\Tnuc$. Within the bubble the dilaton then rolls towards the minimum of the potential on a timescale $\sim1/\Tnuc$ (see \cite[App. A]{Baldes:2020kam}). The dilaton condensate decays into the SM following bubble collision.
For $\msigma \gg m_{h}$ the dilaton dominantly decays into the four components of the Higgs doublet\footnote{Or equivalently --- by virtue of the Goldstone equivalence theorem \cite{Chanowitz:1985hj,Peskin:1995ev} --- into the physical Higgs boson and longitudinal modes of the WW and ZZ channels, in the broken electroweak phase.}
	\begin{equation}
	\Gamma_{\dilaton} \approx \frac{\msigma^3}{8 \pi Z^2 \dilatonvev^2}.
	\end{equation}
Hence the decay rate is much larger than the Hubble factor, $H \sim \dilatonvev^2/M_{\rm pl}$, so that we can neglect the short period of time the dilaton redshifts as matter during its coherent oscillations. Therefore, all the vacuum energy of the transition is efficiently converted into radiation at reheating
	\begin{align}
	\label{eq:Treh_def}
	&\frac{\pi^2}{30} \gSM \Treh^4 \simeq Z^2 \frac{\msigma^2 \dilatonvev^2}{16} +\frac{\pi^2}{30} \left( \gSM + g_{\mathsmaller{\rm q}} +45 N^2/4 \right)\Tnuc^4, \\
	&\implies \quad \Treh \simeq \Big(\frac{106.75 + g_{\mathsmaller{\rm q}} +45 N^2/4}{106.75} \Big)^{1/4} \,  \Tstart  \overset{N\gg1}{\simeq } 0.6 \sqrt{N} \,  \Tstart,
	\end{align} 
	with $\Tstart$ defined in Eq.~\eqref{eq:Tstart_def}.
	
\paragraph{Dilution of the quarks.}
As a result of the supercooling period, the dark quark abundance is diluted according to~\cite{Wainwright:2009mq,Konstandin:2011dr,Hambye:2018qjv}
	\begin{equation}
	\label{eq:supercoolDM}
	Y_{\TC} ^{\mathsmaller{\rm SC}} =  Y_{\TC}^{\rm eq} \left( \frac{ \Tnuc}{ \Tstart} \right)^3 \frac{\Treh}{T_{\rm start}}.
	\end{equation}
This can be derived by using standard accounting of entropy factors before and after the PT.
The number of e-foldings of the late inflationary phase is given by
	\begin{equation}
	\label{eq:Ne_def}
	N_{e} \equiv \log \left( \frac{\Tstart}{\Tnuc} \right)
	\overset{N\gg1}{\simeq } \log \left( \left[ \frac{Z m_{\sigma}}{\sqrt{6}\pi N f} \right]^{1/2} \frac{f}{\Tnuc} \right),
	\end{equation}
where in the last equality we have used Eq.~(\ref{eq:Tstart_def}).

\subsection{Enhancement by string breaking}
\label{sec:Enhancement of the DM supercool abundance by gluon string breaking}

\begin{figure}[t]
\begin{center}
\includegraphics[width=350pt]{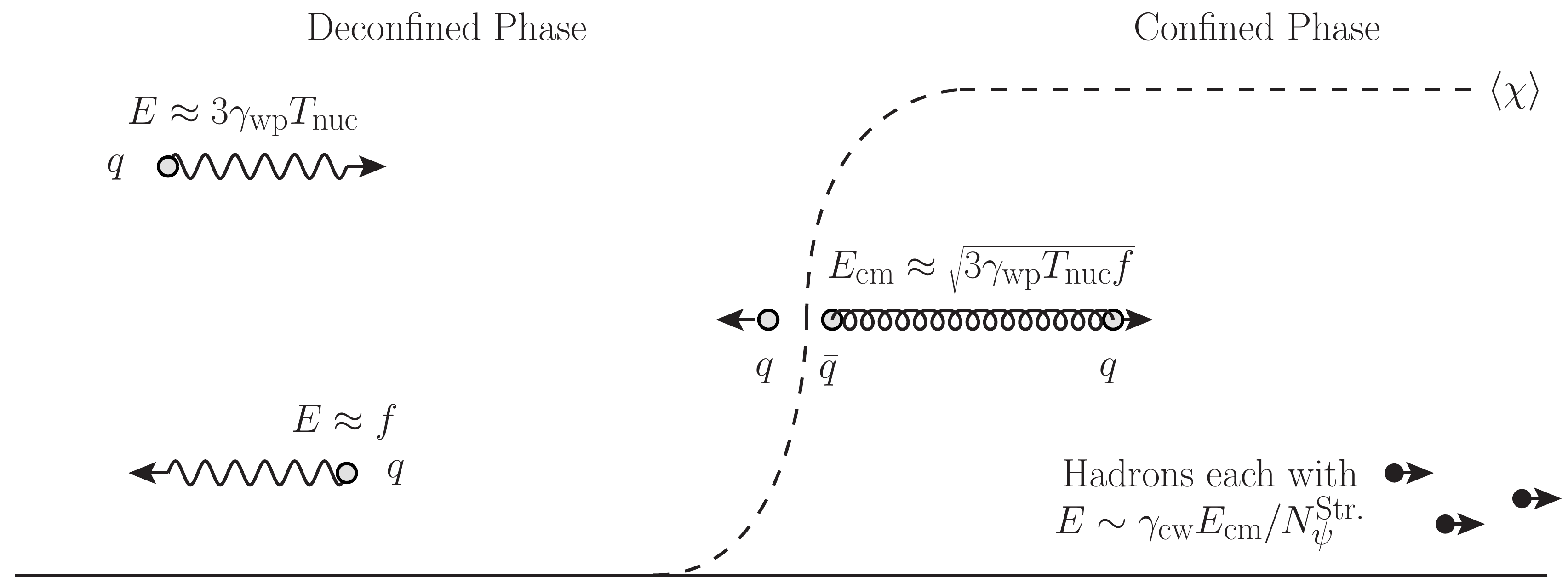}
\end{center}
\caption{\it \small The string formation and fragmentation mechanism as seen in the wall frame. Highly energetic quarks impinge on the wall, leading to a string-like flux configuration. Its centre-of-mass energy is set by geometric mean of the initial quark energy and the energy of the quark ejected from the end of the string. The latter is taken to be of the order of the strong scale. The fragmentation of the string forms hadrons in the confined phase. The fact that the string  is attached to the moving wall acts as a catapult for said hadrons which are then boosted with respect to the plasma frame.}
\label{fig:wall_sketch} 
\end{figure}

\paragraph{String formation.}
A binding potential is switched on when, at $\Tnuc$, the composite sector confines. The potential increases approximately linearly with the distance between the dark techniquanta~\cite{Thorn:1979gv, Greensite:1988tj, Poulis:1996iv, Ko:1999yx, Greensite:2001nx, Bardakci:2002xi, Greensite:2003xf, Greensite:2003bk,Trawinski:2014msa}
	\begin{equation}
	\label{eq:linear_pot}
	E_{\rm TC} \simeq \fconf^2 \, d.
	\end{equation}
The overall picture of the string fragmentation mechanism we will now discuss is illustrated in Fig.~\ref{fig:wall_sketch} (more detail of the modelling can be found in~\cite{Baldes:2020kam}). After the supercooling period, the inter-quark distance is $d \simeq n^{-1/3} \simeq 1/\Tnuc$, much larger than the confinement distance $1/\fconf$. The confinement scale seen by the quarks when crossing the bubble wall is close to the zero temperature value because the dilaton rolls quickly to its minimum. The flux confines in a string-like configuration, starting at the dark quark and terminating at the bubble wall. This minimizes the energy compared to if a flux line were to connect two nearest neighbour quarks. 

\paragraph{String fragmentation.}
As the initial quark further penetrate inside the bubble, the flux string fragment into dark sector hadrons. Conservation of colour requires the ejection of a dark quark from the end of the string back into the deconfined phase, see Fig.~\ref{fig:wall_sketch}. Its energy is estimated dimensionally as $E_{\rm ej,w} \sim f$ in the wall frame. The energy in the string center-of-mass frame is found to be \cite{Baldes:2020kam},
	\begin{equation}
	\ECM \approx \sqrt{3 \gwp \Tnuc f},
	\end{equation}
where $\gwp$ is the bubble-wall Lorentz factor in the plasma frame. For values $\gwp > f/\Tnuc$ it is energetically possible for the quarks to enter the bubble. 

\paragraph{Bubble wall velocity.}
In the run-away regime, which will apply in our scenario, one finds the Lorentz factor $\gwp$ is the ratio of the bubble size to the bubble size at nucleation. At collision time the former is set by the timescale of the transition, denoted $\beta^{-1}$ to match onto the literature, and the latter is typically $\sim \Tnuc^{-1}$. Thus  close to bubble-wall collision, \emph{i.e.}~when the majority of the volume is changing phase, we have 
	\begin{equation}
	\gwp \approx \frac{ \sqrt{48} \, \Tnuc M_{\rm pl} }{ \beta/H Z m_{\sigma} f},
	\end{equation} 
where for typical supercooled phase transitions we have $\beta/H \sim 10$.

\paragraph{Number of string fragments.}
The fragmentation of the string produces a number of composite states, $N_{\psi}^{\rm Str.}$, which on general grounds, we expect $N_{\psi}^{\rm Str.} = \mathcal{P}(\mathrm{log}[E_{\rm cm}/m_{\ast}])$, where $\mathcal{P}$ is a polynomial function, and $m_{\ast} \equiv g_{\ast} f$ is the mass scale of hadrons with $1 \lesssim g_{\ast} \lesssim 4 \pi$. (As in QCD, where the multiplicity of final state particles grows as a polynomial function of $\log{[\sqrt{s}/ \mathrm{GeV}]}$, with current data being well fitted by a cubic polynomial with $\mathcal{O}(1)$ coefficients~\cite{GrosseOetringhaus:2009kz,Kumar:2012hx}.) The yield of composite states, $\psi$, after string fragmentation is
	\begin{equation}
	Y_{\psi} ^{\mathsmaller{\rm Str.}} \approx Y_{\TC}^{\rm eq} \left( \frac{ \Tnuc}{ \Tstart} \right)^3 \frac{\Treh}{T_{\rm start}} \, \mathcal{P}\left(\mathrm{log}\left[\frac{ E_{\rm cm} }{ m_{\ast} }\right]\right).
	\end{equation}
As dark gluons can also initiate strings, we include their contribution to the degrees-of-freedom in the above equation.  Note there is significant model uncertainty as to the precise form of $\mathcal{P}$. Nevertheless, as long as certain assumptions hold, the yield following the next step of deep inelastic scattering becomes insensitive to $\mathcal{P}$, as we shall now discuss.

\subsection{Enhancement by deep inelastic scattering}

\paragraph{Gluon string catapult.}
The Lorentz factor of the string COM frame in the wall frame is \cite{Baldes:2020kam}
	\begin{equation}
	\gcw \approx \sqrt{\frac{3 \gwp T_{\rm nuc}}{f}} \approx \sqrt{ \frac{ 21 M_{\rm pl}\Tnuc^{2} }{ \beta/H Z m_{\sigma} f^{2}}  }.
	\end{equation}
And the Lorentz factor of the string centre-of-mass frame in the plasma frame is 
	\begin{equation}
	\gcp \approx \frac{1}{2} \frac{ \gwp }{ \gcw}  \approx \sqrt{ \frac{M_{\rm pl}}{ \beta/H Z m_{\sigma}}}.
	\label{eq:gcp}
	\end{equation}
Therefore, the $N_{\psi}^{\rm Str.}$ composite states formed after fragmentation of the string, are catapulted with the Lorentz boost factor in Eq.~\eqref{eq:gcp} relative to the plasma frame. 

\paragraph{String fragment energy.}
Following the string fragmentation, the average energy of the hadrons (which we denote population A) as measured in the plasma frame reads
	\beq
	E_\text{A,p} \approx 2 \gcp \frac{ \ECM}{N_{\psi}^{\rm Str.}(\ECM)} \approx \frac{\gwp f}{N_{\psi}^{\rm Str.}(\ECM)}.
	\label{eq:EAp}
	\eeq
Similarly the energy of the ejected quarks (population B) is
	\beq
	E_\text{B,p} \approx \gwp f.
	\label{eq:EBp}
	\eeq
These form a dense shell around the bubble and confine without the formation of extended strings (due to their close distance) when entering the opposing bubble.

\paragraph{Cascade of deep inelastic scattering.}
The kinetic energy of these populations of particles are dissipated through scatterings with the preheated plasma coming from the decay of the dilaton condensate after bubble collision. The energy of the particles in the preheated plasma is set by the oscillation frequency of the dilaton condensate $E_{\rm prh} \approx m_{\sigma}$. Scatterings of the boosted composite resonances with the preheated plasma are energetic enough to produce additional composite states provided $\sqrt{s} > m_{\ast}$. Above this threshold, a majority of the energy is dissipated through deep inelastic scattering, provided that $m_{\sigma}^{4} \ll \Delta V$, see~\cite[Sec.~8.2]{Baldes:2020kam}.  In our scenario this translates to $4 m_{\sigma} < Zf$ which coincides with the region of significant supercooling, see Fig.~\ref{fig:IB_supercooling_nuc_temp}. Otherwise, dissipation through elastic scattering dominates.

\paragraph{DM abundance after DIS.}
Assuming the kinetic energy above threshold is dissipated through a cascade of DIS, we find the DM to be independent of $\mathcal{P}$, and to read~\cite{Baldes:2020kam}
	\begin{equation}
	\YDM ^{\mathsmaller{\rm DIS}} \approx
	\, \frac{ 0.43 \, \mathrm{eV}}{ m_{\ast}} \frac{\mathrm{BR}}{10^{-6}}
	\, \frac{1}{Z^{5/2}}
	\, \frac{g_\TC}{130}
	\, \frac{4\pi}{g_*}
	\, \left( \frac{10}{\beta/H} \right) 
	\, \left( \frac{0.2}{m_{\sigma}/f} \right)^{3/2}
	\, \left(\frac{\Tnuc/f}{10^{-5.9}}\right)^{\!4}\,.
	\label{eq:DIS}
	\end{equation}
Here we have conveniently normalised to the observed DM density $\YDM\MDM = 0.43$ eV~\cite{Aghanim:2018eyx} with the expectation that $\MDM \approx m_{\ast}$. We have also included a model dependent pseudo-branching fraction, $\mathrm{BR}$, which takes into account that not only DM particles are created in the string fragmentation and DIS but also unstable composite states which decay into the SM. (Alternatively, for secluded dark sectors, the unstable states could decay to dark radiation rather than to the SM). The quantity $\mathrm{BR}$ is not a true decay branching fraction, but seeks to capture the overall suppression of DM production. This suppression may be significant, e.g.~if DM is a heavy baryon of the strong sector, then we expect the light dilaton to be produced much more copiously in the string fragmentation and DIS. Taking into account various possibilities, typical estimates give $\mathrm{BR} \sim 10^{-6}$ to $10^{-3}$~\cite[App. C]{Baldes:2020kam}.

\subsection{Thermal effects}

\label{sec:sub-thermal-abundance}

\begin{figure}[p]
\centering
\vspace{-1.5cm}

\raisebox{0cm}{\makebox{\includegraphics[height=0.49\textwidth, scale=1]{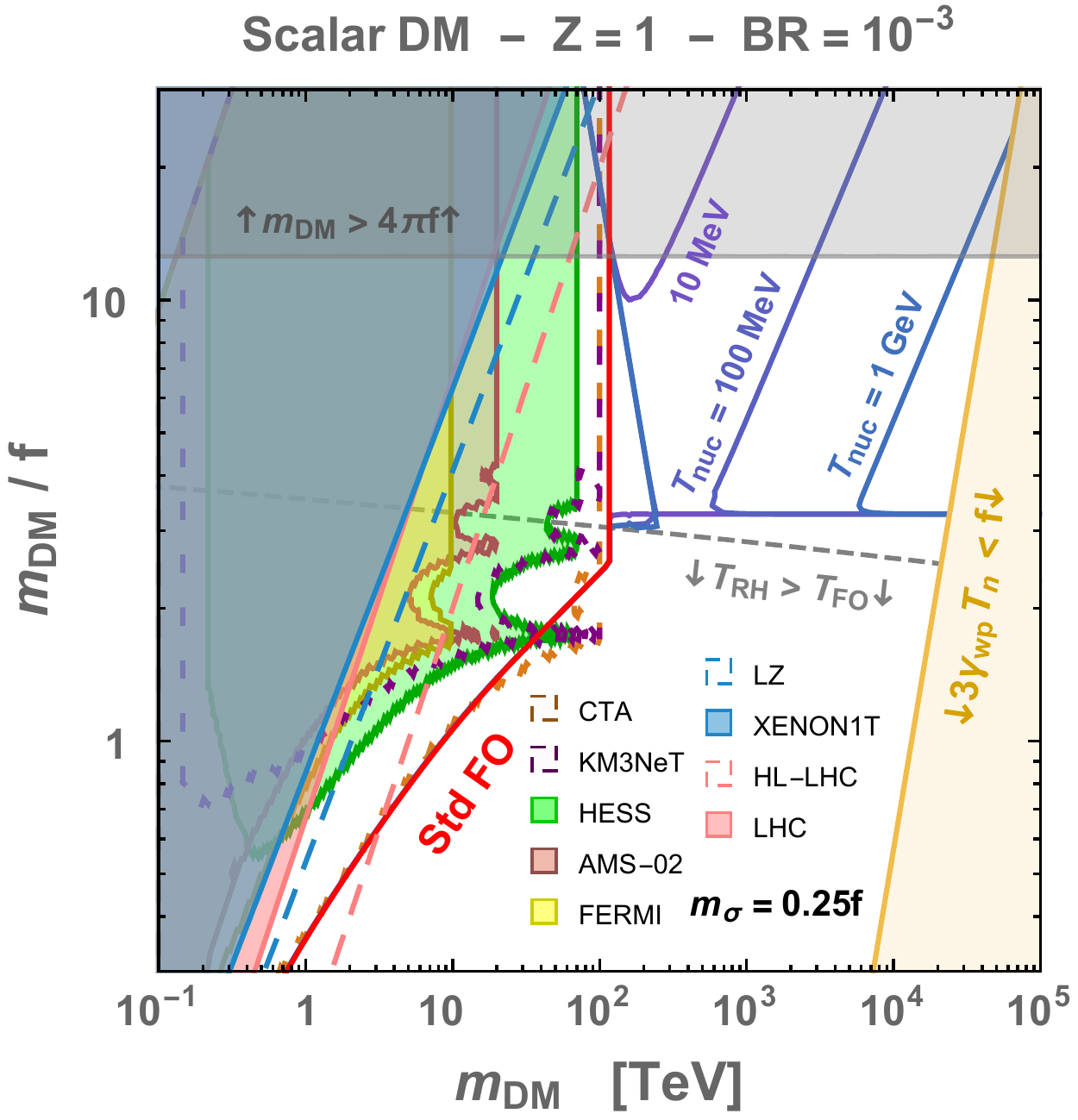}}}
\raisebox{0cm}{\makebox{\includegraphics[height=0.49\textwidth, scale=1]{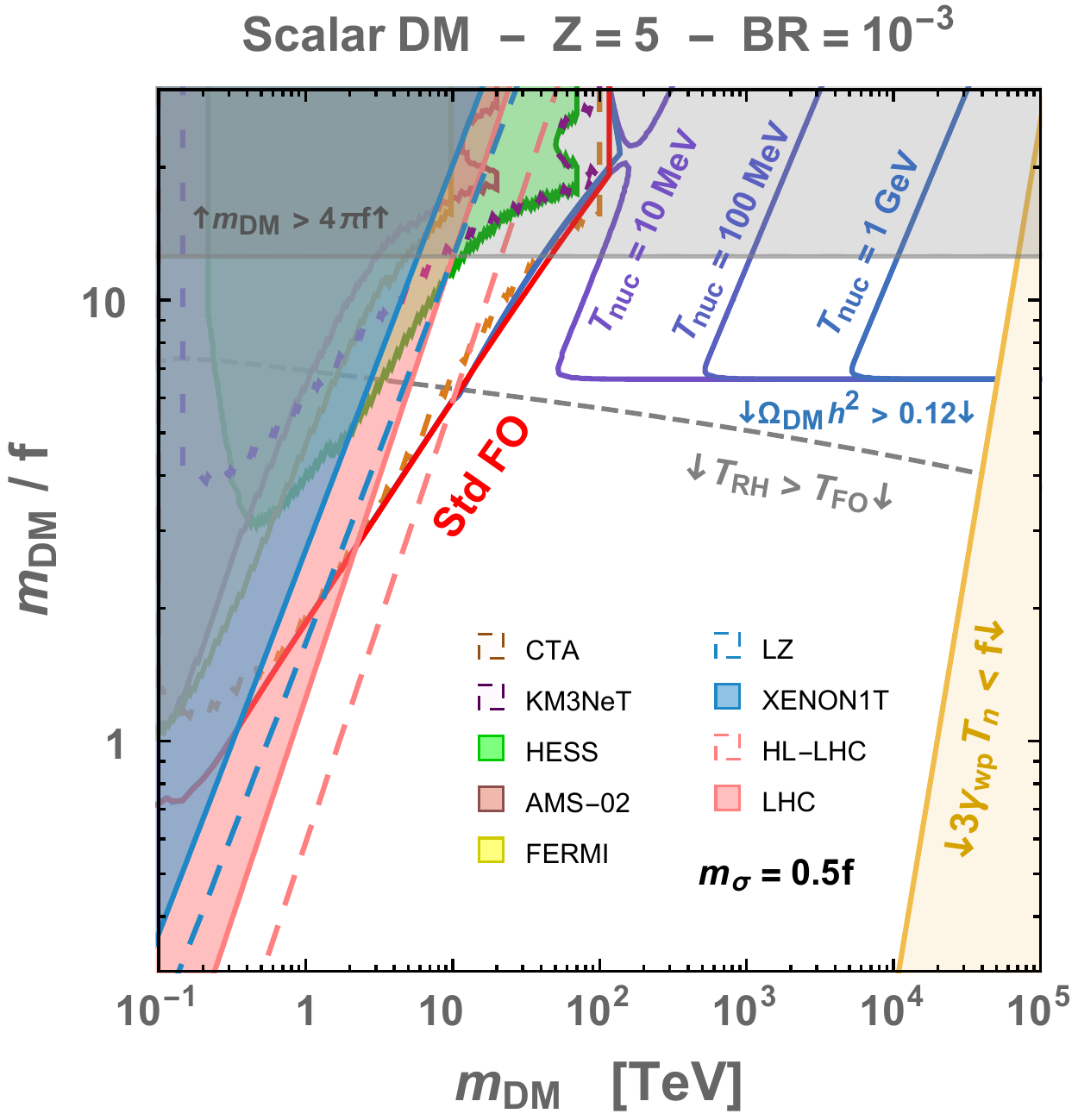}}}
\vspace{.5 cm}

\raisebox{0cm}{\makebox{\includegraphics[height=0.49\textwidth, scale=1]{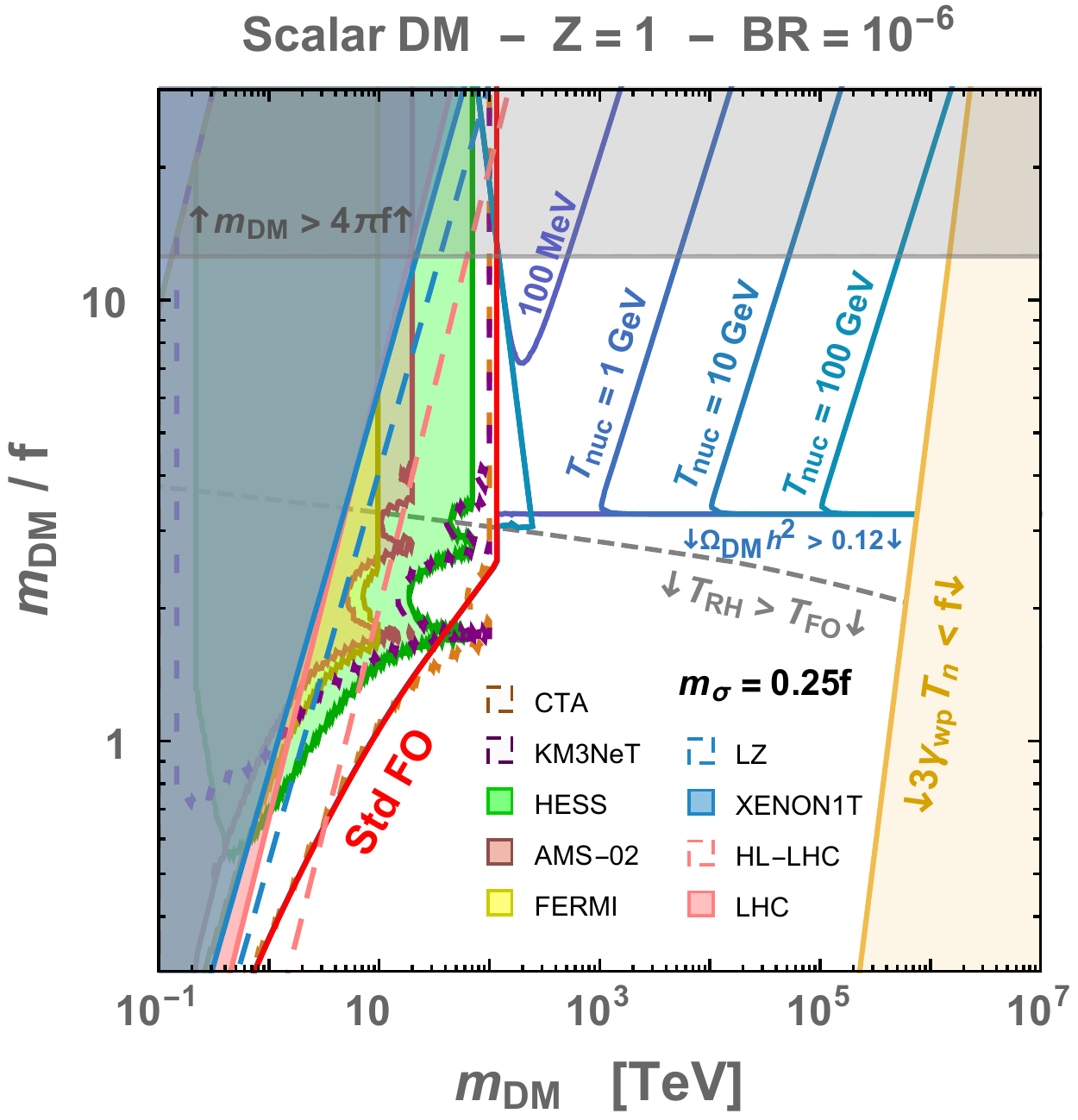}}}
\raisebox{0cm}{\makebox{\includegraphics[height=0.49\textwidth, scale=1]{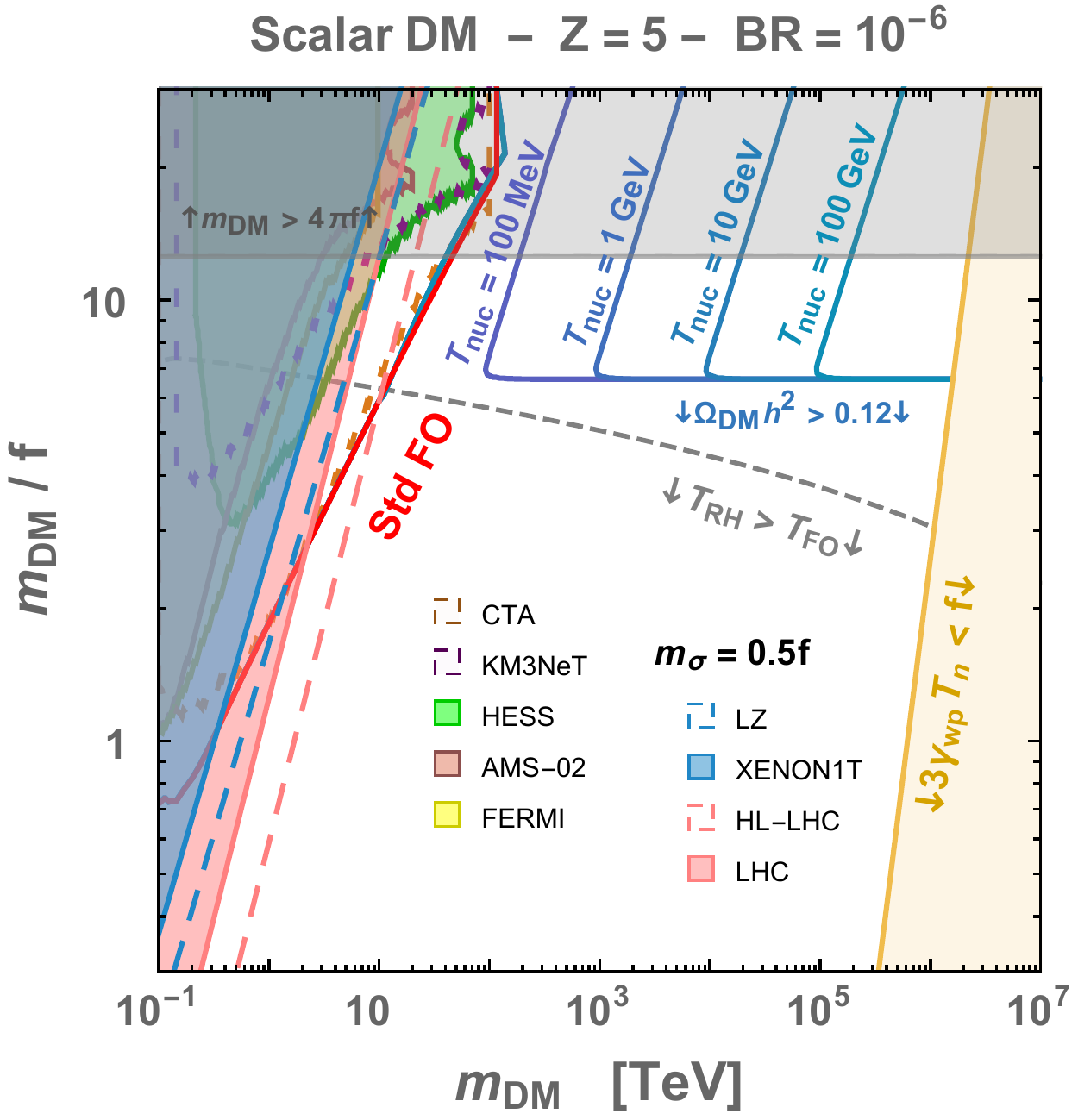}}}

\caption{\it \small \textbf{Supercooled dilaton-mediated scalar DM,}
for branching ratio of DM production (defined in Eq.~\eqref{eq:DIS}) $BR = 10^{-3}$ (top) and $BR = 10^{-6}$ (bottom), and for wave-function renormalization (defined in Sec.~\ref{par:Znormalization}) $Z=1$ (left) and $Z=5$ (right).
When $\MDM \gtrsim 100 ~\mathrm{TeV}$, the unitarity bound on the DM annihilation cross-section leads to DM overproduction, cf. Sec.~\ref{sec:unitary-bound}.
However, in presence of supercooled confinement, DM can have a larger mass, cf. Sec.~\ref{sec:SupercoolDM}. This is displayed by the lines where the correct relic abundance is reproduced for different values of the nucleation temperature $\Tnuc$ (purple to blue).
Below the horizontal blue line the universe is overclosed, as can be understood by the dashed gray line lying very close to it, where the reheating temperature $\Treh$, Eq.~\eqref{eq:Treh_def}, is larger than the would-be freeze-out temperature $T_{\rm FO}$, Eq.~\eqref{eq:x_FO}.
For masses larger than $\sim10^6~\mathrm{TeV}$, the techniquanta of the confining sector do not have enough kinetic energy to penetrate the bubbles~\cite{Baldes:2020kam},  and our picture breaks down (yellow regions).
If DM is a bound-state of the strong sector with confining scale $f$, we do not expect its mass to be $m_{\DM} \gtrsim 4\pi f$ (grey regions). We also display in red the line where the DM abundance would be obtained via the standard freeze-out mechanism, if there was no supercooling.
The current constraints from direct-detection, indirect-detection, and the LHC, displayed with shadings as in Fig.~\ref{fig:stdFO_pheno_msigma_p25a},
 do not probe the regions opened by the presence of supercooling. The vector and fermion cases are shown in Figs.~\ref{fig:DM_SC_abund_vector} and~\ref{fig:DM_SC_abund_fermion}, respectively.
 }
\label{fig:DM_SC_abund}
\thisfloatpagestyle{empty}
\end{figure}

\paragraph{Boltzmann equation.}
Finally, we must determine any non-negligible changes to the yield due to $2 \leftrightarrow 2$ scattering with the thermal bath after supercooling and reheating. (Assuming the DM annihilation products --- or equivalently the  initial states of the inverse annihilation --- consist of relativistic bath particles in thermal equilibrium, as we do here, our estimates also immediately capture any $2 \leftrightarrow N$ type processes because of detailed balance.) Following~\cite{Hambye:2018qjv} we refer to this as the sub-thermal abundance. In this epoch, the DM abundance again evolves according to the Boltzmann equation, Eq.~(\ref{eq:Bmann}), introduced in Sec.~\ref{sec:DM_abundance_freeze-out}. However, the initial conditions are now given by $x_{\rm RH} = \MDM/\Treh$ and $Y_{\mathsmaller{\rm DM}} ^{\mathsmaller{\rm DIS}}$, the abundance after string breaking and DIS estimated in Eq.~\eqref{eq:DIS}. 

We integrate from $x_{\rm RH}$ to $\infty$ and obtain the final DM abundance as a sum of the contribution in Eq.~\eqref{eq:DIS} together with sub-thermal correction. For simplicity, we again neglect the change in velocity dependence arising from the Sommerfeld enhancement for our analytical estimates, but include it in our numerical calculations used for the plots. It is useful to introduce $Y_{\mathsmaller{\rm DM}}^{\rm \mathsmaller{FO}} = Y _{\mathsmaller{\rm DM}}^{\infty}$, i.e.~the standard freeze-out abundance obtained when $Y_{\mathsmaller{\rm DM}} ^{\mathsmaller{\rm DIS}} \gg Y_{\mathsmaller{\rm DM}}^{\rm \mathsmaller{eq}}$. 

\paragraph{Standard freeze-out.}
If the reheating temperature is above the freeze-out temperature, $\Treh > T_{\rm FO}$, the DM abundance quickly relaxes to thermal equilibrium and the standard FO abundance in Eq.~\eqref{eq:Y_DM_FO} is recovered. However, if $\Treh < T_{\rm FO}$, the usual freeze-out mechanism is not recovered. We discuss this case next.

\paragraph{Sub-freeze-out.}
If the DM yield after supercooling is larger than the yield at thermal equilibrium, i.e. if
\begin{equation}
Y_{\mathsmaller{\rm DM}} ^{\mathsmaller{\rm DIS}} \gtrsim Y_{\mathsmaller{\rm DM}}^{\rm \mathsmaller{EQ}}(x_{\rm RH}), \qquad \text{with}\quad Y_{\mathsmaller{\rm DM}}^{\rm \mathsmaller{EQ}}(x) = \frac{45 \gDM}{4\sqrt{2}\,\pi^{7/2} \gSM}x^{3/2} e^{-x},
\end{equation}
then thermal effects annihilate a bit of DM and the final yield is given by 
	\begin{equation}
\label{eq:sub-thermal-bottom-right}
	\frac{1}{Y_{\mathsmaller{\rm DM}}} =
                \begin{dcases}
                   \frac{1}{Y_{\mathsmaller{\rm DM}} ^{\mathsmaller{\rm DIS}}} + \frac{ \lambda_s}{x_{\rm RH}}, \qquad \text{for s-wave,}\\
                  \frac{1}{Y_{\mathsmaller{\rm DM}} ^{\mathsmaller{\rm DIS}}} + \frac{ \lambda_p}{2x_{\rm RH}^2}, \qquad \text{for p-wave.}
                \end{dcases}
	\end{equation}
	
\paragraph{Sub-freeze-in.}
If instead $Y_{\mathsmaller{\rm DM}} ^{\mathsmaller{\rm DIS}} \lesssim Y_{\mathsmaller{\rm DM}}^{\rm \mathsmaller{EQ}}(x_{\rm RH})$, thermal effects produce a bit more of DM and the final yield is given by
	\begin{equation}
	\label{eq:sub-thermal-top-left}
	Y_{\mathsmaller{\rm DM}} =
                \begin{dcases}
                Y_{\mathsmaller{\rm DM}} ^{\mathsmaller{\rm DIS}} +  \lambda_s \frac{2025 \gDM^2}{128 \pi^7 \gSM^2} e^{-2x_{\rm RH}}(1+2x_{\rm RH}),
		   \qquad \text{for s-wave,}\\ 
               Y_{\mathsmaller{\rm DM}} ^{\mathsmaller{\rm DIS}} + \lambda_p \frac{2025 \gDM^2}{64 \pi^7 \gSM^2} e^{-2x_{\rm RH}}, \qquad \qquad \qquad \quad \; \text{for p-wave.}
                \end{dcases}\\
	\end{equation}
	
\paragraph{Heavy thermal Dark Matter beyond the unitarity limit.}
We show the final DM relic abundance lines in Fig.~\ref{fig:DM_SC_abund}. We can see that the introduction of a period of supercooling, ending at a temperature $T_{\rm nuc}$ before the strong sector confines --- shown by the purple to blue lines --- opens the parameter space of DM to masses beyond the unitarity bound at $100$~TeV, shown by the red line. As shown explicitly in Fig.~\ref{fig:sketch_intro}, the DM abundance along the horizontal blue-to-purple line is set by sub-freeze-in, Eq.~\eqref{eq:sub-thermal-top-left}, while the vertical lines are set by sub-freeze-out, Eq.~\eqref{eq:sub-thermal-bottom-right}. Note that whenever the reheating temperature is lower than the freeze-out temperature $\Treh < T_{\rm FO}$, the DM abundance differs from standard freeze-out even for vanishing supercooling (large $\Tnuc$), cf.~Eq.~\eqref{eq:sub-thermal-bottom-right}.
The limitation is set by the yellow region in Fig.~\ref{fig:DM_SC_abund} where the dark quarks have not enough kinetic energy to penetrate inside the bubble and our picture breaks down. 

\paragraph{Impact of dilaton kinetic terms normalization $Z$.}
We comment that for a large renormalization of the dilaton kinetic term equal, $Z=5$, supercooling can be relevant down to values of $\MDM$ which are smaller, by a factor of a few, with respect to the case $Z=1$.
This is due to the suppression of DM annihilations, cf. Eqs.~\eqref{eq:cross-section-DM-WW}, \eqref{eq:cross-section-DM-ss} and \eqref{eq:alpha_def}, and is visible in Fig.~\ref{fig:DM_SC_abund}: the red line where the standard relic abundance would be reproduced is moved to smaller masses, leaving more space for the non-thermal abundance reproduced on the supercooled lines in blue.
The relevance of this comment is limited to natural values of the compositeness scale, $\dilatonvev \lesssim 10$~TeV.
We also note that one cannot increase $Z$ arbitrarily, as this also increases $\Treh$.
Indeed, we checked that for $Z \gtrsim 10$ the supercool parameter space is pushed to the region $\MDM \gtrsim 4\pi f$, which is both unexpected and where our EFT breaks down.

\FloatBarrier
\subsection{Phenomenology}

\begin{figure}[th!]
\begin{center}
\includegraphics[width=220pt]{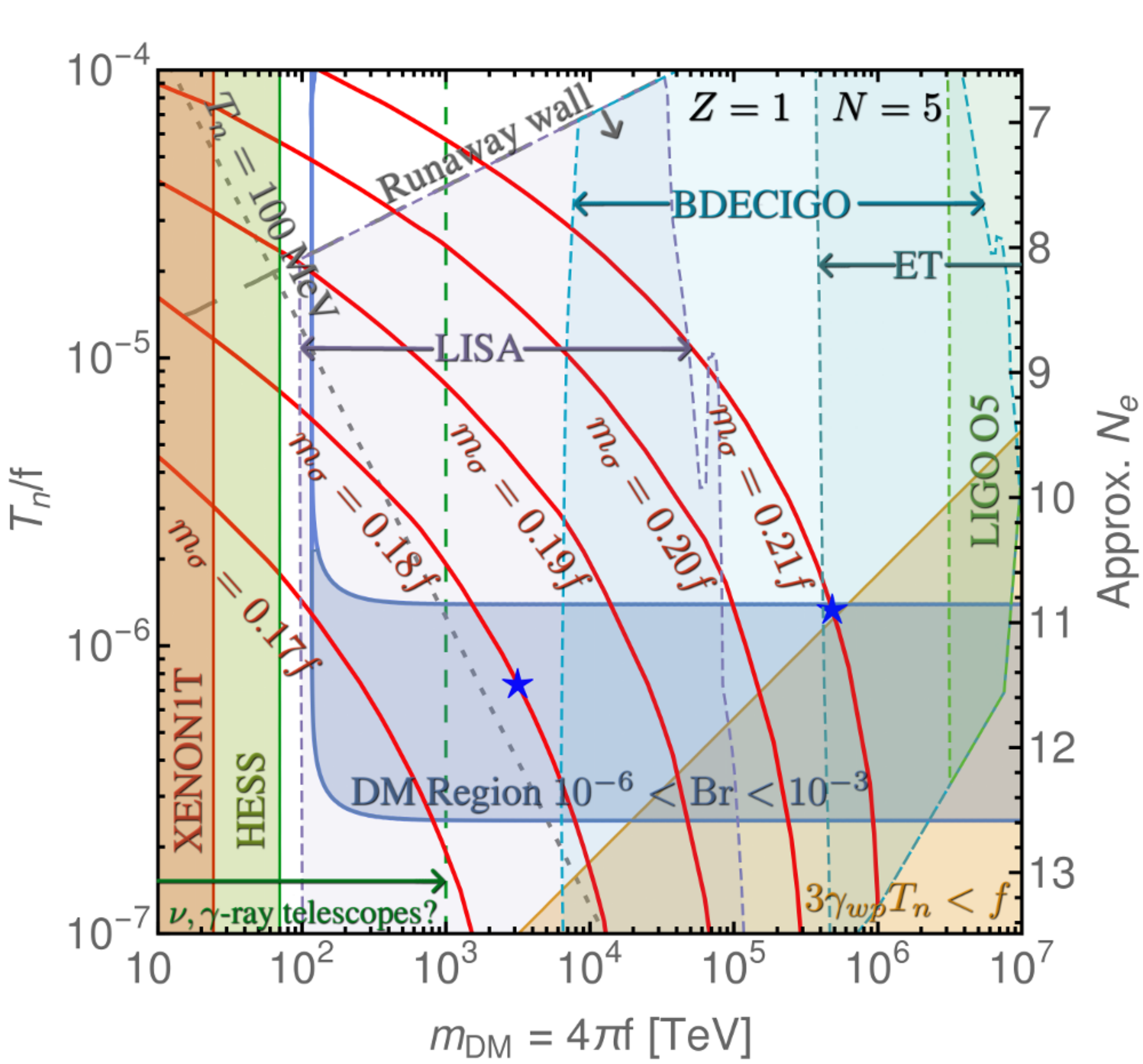} 
\includegraphics[width=220pt]{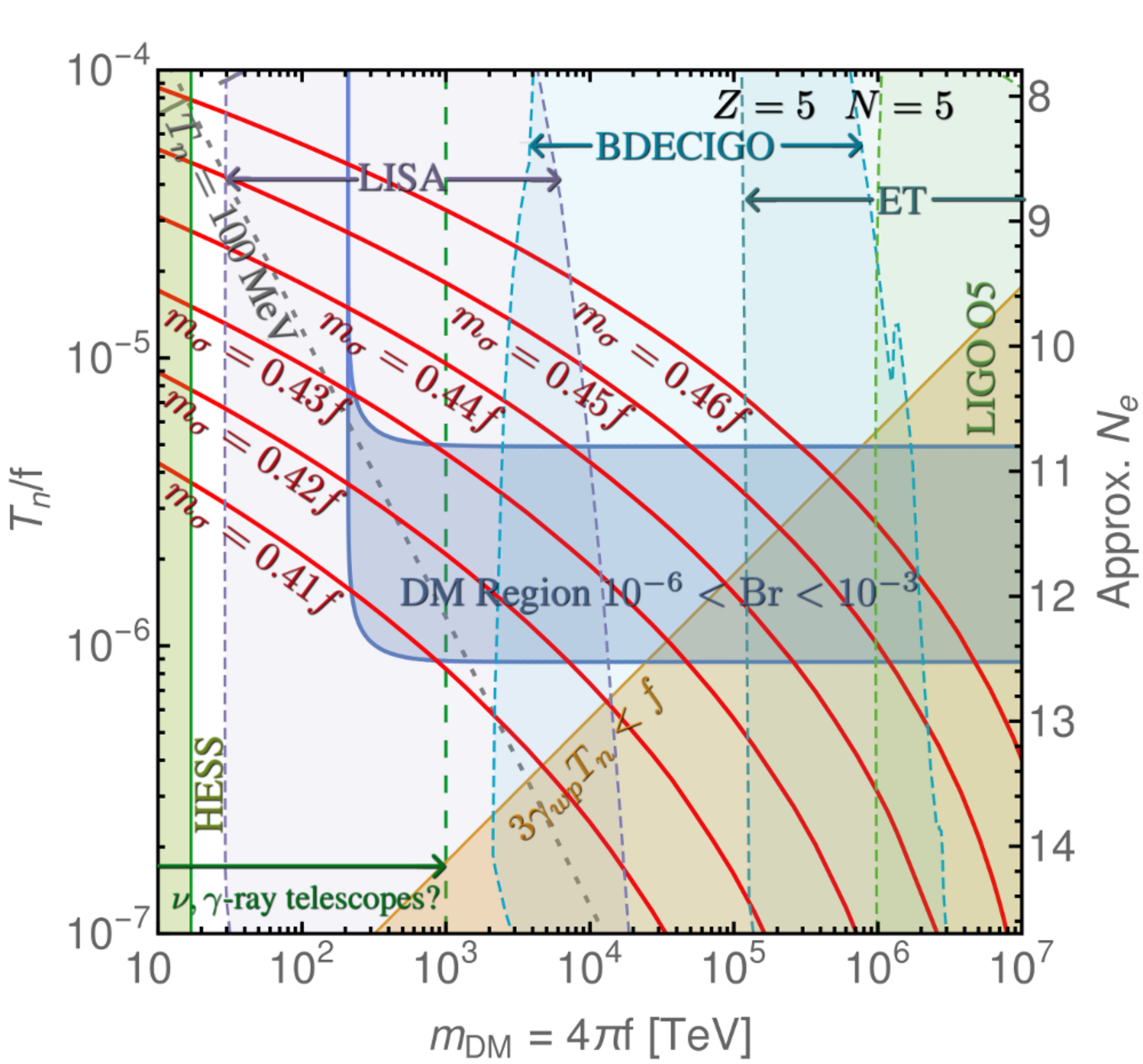}
\end{center}
\caption{\it \small \textbf{Left:} The parameter space for $Z=1$ and $\MDM = 4\pi f$ showing the required supercooling to match the relic abundance in the dark blue region (standard thermal freeze-out is recovered where the dark blue region morphs into a vertical line). The supercooling is achieved for $m_{\sigma} \sim 0.2f$ with the precise mass of the dilaton depending on the scale $f$. The red contours terminate when the universe becomes trapped in the metastable state.
In the yellow region in the bottom-right corner the quarks are insufficiently energetic to enter the first bubble they encounter. The stars indicate two benchmark points for which we show the gravitational wave spectrum in the following section. Regions testable with LISA, BDECIGO, ET $(\mathrm{SNR}_{\rm FGL} > 10^4)$ and LIGO O5 $(\mathrm{SNR}_{\rm FGL} > 10)$ are shown. (The gap between LISA and ET closes for $\mathrm{SNR}_{\rm FGL} = 10^2$.) The GW signal has only been calculated in the runaway regime, so that it is inaccurate in the top-left corner. Current direct and indirect detection limits are shaded in red and green, respectively. The LZ reach is not indicated to avoid clutter, as it falls in the region already excluded by HESS.
The presumed reach of future $\gamma$-ray telescopes to DM annihilations with cross sections close to the unitarity bound, as predicted in our model, is indicated with a dashed green line. \textbf{Right:} The parameter space for $Z=5$. The required supercooling is achieved for $m_{\sigma} \sim 0.45f$.
The number of e-folds has a small log dependence on $m_\sigma$, Eq.~\eqref{eq:Ne_def}, such that the approximate $N_e$ is very accurate for both plots.}
\label{fig:summary} 
\end{figure}

\paragraph{Comparison of experimental coverage.}
Finally, we combine our results about the DM relic abundance with a calculation of the nucleation temperature in Fig.~\ref{fig:summary}. This shows that large areas of parameter space of heavy DM are possible in this scenario provided the dilaton is somewhat light compared to $f$, namely for $m_\sigma \sim 0.2f$ for $Z=1$ and $m_\sigma \sim 0.45f$ for $Z=5$. 
These regions of heavy DM are out of reach of direct detection and collider probes. But they return a strong stochastic gravitational wave background (SGWB), details of which are given in the following section, which leads to a detectable signal-to-noise ratio in LISA, BDECIGO and/or ET (above astrophysical foregrounds).
Projected sensitivities of future $\gamma$-ray and neutrino telescopes, such as CTA~\cite{CTAConsortium:2018tzg}, LHAASO~\cite{Bai:2019khm}, SWGO~\cite{Schoorlemmer:2019gee,Viana:2019ucn} and KM3NeT~\cite{KM3NeT}, typically cut plots at around 100 TeV. Naively extrapolating their results to higher $\MDM$, however, shows such instruments are sensitive up to $\MDM \approx 1000$ TeV, for DM annihilations in the Milky Way with cross section close to the unitarity bound, see Sec.~\ref{par:telescopes_1PeV} for more details. We tentatively also indicate this region in our plots.

\paragraph{Rooms for refinement.}
We wish to emphasise that there are large uncertainties both in our determination of $Y_{\mathsmaller{\rm DM}}^{\mathsmaller{\rm SC}}$, as it involves non-perturbative physics, and our calculation of $T_{\rm nuc}$ in Sec.~\ref{sec:nucleation-temperature}, given the various approximations we have made in modelling the effective potential. Furthermore, QCD corrections should be added if $\Tnuc \lesssim 100$ MeV, indicated by a dashed line in Fig.~\ref{fig:summary}. This adds further model dependence, as the QCD confinement scale can be suppressed in the high temperature phase of the new strong sector, and any QCD corrections to the effective potential come with their own uncertainties~\cite{vonHarling:2017yew,Baratella:2018pxi}.

\section{Stochastic gravitational wave background}
\label{sec:SGWB  prediction}

Supercooled cosmological phase transitions have generated much enthusiasm due to the prediction of a large Stochastic Gravitational Wave Background (SGWB) detectable by future GW interferometers, see \cite{Gouttenoire:2022gwi} for a review. This was first discussed in the context of warped geometry \cite{Randall:2006py,Nardini:2007me,Hassanain:2007js, Konstandin:2010cd, Konstandin:2011dr, vonHarling:2017yew,Fujikura:2019oyi,Bunk:2017fic, Dillon:2017ctw, Megias:2018sxv,Megias:2020vek, Agashe:2020lfz} that relates by holography to  nearly-conformal strong dynamics~\cite{Bruggisser:2018mus, Bruggisser:2018mrt, Baratella:2018pxi,DelleRose:2019pgi,vonHarling:2019gme,Agashe:2019lhy,Baldes:2020kam}. Supercooled PTs were also studied in models described by classically conformal dynamics \cite{Jinno:2016knw, Marzola:2017jzl, Marzo:2018nov, Iso:2017uuu, Baldes:2018emh, Prokopec:2018tnq, Brdar:2019qut,Ellis:2019oqb, Ellis:2020awk,Ellis:2020nnr}.
 In contrast, non-supercooled confining PTs have a weaker SGWB, while still possibly detectable~\cite{Schwaller:2015tja, Aoki:2017aws,Aoki:2019mlt,DeCurtis:2019rxl,Azatov:2020nbe,Bigazzi:2020avc,Ares:2020lbt,Huang:2020crf, Halverson:2020xpg, Reichert:2021cvs}.

\subsection{GW from bubble collision}

\paragraph{PT parameters.}
The SGWB depends on the bulk parameters of the transition, which can be taken to be the nucleation temperature $T_{\rm nuc}$, the wall velocity $v_{w}$, the ratio of the vacuum energy density released in the transition to that of the radiation bath $\alpha_{\rm PT}$, the duration of the phase transition $\beta^{-1}$ and the efficiency of the energy transfer from the vacuum energy to the scalar field gradient $\kappa_{\phi} $:\footnote{A more correct definition of $\beta$ would be $\beta \equiv -(1/\Gamma)(d\Gamma/dt)|_{\rm nuc}$ where $\Gamma$ is the tunneling rate defined in Eq.~\eqref{eq:tunneling_rate}. Our choice, $\beta \equiv (dS/dt)|_{\rm nuc}$, is of course, more conservative when finding the GW amplitude. Furthermore, estimates of $\beta/H \approx 1$ coming from the nucleation temperature tend to become inaccurate and bubble percolation must instead be taken into account.} 
	\begin{align}
	\alpha \equiv \frac{\Delta V}{\rho_{\rm rad}(\Tnuc)} , \qquad  \qquad
		\beta \equiv \frac{dS}{dt}\Big|_{\rm nuc}= -H_* T \frac{dS}{dT} \Big|_{\rm nuc} , \qquad  \qquad
	\kappa_{\phi} \equiv \frac{\rho_{\phi}}{\Delta V}.
	\end{align}
Here $\Delta V$ is the zero-temperature energy difference between the false vacuum and the true vacuum, Eq.~\eqref{eq:vacuum_energy}. Furthermore, $\alpha_{\rm PT}  \gg 1$, $\kappa_{\phi} \simeq 1$, $v_{w} \simeq 1$, and $\beta$ follows from our calculation of $T_{\rm nuc}$.

\paragraph{Choice of modeling.}
In \cite{Baldes:2020kam}, we show that in the regions of interest for DM, the bubbles collide in the runaway regime, therefore, all the vacuum energy is transferred to the gradient of the scalar field.
Here we adopt the bulk flow model~\cite{Jinno:2017fby,Konstandin:2017sat} which shows an IR enhancement of the GW spectrum due to the long-lasting free propagation of the shells of anisotropic energy-momentum tensor after the collision (but also see possible subtleties raised in~\cite{Lewicki:2020jiv,Lewicki:2020azd}). The IR enhancement has been observed in lattice simulations \cite{Cutting:2020nla} where it has been shown to be larger for thick-walled bubbles, which are found in the close-to-conformal vacuum type transitions studied in this work (also see \cite[App. A]{Baldes:2020kam}).\footnote{In contrast, for thin-walled bubbles, after collision the scalar field can be trapped back in the false vacuum, so that instead of propagating freely, the shells of energy-momentum tensor dissipates via multiple bounces of the walls \cite{Konstandin:2011ds,Jinno:2019bxw,Lewicki:2019gmv}.}
We use the estimate of the differential SGWB per logarithmic frequency interval for $v_{w} \simeq 1$ as given in~\cite{Konstandin:2017sat}
	\begin{equation}
	 h^2\Omega_{\rm \phi} = 1.67 \times 10^{-5}\,  \left(\frac{H_*}{\beta} \right)^2\, \left( \frac{\kappa_{\phi} \alpha_{\rm PT}}{1+\alpha_{\rm PT}} \right)^2  \,  \left( \frac{0.0866 v_{\rm w}^3}{1 + 0.354v_{\rm w}^2}\right) \bigg( \frac{ 100 }{ g_{\rm SM} }\bigg)^{1/3} \; S_{\phi}(\nu),
	\end{equation} 
where $S_{\phi}(\nu)$ describes the spectral shape
	\begin{equation}
	\label{eq:spectral_shape_scalar}
	S_{\phi}(\nu) = \frac{ 3\nu_{\phi}^{2.1}\nu^{0.9} }{2.1\nu_{\phi}^3+0.9\nu^{3}},
	\end{equation}
and the peak frequency is
	\begin{equation}
	\nu_{\phi} = 2.62 \times 10^{-3} ~ \text{mHz} \; \left( \frac{ 1.24 }{ 1 - 0.047v_{w} + 0.58v_{w}^2} \right)   \left( \frac{\beta}{H_*} \right) \left( \frac{\Treh}{100~\text{GeV}}  \right)\bigg( \frac{ g_{\rm SM} }{100}\bigg)^{1/6},
	\end{equation}
and  in the above we assume the bath is dominated by $g_{\rm SM}$ following reheating. Compared to the envelope approximation \cite{Kamionkowski:1993fg,Huber:2008hg,Jinno:2016vai}, the bulk flow model predicts a softer IR and harder UV spectrum for supercooled transitions, while the peak frequency and amplitude remains relatively unchanged. In addition to the above, for redshifted frequencies smaller than $\nu_{\ast} = H_{\ast}/(2\pi)$ we impose an $\nu^{3}$ scaling as required by causality~\cite{Durrer:2003ja,Caprini:2009fx,Cai:2019cdl,Hook:2020phx}.

\subsection{Signal-to-noise ratio}
\paragraph{Most optimistic SNR.}
One can check that the entire supercooled region of this model returns observable signatures at current or future interferometers. We use the signal-to-noise ratio~\cite{Thrane:2013oya,Caprini:2019pxz}\footnote{Equation~\eqref{eq:snr} assumes the signal is sub-dominant to the noise in the detector. In the case in which the signal is large compared to the noise, the SNR can be roughly approximated by making the replacement $\Omega_{\rm sens}^2 \to 2\Omega_{\rm GW}^2 + 2\Omega_{\rm GW}\Omega_{\rm sens} + \Omega_{\rm sens}^2$ in the denominator of Eq.~\eqref{eq:snr}~\cite{Allen:1997ad,Kudoh:2005as,Brzeminski:2022haa}. The SNR eventually saturates to a maximum value of $\sim 10^{3}$ for LISA,  $\sim 10^{4}$ for BDECIGO, and $\sim 10^{5}$ for ET. As these are well inside the detectable regimes of parameter space, our qualitative results are not affected by assuming the simple noise-dominated SNR used in the literature.}
	\begin{equation}
	\mathrm{SNR} = \sqrt{t_{\rm obs} \int \left(\frac{\Omega_{\rm GW}(\nu)}{\Omega_{\rm sens}(\nu)}\right)^{2} d\nu}, 
	\label{eq:snr}
	\end{equation}
where $t_{\rm obs}$ is the observation time. (We set $t_{\rm obs} = 3$ years for LISA and BDECIGO, 10 years for ET, 30 days for LIGO Hanford-Livingston O1~\cite{TheLIGOScientific:2016dpb}, 99 days for Hanford-Livingston O2~\cite{LIGOScientific:2019vic}, 168 days for Hanford-Livingston O3~\cite{Abbott:2021xxi}, 160 days for Hanford-Virgo and Livingston-Virgo O3~\cite{Abbott:2021xxi}, and  1 year for LIGO O5.) Here $\Omega_{\rm sens}(\nu)$ encodes the sensitivity of the detector,
	\begin{equation}
	h^2\Omega_{\rm sens}(\nu) = \frac{2\pi^{2}\nu^{3}}{3H_{100}^2}S_{n}(\nu),
	\end{equation}
where $S_n$ is the noise spectral density, either for LISA~\cite{Caprini:2019pxz}, BDECIGO~\cite{Isoyama:2018rjb,Breitbach:2018ddu}, or ET~\cite{Hild:2010id}.\footnote{There is a factor of two difference in the formula for $\Omega_{\rm sens}(\nu)$ in~\cite{Caprini:2019pxz}, due to a difference in the definition of $S_n$, which we take into account.} For LIGO and Virgo cross correlations we use~\cite{Thrane:2013oya}
	\begin{equation}
	h^2\Omega_{\rm sens}(\nu) = \frac{\pi^{2}\nu^{3}}{3H_{100}^2}\sqrt{ \frac{ 2  S_{n}^{\rm i}S_{n}^{\rm j} }{ \Gamma_{\rm ij}^{2} } },
	\end{equation}
where $S_{n}^{\rm i}$ and $S_{n}^{\rm j}$ are the noise spectral densities for the interferometers, and $\Gamma_{\rm ij}$ is the overlap reduction function\footnote{For the $S_{n}^{\rm i}$ see: \\ https://dcc.ligo.org/LIGO-T1600302/public, \\ https://dcc.ligo.org/LIGO-T1500293-v13/public, \\  https://dcc.ligo.org/LIGO-T1800042-v5/public. \\ For $\Gamma_{\rm ij}$ see  https://dcc.ligo.org/public/0022/P1000128/026/figure1.dat.}, and the cross correlation SNRs are combined in quadrature. Using the above SNR, the usual power-law-integrated sensitivity curves can be constructed~\cite{Thrane:2013oya}. Performing this exercise, we see our results agree with~\cite{Caprini:2019pxz} for LISA, \cite{Breitbach:2018ddu} for BDECIGO, and~\cite{Gouttenoire:2019kij} for ET. However, we find our curves are approximately a factor of five more sensitive for LIGO-Virgo O3 and O5 compared to those found in the full LIGO-Virgo collaboration analysis~\cite{LIGOScientific:2019vic,Abbott:2021xxi} (although the shapes match). We therefore weaken our respective SNRs for LIGO-Virgo by these factors, in order to conservatively mimic the complete analysis. 

\begin{figure}[t]
\begin{center}
\includegraphics[width=220pt]{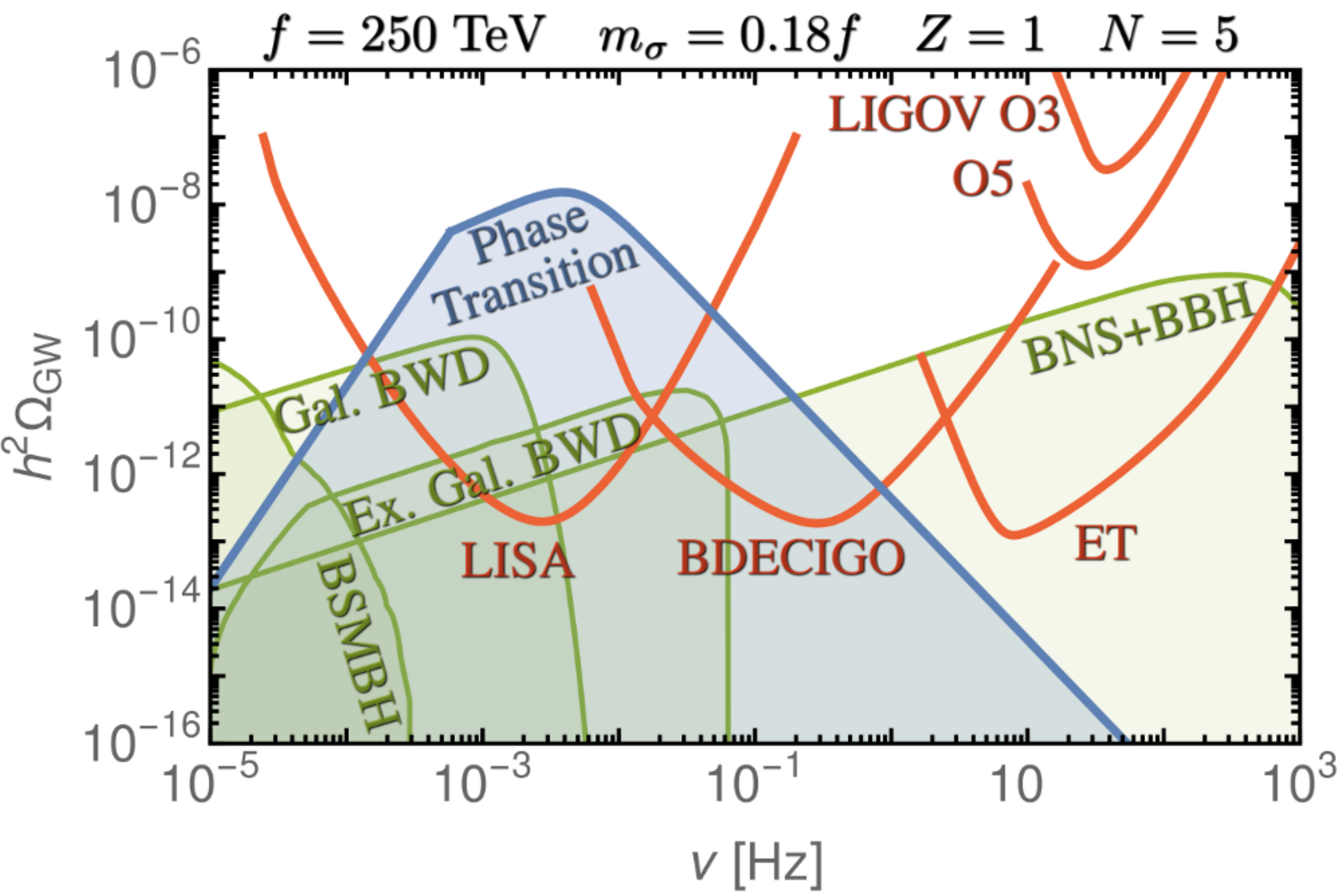} \;
\includegraphics[width=220pt]{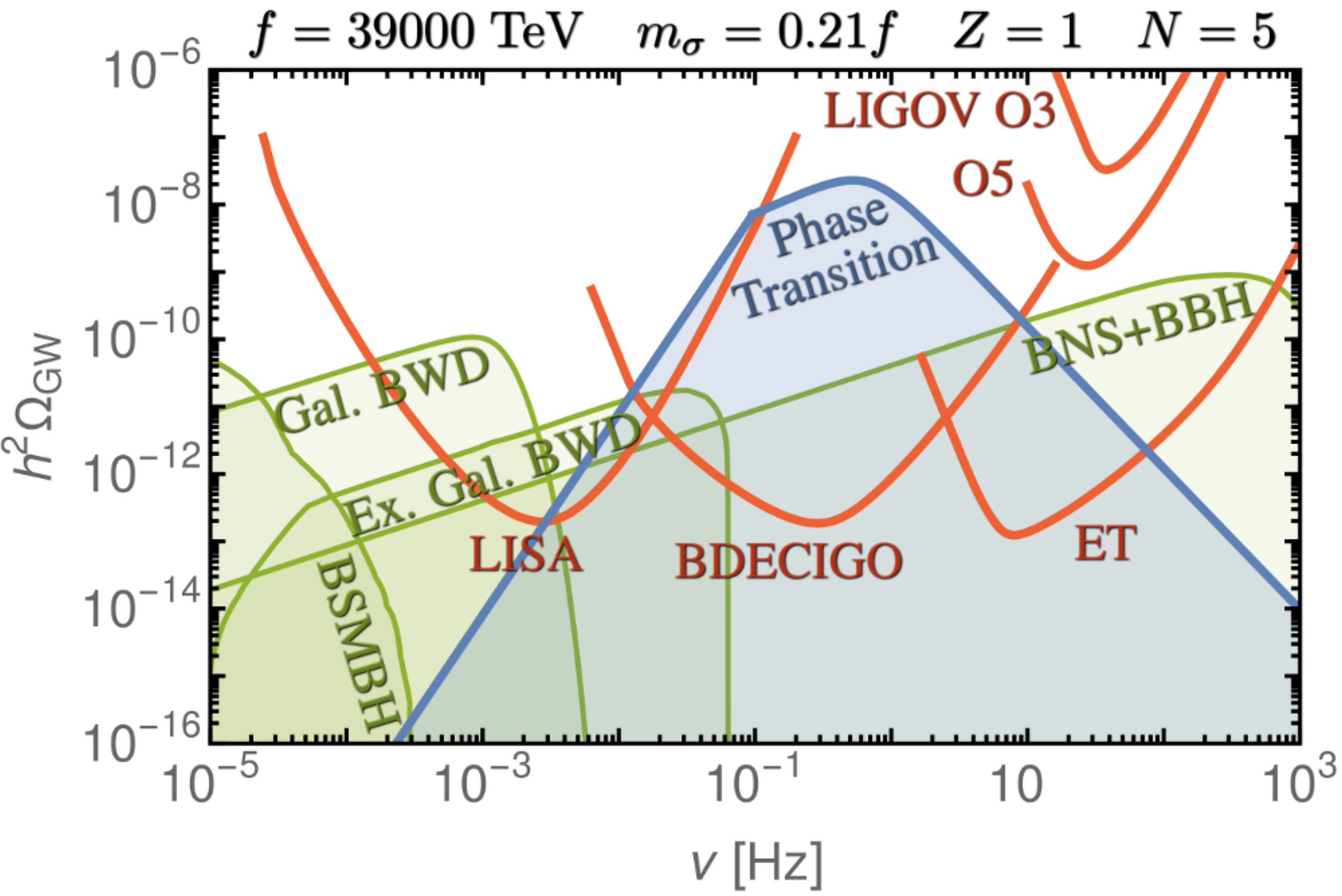}
\end{center}
\caption{\it \small The stochastic gravitational wave background sourced by the bubble collisions following the supercooled phase transition for the two benchmark points. Foregrounds due to super-massive black hole, galactic and extragalactic white dwarf, neutron star, and black hole binaries are also shown~\cite{Rosado:2011kv,Cornish:2018dyw,Abbott:2021xxi}. Foreground subtraction techniques continue to be developed~\cite{Cornish:2018dyw,Cutler:2005qq,Adams:2010vc,Adams:2013qma,Sachdev:2020bkk,Flauger:2020qyi,Boileau:2020rpg,Martinovic:2020hru}. The power-law integrated sensitivity curves SNR $= 10$, defined in Eq.~\eqref{eq:snr}, of various current and future detectors are also shown, constructed using the projected sensitivities described in the text.}
\label{fig:GW_Example} 
\end{figure}

\begin{figure}[t]
\begin{center}
\includegraphics[width=220pt]{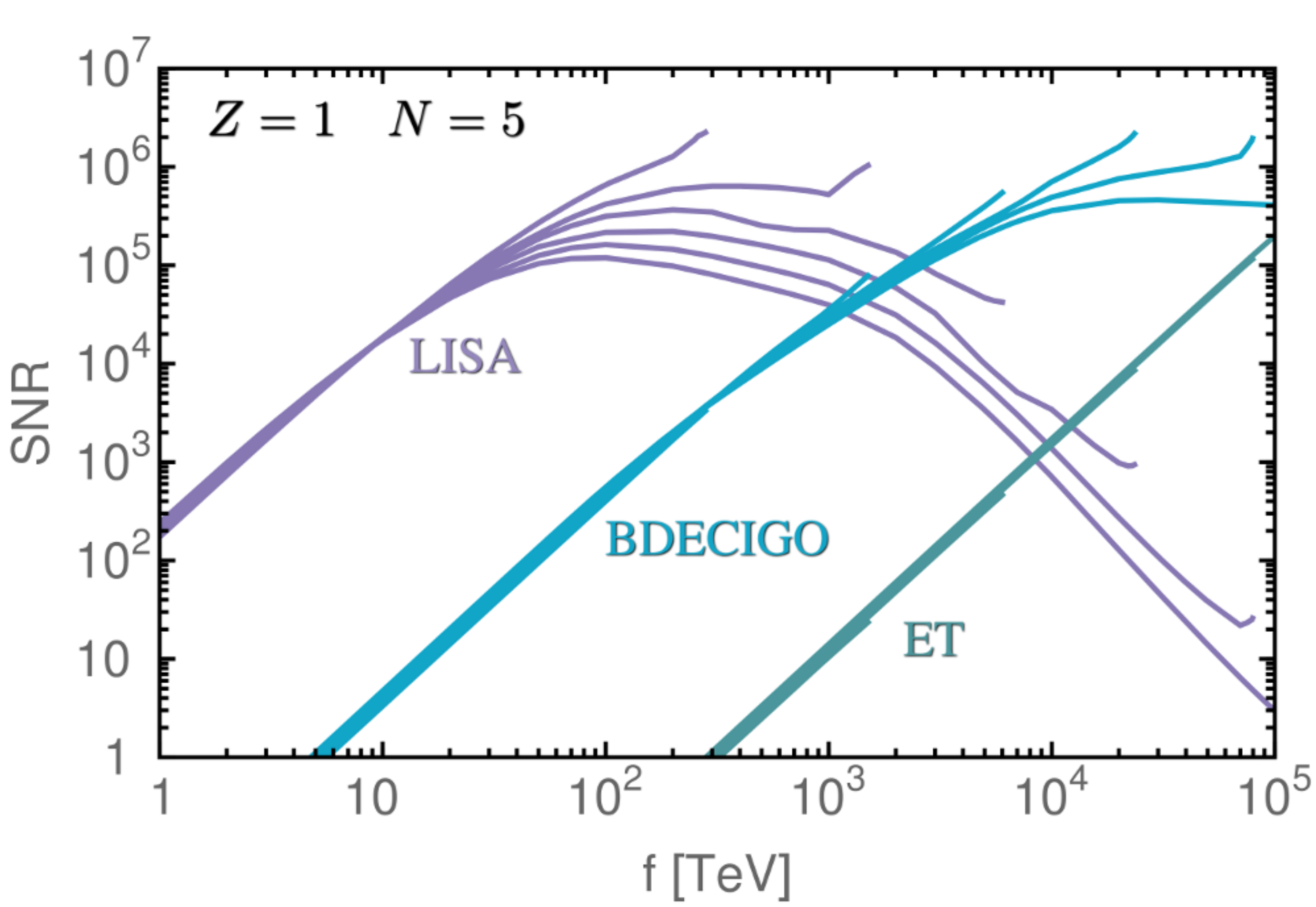} \;
\includegraphics[width=220pt]{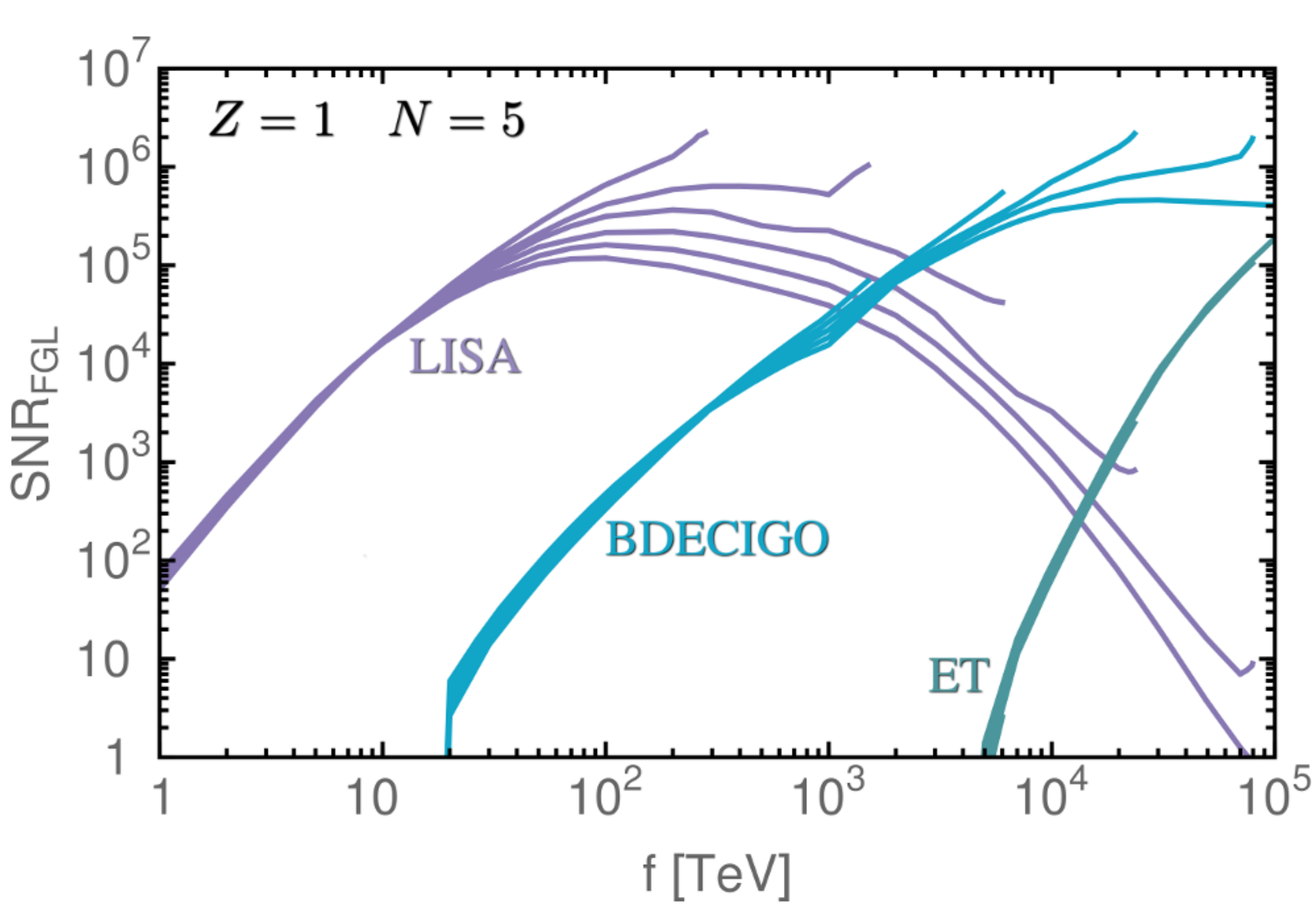}
\end{center}
\caption{\it \small The signal-to-noise ratio estimated for future measurements at LISA, BDECIGO, and ET. The curves have been calculated for $Z=1$ and $m_{\sigma}=0.17f$, $0.18f$, $0.19f$, $0.20f$, $0.21f$, $0.22f$ (from top to bottom). The curves terminate when the universe becomes stuck in the false vacuum. The plot on the right assumes degradation of the achievable signal due to white dwarf and neutron star binaries, cf. Eq.~\eqref{eq:snrfgl}. Even without foreground subtraction, the entire parameter space of this scenario can be tested with a combination of these interferometers.}
\label{fig:GW_SNR} 
\end{figure}

\paragraph{Foreground-limited SNR.}
The above acts as a simple measure as to whether a signal is detectable. Full reconstruction requires more advanced techniques which we do not seek to emulate here but detailed analysis shows some signal is recoverable provided $\mathrm{SNR} > 10$~\cite{Caprini:2019pxz}. Furthermore, there is an issue of contamination by astrophysical foregrounds, $\Omega_{\rm FG}(\nu)$. Dealing with such confusion noise requires more advanced techniques still under development~\cite{Flauger:2020qyi,Boileau:2020rpg,Martinovic:2020hru}. In absence of such a treatment, but to still show that the phase transitions studied here are detectable, we define a foreground limited SNR
	\begin{equation}
	\mathrm{SNR}_{\rm FGL} = \sqrt{t_{\rm obs} \int \left(\frac{ \mathrm{Max} [0,\Omega_{\rm GW}(\nu)-\Omega_{\rm FG}(\nu)]} {\Omega_{\rm sens}(\nu)} \right)^{2} d\nu}.
	\label{eq:snrfgl}
	\end{equation}
The first foreground we take into account here corresponds to the component of galactic white dwarf binaries which cannot be subtracted after four years of LISA datataking~\cite{Cornish:2018dyw} (with the sign error correction noted in~\cite{Schmitz:2020rag}). We also include extragalactic white dwarf binaries~\cite{Rosado:2011kv}, together with the binary black hole and binary neutron stars from the median value shown in~\cite{Abbott:2021xxi} (extrapolated to lower frequencies by assuming a $\propto \nu^{2/3}$ scaling). For other estimates see~\cite{Farmer:2003pa,Regimbau:2011rp,Perigois:2020ymr}. Note the binary neutron star and binary black hole signals could to some extent also be subtracted~\cite{Cutler:2005qq,Adams:2010vc,Adams:2013qma,Sachdev:2020bkk}. Example SGWB spectra for the two benchmark points, together with power-law-integrated sensitivity curves, and astrophysical foregrounds are shown in Fig.~\ref{fig:GW_Example}. Estimates of the signal-to-noise ratio are displayed in Fig.~\ref{fig:GW_SNR} showing our scenario is testable at upcoming experiments. Similarly, we have shown areas with large values of $\mathrm{SNR}_{\rm FGL}$ in our summary plots in Fig.~\ref{fig:summary}. Current limits do not constrain our scenario~\cite{Romero:2021kby}.

\subsection{Analytical estimate}

\paragraph{Supercooled PTs have large bubbles.}
The scenario is generically testable because small values $\beta/H \sim \mathcal{O}(10)$ are typical of supercooled PTs from nearly-conformal potentials~\cite{Randall:2006py, Nardini:2007me, Konstandin:2011dr, Baldes:2018emh, Baratella:2018pxi,Gouttenoire:2022gwi}. As a result, the bubble radius at collision, $r_{\rm coll} \sim v_w\,\beta^{-1}$, is larger and the typical GW signal is enhanced. 

\paragraph{Beta parameter in the thick-wall limit.}
Along with our numerical determination above, this is well illustrated by making use of an analytical estimate for the beta parameter, starting with the $O(4)$ action in the thick wall limit 
\begin{equation}
S_4 =\frac{   c_* \, 2 \pi^2 Z^2 f^2}{m_{\sigma}^2\log{(T_c/T)}},
\end{equation}
following a derivation given in App. B. Next, we characterize the duration of the supercooling by the number of e-foldings $N_e \equiv \mathrm{log}(T_{\rm start}/\Tnuc) $ where $T_{\rm start}$ is the temperature when the inflation stage starts, defined in Eq.~\eqref{eq:Tstart_def}. From this, we obtain 
	\begin{equation}
	\frac{\beta}{H_*}  \approx 10 \left( \frac{0.2 Zf}{ m_{\sigma}} \right)^2 \left(\frac{10}{N_{\mathsmaller{\rm e}}}\right)^2,
	\end{equation}
showing approximate analytical agreement with our numerical result.
  
\subsection{Comments}
Three caveats to our conclusions are in order:
\begin{itemize}  
\item \textbf{QCD catalysis.} We again caution that the eventual inclusion of QCD effects could change the predicted behaviour of the PT and hence GW signal if $\Tnuc \lesssim 100$~MeV~\cite{vonHarling:2017yew}. On the other hand, if QCD confinement is delayed due to the altered $\beta$ function from the BSM field content, then these effects may be completely negligible.
\item \textbf{Matter-dominated era.} The rapid decay of the dilaton means there is no early matter dominated phase following the phase transition, which would lead to additional redshifting and a weaker GW signal~\cite{Barenboim:2016mjm, Baldes:2018emh, Cai:2019cdl, Guo:2020grp, Ellis:2020nnr, Hook:2020phx}, as occurs the Coleman-Weinberg potential of radiative symmetry breaking studied in~\cite{Hambye:2018qjv}.

The rapid decay in our scenario arises due to the dilaton coupling generating the kinetic term of the SM Higgs, $(\sigma/f) (D_{\mu}H)^2$, which is present due to our implicitly assuming the Higgs is a pNGB of the same sector that breaks scale invariance.

In the Coleman-Weinberg example, the reheating instead occurs through a Higgs portal interaction of the form, $\lambda_{h \sigma}\sigma^{2}|H|^{2}$, where in order to generate the Higgs mass term, one needs $\lambda_{h \sigma} \sim (v_{\rm H}/f)^{2} \ll 1$ in the limit $f \gg v_{\rm H}$. Hence, in this case, the decay of the condensate is suppressed compared to Hubble following the PT.

\item \textbf{Hidden sector confinement.} The above construction, in particular the dilaton interactions with the SM, Eq.~\eqref{eq:eff_lag_dilaton}, is that of a composite Higgs.

One may instead wish to consider a scenario in which the EW sector is external to the confining dark sector CFT. The scenario remains viable with the following changes:
	\begin{enumerate}
\item To avoid matter domination after the phase transition, which falls outside the scope of our calculation, one requires rapid decay of the dilaton. This can be achieved by having the dilaton decay to lighter hidden sector degrees of freedom, as any portal coupling to the SM is now expected to be small (in analogy to the Coleman-Weinberg example).
\item Similarly, along these lines, direct detection constraints are severely weakened due to the suppressed portal coupling.
\item Indirect detection is also modified, with additional steps in the cascade possible if all light hidden sector degrees of freedom eventually decay to the SM. This leads to softer spectra and constraints that could be weaker or stronger, depending on the DM mass.
If the dilaton decay products instead cascades into dark radiation there is, of course, no measurable indirect detection signal apart from possible changes to $N_{\rm eff}$. (Leaving only gravitational waves from the phase transition.) 
\end{enumerate}
\end{itemize}

\section{Summary and Outlook}
\label{sec:outlook} 

We have considered Dark Matter as a composite state of a new confining sector where the dilaton, the pNGB associated with approximate scale symmetry (corresponding to the radion in the 5D dual description, see App.~\ref{app:5D} for the dictionary), mediates the interactions of DM to the Standard Model.
We have provided a  detailed analysis of this well-defined scenario, improving over ~\cite{Bai:2009ms,Blum:2014jca,Efrati:2014aea,Kim:2016jbz,Fuks:2020tam} by including i) a non-canonical kinetic term for the dilaton, ii) recent experimental constraints, iii) the possibility that DM is also a pNGB, see Figures~\ref{fig:stdFO_pheno_msigma_p25a} and~\ref{fig:stdFO_pheno_msigma_p25}.
 
This scenario naturally predicts a supercooled phase transition in the early universe. This induces a stage of inflation that dilutes the pre-existing abundance of any particle, and so allows for thermal DM masses well beyond the Griest and Kamionkowski unitarity bound of O(100) TeV~\cite{Griest:1989wd,Gouttenoire:2022gwi}.\footnote{Other ways to evade this bound are, for example, DM dilution after a matter era \cite{Giudice:2000ex,Patwardhan:2015kga,Berlin:2016vnh,Berlin:2016gtr,Bramante:2017obj,Hamdan:2017psw,Allahverdi:2018aux,Cirelli:2018iax,Contino:2018crt,Chanda:2019xyl,Contino:2020god,Garani:2021zrr}, or having a dark sector being much cooler than SM \cite{Heurtier:2019eou,Hambye:2020lvy}, DM becoming heavy only after freezing-out \cite{Davoudiasl:2019xeb}, DM annihilating with one spectator field \cite{Kramer:2021hal,Bian:2021vmi} or with many of them \cite{Kim:2019udq}, DM forming an extended object which undergoes a second annihilation stage \cite{Harigaya:2016nlg, Mitridate:2017oky,Geller:2018biy}. Mechanisms involving phase transitions include the possibility of a short inflationary stage associated with perturbative DM mass generation~\cite{Hambye:2018qjv}, DM filtered \cite{Baker:2019ndr,Chway:2019kft,Ahmadvand:2021vxs} or squeezed-out \cite{Asadi:2021pwo,Asadi:2021yml,Gross:2021qgx} by non-relativistic bubble wall motion, DM produced by elastic bubble-bubble collisions \cite{Falkowski:2012fb} or perturbative plasma interactions with relativistic walls \cite{Azatov:2021ifm}. (In contrast with these last possibilities, the mechanism which we studied in \cite{Baldes:2020kam} and which we review in the present work, relies on a strongly interacting theory, with relativistic bubble walls, which come from effective potentials which result in inelastic bubble wall collisions.)}

DM production in this scenario was studied in~\cite{Baldes:2020kam}, which pointed out novel dynamical features and proposed a way to model them, that affected the DM abundance by orders of magnitudes with respect to the previous (wrong) treatment. In this paper we have built on the results of~\cite{Baldes:2020kam} (synthesised in Sec.~\ref{sec:SupercoolDM}) by including in our description the dilaton. 
We made progress in two respects:

\begin{itemize}
\item[1.] We have computed the nucleation temperature $\Tnuc$, where the universe transits to the confined vacuum via bubble nucleation, as a function of the dilaton mass ($\Tnuc$ was treated instead as a free parameter in~\cite{Baldes:2020kam}).
The essential message can be read in Figure~\ref{fig:sketch_intro}, which is a simplified version of
Figs.~\ref{fig:DM_SC_abund}, \ref{fig:DM_SC_abund_vector} and \ref{fig:DM_SC_abund_fermion}.
Novelties with respect to previous works include:
\begin{itemize}
\item[1.1] Unlike in previous studies of DM yields from supercooling, which used classically scale invariant potentials~\cite{Hambye:2018qjv}, we found reheating following the PT is generally efficient in this scenario. This is desirable aspect from the viewpoint of achieving a strong signal of GWs, as a period of matter domination following the PT leads to a suppression of the GW signal~\cite{Barenboim:2016mjm, Baldes:2018emh, Cai:2019cdl, Guo:2020grp, Ellis:2020nnr, Hook:2020phx}.
\item[1.2] On a more technical side, we have numerically computed $\Tnuc$ and investigated its dependence on the dilaton wavefunction renormalization $Z$, see e.g. Fig.~\ref{fig:IB_supercooling_nuc_temp}, and we have quantified in detail the $O(3)$ vs $O(4)$ contributions to nucleation together with the uncertainty of the analytical and other approximations, see App.~\ref{app:O(3)_vs_O4}.
\end{itemize}
\item[2.] We have studied the signals of the model at gravitational waves interferometers, colliders, direct and indirect detection experiments, for the cases where DM is a heavy scalar, fermion or vector resonance of the strong sector, or a pNGB. The progress with respect to previous literature can be summarised as follows:
\begin{itemize}
\item[2.1] To the best of our knowledge, this is the first quantitative analysis of the interplay of gravitational wave interferometers with other telescopes in testing models of heavy DM.
The summary of our results is shown in Fig.~\ref{fig:summary}: while only GWs will potentially test all the parameter space up to $\MDM \sim 10^6$~TeV, for DM masses of hundreds of TeV telescopes will provide a complementary access to the same models, realising a new kind of multimessenger approach to heavy DM.

More information on the gravitational wave signals is reported in Sec.~\ref{sec:SGWB  prediction} (see e.g. Fig.~\ref{fig:GW_Example} for two spectra), and a lot more details on the DM phenomenology are provided in Figs.~\ref{fig:DM_SC_abund}, \ref{fig:DM_SC_abund_vector} and \ref{fig:DM_SC_abund_fermion} in the supercooled case, and in Figs.~\ref{fig:stdFO_pheno_msigma_p25a}, \ref{fig:stdFO_pheno_msigma_p25}, \ref{fig:stdFO_pheno_app} and~\ref{fig:stdFO_pNGB_pheno_app} in the standard freeze-out one. Note that our results also constitute both an improvement and an update in the field of standard frozen-out composite DM, of particular relevance if the composite sector is motivated by the electroweak hierarchy problem.
\item[2.2] We have calculated and reported a large number of quantities necessary to determine the signals of these models, in some cases completing and extending previous literature, for example with the dependence on $Z$ of these results.
In particular, the interested reader can find the DM and dilaton couplings in Sec.~\ref{sec:DM in Composite Higgs}, the non-relativistic DM potential relevant for the computation of Sommerfeld enhancement (including the case of pNGB DM) in App.~\ref{app:NR_potential}, the DM cross sections for annihilation in App.~\ref{app:ann_cross-section} and for direct detection in App.~\ref{app:DD_cross-section}, the dilaton decay widths in App.~\ref{app:dilaton_decay_width}, and finally Gravitational Waves predictions in Sec.~\ref{sec:SGWB  prediction}.
\end{itemize}
\end{itemize}

Although we have assumed for definiteness the Higgs to be a pNGB from the same sector of DM and the dilaton, the results summarised at point 1., as well as the important message that GWs and telescopes could interplay in testing these models, are not modified if instead the Higgs belong to another sector, as long as the dilaton decay width is not too suppressed.

In our calculation, a large number of uncertainties have entered. These include the modelling of the non-perturbative physics of string formation and breaking, deep inelastic scattering in the early universe, QCD effects in the effective potential at low temperatures, and the uncertainties in the predictions of the stochastic GW background. These aspects need to be improved in order to extract precise information about these models at upcoming GW observatories such as LISA, BDECIGO, and the ET, as well as at future colliders, DD experiments, and telescopes like KM3NeT, CTA, LHAASO, SWGO etc. The promising phemonenological findings of this paper motivate and encourage such a refinement.

Finally, our results adds to the motivation (see e.g.~\cite{Cirelli:2018iax}) to employ current and future telescope data to test models of annihilating heavy Dark Matter. Their complementarity with GW interferometers, studied in this paper, could open a new multimessenger approach in the quest for DM.


\medskip

\section*{Acknowledgements}

We thank Kallia Petraki for clarifications regarding Sommerfeld enhancement, and Camilo Garcia Cely, Andreas Goudelis, Oleksii Matsedonskyi and Andrea Tesi for additional useful discussions. IB, YG and FS are grateful to GGI for hospitality and partial support during the completion of this work. YG thanks the commune of Rozier-en-Donzy for hospitality during the COVID pandemic.

\paragraph*{Funding information}
The work of YG and FS was partly supported by a PIER Seed Project funding (Project ID PIF-2017-72). YG is grateful to the Azrieli Foundation for the award of an Azrieli Fellowship. IB is a postdoctoral researcher of the F.R.S.--FNRS with the project ``\emph{Exploring new facets of DM}." This work was partly supported by the Deutsche Forschungsgemeinschaft under Germany Excellence Strategy - EXC 2121 ``Quantum Universe'' - 390833306''.
\medskip

\appendix

\section{5D dual picture}
\label{app:5D}

\stoptocwriting

\subsection{Dilaton in warped extra dimensions}
\paragraph{A warped extra dimension.}
The five dimensional (5D) line-element in warped extra dimensions (WED) reads~\cite{Randall:1999ee}
	\beq
	ds^2 = e^{-2\,k\,T(x)\,|\phi|} g_{\mu\nu} dx^\mu dx^\nu + T(x)^2 d\phi^2
	\eeq
where $\phi$ is an angular dimensionless variable and where, in units where $c=1$ and denoting length dimensions by $L$, $[T(x)] = [k]^{-1} = L$.
The WED solution to the hierarchy problem of the Fermi scale is tied to the values~\cite{Randall:1999ee}
	\beq
	\langle T\rangle \equiv r_c \simeq \mathcal{O}(10)/k, \qquad k = \mathcal{O}(\MPl),
	\eeq
where the latter size is in units of $\hbar=1$ ($[\MPl] = [\text{vev}]$). One  then has
	\beq
	\MPl^2 = \frac{M_5^3}{k} (1-e^{-2 k r_c}) \simeq \frac{M_5^3}{k},
	\eeq
where $M_5$ is the 5D Planck scale.\footnote{$S_5 = \int d^4x \int_{-\pi}^{\pi}d\phi \sqrt{-G} (2 M_5^3\,R - \Lambda)$,
with $G$ the 5D metric, $R$ the 5D Ricci scalar, and $[M_5] = [k] [\text{vev}]^2$.
}

\paragraph{Stability of the IR brane.}
Let us now define the position of the IR brane $z$ and the running mass scale $\mu$ as
	\beq
	z(x) \equiv \frac{1}{\mu(x)} \equiv \frac{1}{k\,e^{-\pi k T(x)}} , \qquad z_1^{-1} \equiv \langle \mu \rangle = k \,e^{- \pi k r_c} \simeq \text{TeV}.
	\eeq
The Goldberger and Wise potential~\cite{Goldberger:1999uk} that stabilises the IR brane at $z = z_1$ reads~\cite{Rattazzi:2003ea,Nardini:2007me}
	\beq
	V_\text{GW}
	\simeq v_1^2 \mu^4 \left((4+2\delta)\Big[ 1- \frac{v_0}{v_1}\Big(\frac{\mu}{k}\Big)^\delta \Big]^2  - \delta_a \right),
	\eeq
where $v_0$ and $v_1$ are the vevs of the 5D scalar field on the UV and IR brane in units of $m^{3/2}$ and $\delta$, $\delta_a$ are small dimensionless parameters. This potential has extrema at
	\beq
	\langle \mu \rangle = 0,\qquad \langle \mu \rangle = k \Big(\frac{v_1}{v_0}\Big)^\frac{1}{\delta} X_\pm^\frac{1}{\delta}, \quad X_\pm = \frac{4+\delta \pm \sqrt{4 \delta_a  + \delta^2}}{4+ 2\delta},
	\eeq
where $X_+$ $(X_-)$ is a minumum for $\delta > 0$ $(\delta < 0)$. By choosing $\delta_a > 0$ one guarantees that the minimum $\langle \mu \rangle \neq 0$ is deeper than the minimum at $\langle \mu \rangle = 0$. An IR-UV hierarchy $\langle \mu \rangle/k \approx 10^{-15}$, as needed to solve the big hierarchy problem, is achieved for a moderate hierarchy between the vevs on the UV and IR branes and for a moderately small $\delta$. 

\paragraph{4D kinetic terms.}
The action for the 5D Ricci induces the following 4D kinetic term for $\mu$~\cite{Rattazzi:2003ea}
	\beq
	\mathcal{L}_\text{kin} = 12 \frac{M_5^3}{k}  (\partial^\mu (e^{-\pi k T(x)}))^2
	= 24 \frac{M_5^3}{k^3} \frac{(\partial_\mu \mu)^2}{2}.
	\eeq

\paragraph{Masses of the 4D resonances.}
Finally, the 4D masses $m_{kk}$ of KK-modes, i.e. the 4D composite resonances, read~\cite{Rattazzi:2003ea}
	\beq
	m_{kk} \simeq (n+\frac{1}{4}) \frac{\pi}{z_1} \approx \frac{1}{z_1} = \langle \mu \rangle \equiv g_\chi f\,,
	\label{eq:mKK}
	\eeq
where the last equality defines a 4-D strong coupling $g_\chi$ and a scale $f$, whose notation anticipates the connection with the 4D picture considered in the main body of this paper.

\subsection{From 5D to 4D}
\paragraph{4D CFT viewed as holographic dual of warped AdS$_5$.}
We rely now on the 4D holographic picture of WED~\cite{ArkaniHamed:2000ds,Rattazzi:2000hs} by employing the relation
	\beq
	\Big(\frac{M_5}{k}\Big)^3 \equiv \frac{N^2}{16 \pi^2} \equiv \frac{1}{g_\chi^2},
	\eeq
where $g_\chi$ is a 4D coupling that for simplicity we identify with the one of Eq.~(\ref{eq:mKK}) and where
we have made a choice for an order one ambiguity in the correspondence.
In terms of the canonically normalised field $\chi $ (now with $[\chi] = [\text{vev}]$),
	\beq
	\chi(x) = \frac{\sqrt{24}}{g_\chi} \mu(x)\,,
	\eeq
and choosing for definiteness $\delta < 0$, one then has (in WED $[v_1^2] = [\hbar] = [m]^{-2} [\text{vev}]^2$))
	\beq
	\mathcal{L}
	=  \frac{1}{2} (\partial_\mu \chi)^2 - \frac{4 v_1^2 \, g_\chi^4}{24^2} \chi^4 \, \left( (1+\delta/2)\Big[ 1- X_- \Big(\frac{\chi}{\langle \chi\rangle}\Big)^\delta \Big]^2  - \delta_a \right),
\qquad \langle \chi\rangle = \frac{\sqrt{24}}{g_\chi} \langle \mu\rangle = \sqrt{24} f\,.
	\eeq
\paragraph{Normalization of dilaton kinetic terms $Z$.}
In the notation of the main text, WED then predicts~\cite{Rattazzi:2000hs}
	\beq
	Z = \sqrt{24}\,
	\eeq
as well as
	\beq
	m_\sigma^2 = - 4 \delta c_\chi \, g_\chi^2 \frac{f^2}{Z^2},
	\qquad \Delta V = V(\chi = 0) -  V(\chi = Z f) \simeq \left(\frac{ Z m_\sigma f}{4}\right)^2,
	\eeq
where both expressions are valid up to relative $O(|\delta|)$, and where we have defined the order one number $c_\chi = 4 v_1^2 g_\chi^2 (\sqrt{\delta_a}- \delta_a/2)$.
We have then recovered all the quantities of interest for our 4D study in the main text.

\resumetocwriting

\section{Analytical tunneling approximations}
\label{app:O(3)_vs_O4}

\stoptocwriting

\subsection{The thick wall limit}
\label{sec:tunneling_rate}

\paragraph{Minimal gradient energy.}
When the potential barrier is small compared to the potential energy difference, the wall thickness is as large as possible in order to minimize the gradient energy. This is the thick-wall limit. 
\paragraph{$O(4)$-bounce action.}
The $O(4)$-symmetric bounce action reads
\begin{equation}
S_4 = \pi^2\, R^3 \,\delta R\, \left( \frac{\delta \phi}{\delta R} \right)^2 + \frac{\pi^2}{2}\, R^4 \,\bar{V}
\end{equation}
where $\delta \phi$ is the field excursion while $\delta R$ is the wall thickness. In the thick-wall limit, we can set $\delta R = R$. The quantity $\bar{V}<0$ is the potential energy averaged in the bubble volume. We approximate it as $\bar{V} = V(\phi_*) - V(\phi_{\rm FV})$ where $\phi_*$ is the initial field value, at the center of the bubble.\footnote{In the thick-wall limit, the initial field value $\phi_*$ is generally close to the zero of the potential $\phi_{\rm zero}$ defined by $V(\phi_{\rm zero}) \equiv V (\phi_{\rm FV})$.} Additionally, we assume (without lack of generality) that the false vacuum energy and the corresponding field value (at asymptotic distance) are zero, $V(\phi_{\rm FV})=0$ and $\phi_{\rm FV}=0$, such that we can write $\bar{V} = V(\phi_*) $ and $\delta \phi = \phi_*$. Hence, we obtain
\begin{equation}
S_4 = \pi^2 R^2 \,\delta R\, \phi_*^2 + \frac{\pi^2}{2} \,R^4 \,V(\phi_*),
\label{S_4_thick_wall_before_R_inj}
\end{equation}
where $\phi_*$ is understood as the value which minimizes the bounce action.
Upon solving for the  critical bubble radius $R_c$, solution of $\delta S_4 / \delta R = 0$, and reinjecting into Eq.~\eqref{S_4_thick_wall_before_R_inj}, we obtain \cite{Anderson:1991zb, Randall:2006py, Nardini:2007me, Gouttenoire:2022gwi}
\begin{equation}
R_c^2 = \frac{\phi_*^2}{-V(\phi_*)} \qquad \textrm{and}\qquad S_4 = \frac{\pi^2}{2} \frac{\phi_*^4}{-V(\phi_*)}.
\label{S_4_thick_wall}
\end{equation}
\paragraph{$O(3)$-bounce action.}
Repeating straightforwardly the same steps for the $O(3)$-symmetrical bounce leads to
\begin{equation}
R_c^2 = \frac{\phi_*^2}{-2\,V(\phi_*)} \qquad \textrm{and}\qquad S_3 = \frac{4\pi}{3} \frac{\phi_*^3}{\sqrt{-2\,V(\phi_*)}}.
\label{eq:S_3_thick_wall}
\end{equation}
Whether the tunneling completes via $O(3)$ or $O(4)$ bounce depends on the ratio
\begin{equation}
\frac{S_4}{S_3/T} = \frac{3 \pi}{4}\frac{\phi_*}{\sqrt{-2 V(\phi_*)}} T \sim 2 \,R_c\, T.
\end{equation}

\subsection{Application to the light-dilaton scenario}
\label{sec:light_dilaton}

\subsubsection{High-temperature or low-temperature expansion.}
We now apply the general findings of the previous subsection to the scenario which we study in this paper, where the effective potential in Eq.~\eqref{eq:tot_pot_dil} can be written as
\begin{equation}
V(\chi) - V(0) \simeq \left\{
                \begin{array}{ll}
                 \dfrac{m_{\rm eff}^2}{2} \chi^2 -  \dfrac{\lambda_{\rm eff}(\chi)}{4} \chi^4 \qquad \text{if}~\chi \lesssim T/g_\chi, \vspace{0.5cm} \\
                 b N^2T^4 -  \dfrac{\lambda_{\rm eff}(\chi)}{4} \chi^4 \qquad \text{if}~\chi \gtrsim T/g_\chi ,
                \end{array}
              \right.
              \label{eq:eff_pot_app}
\end{equation}
with 
\begin{align}
&b \equiv \frac{\pi^2}{8}, \label{eq:b_def} \qquad \qquad  m_{\rm eff}^2 \equiv \frac{15}{16} \frac{N^2 g_{\chi}^2}{Z^2}T^2, \\
&\lambda_{\rm eff}(\chi) \equiv   4\frac{c_\chi}{Z^4} g_\chi^2  \Big[1 - \frac{1}{1+\gamma_\epsilon/4} \Big(\frac{\chi}{Z f}\Big)^{\gamma_\epsilon} \Big]\, \simeq \frac{m_\sigma^2}{Z^2 f^2} \mathrm{log} \left(\frac{Zf}{\chi}\right) +\mathcal{O}(\gamma_\epsilon^2). \label{eq:quartic_app}
\end{align}
In the first line of Eq.~\eqref{eq:eff_pot_app}, we expanded the function $J_{b}(m_{\rm \mathsmaller{CFT}}^2/T^2)$ in Eq.~\eqref{eq:finiteTempPot}  in the so-called `high-temperature' limit $J_b(x) \underset{x \to 0}{\sim} -\frac{\pi^4}{45} + \frac{\pi^2}{12}x$, while in the second line we considered the `low-temperature' limit $J_b(x) \underset{x \to \infty}{\sim} 0$. We have expanded the quartic $\lambda_{\rm eff}(\chi) $ in Eq.~\eqref{eq:quartic_app} in the conformal limit $\gamma_\epsilon \to 0$. We recall that the anomalous dimension $\gamma_\epsilon$ controls the beta function of the confining sector and is related to the dilaton mass $m_\sigma$ through
\beq
m_\sigma^2
= - 4\,\gamma_\epsilon\, c_\chi g_\chi^2 \, \frac{f^2}{Z^2}.
\label{eq:msigma_app}
\eeq
\subsubsection{$O(3)$ thick-wall action.}
\paragraph{Formula.}
We can compute the thick-wall approximated $O(3)$-bounce action by injecting Eq.~\eqref{eq:eff_pot_app} into Eq.~\eqref{eq:S_3_thick_wall}, so that we get
\begin{equation}
S_3 \simeq
 \left\{
                \begin{array}{ll}
                c_* \frac{4\pi}{3}~\underset{\dilaton_*}{\textrm{Min}}  \frac{\dilaton^3_*}{\sqrt{-2\left(m_{\rm eff}^2 \chi^2_*/2 - \lambda_{\rm eff}\, \dilaton_*^4/4\right)}}, \qquad ~\,\,\text{if}~\chi_* \lesssim T/g_\chi, \vspace{0.25cm} \\
                 c_* \frac{4\pi}{3}~\underset{\dilaton_*}{\textrm{Min}} \frac{\dilaton^3_*}{\sqrt{-2\left(b \,N^2\,T^4 - \lambda_{\rm eff}\, \dilaton_*^4\right)}}, \qquad \qquad \text{if}~\chi_* \gtrsim T/g_\chi ,
                \end{array}
              \right.
\end{equation}
where $c_*$ is a coefficient which we add and which we fit on the numerical solution. We later set
\begin{equation}
c_* \simeq 2, \label{eq:c_star_def}
\end{equation}
for both the $O(3)$- and $O(4)$-bounce actions.
\paragraph{Minimization.}
After minimization with respect to $\chi_*$, we obtain
\begin{equation}
S_3/T \simeq
 \left\{
                \begin{array}{ll}
                c_* ~ \dfrac{16\pi}{3\lambda_*} \dfrac{m_{\rm eff}}{T}, \vspace{0.25cm} \\
                 c_* ~ \dfrac{4\pi}{\sqrt{3}}\dfrac{ \chi_*/T}{\sqrt{\lambda_*}},
                \end{array}
              \right.
              \qquad
              \chi_* \simeq
 \left\{
                \begin{array}{ll}
                \dfrac{2m_{\rm eff}}{\sqrt{\lambda_*}}, \qquad \qquad \qquad \quad \text{if}~\chi_* \lesssim T/g_\chi, \vspace{0.25cm} \\
                \left(\dfrac{12\,b\,N^2}{\lambda_*}\right)^{\!1/4}~T, \qquad ~\text{if}~\chi_* \gtrsim T/g_\chi .
                \end{array}
              \right.
              \label{eq:S3OverT_2_cases}
\end{equation}
We now evaluate the quantity $g_{\chi}\,\chi_*/T$ for the two lines in Eq.~\eqref{eq:S3OverT_2_cases}, in the conformal limit $\gamma_\epsilon \to 0$
\begin{equation}
   \frac{g_\chi \,\chi_*}{T} \simeq
 \left\{
                \begin{array}{ll}
                77 \left( \frac{5}{N} \right)\left( \frac{10}{N_e} \right)^{\!1/2} \left( \frac{0.25}{m_{\sigma}/f} \right) \gg 1,\vspace{0.25cm} \\
                 12 \left( \frac{5}{N} \right)^{\!1/2}\left( \frac{10}{N_e} \right)^{\!1/4} \left( \frac{0.25}{m_{\sigma}/f} \right)^{\!1/2}  Z^{1/2} \gg 1,
                \end{array}
              \right.
\end{equation}
where $N_e$ is the number of e-fold of supercooling
\begin{equation}
N_e \simeq {\rm log }\left(\frac{f}{T}\right).
\end{equation}
\paragraph{Final result.}
Hence we conclude that the second line of Eq.~\eqref{eq:S3OverT_2_cases} is the correct solution, meaning that the thermal corrections can be neglected at the tunneling point. To be clearer for the reader we report the final solution here
\begin{equation}
\frac{S_3}{T} =   c_* ~ \dfrac{4\pi}{\sqrt{3}}\dfrac{\chi_*/T}{\sqrt{\lambda_*}}, \label{eq:S3_thickwall_light_dilaton}
\end{equation}
with
\begin{align}
&\lambda_* \equiv \lambda_{\rm eff}(\dilaton_*),\label{eq:lambda_star} \qquad \qquad \chi_{*} =   \left(\dfrac{12\,b\,N^2}{\lambda_*}\right)^{\!1/4}~T+\mathcal{O}(\gamma_\epsilon^2).
\end{align}
and $b$, $\lambda_{\rm eff}$, $c_*$ defined in Eq.~\eqref{eq:b_def},~\eqref{eq:quartic_app},~\eqref{eq:c_star_def}.

\subsubsection{$O(4)$ thick-wall action.}
\paragraph{Formula.}
We repeat the procedure for the $O(4)$-bounce by plugging Eq.~\eqref{eq:eff_pot_app} into Eq.~\eqref{S_4_thick_wall}
\begin{equation}
S_4 \simeq
 \left\{
                \begin{array}{ll}
                c_* \frac{\pi^2}{2}~\underset{\dilaton_*}{\textrm{Min}}  \frac{\dilaton^4_*}{-\left(m_{\rm eff}^2 \chi^2_*/2  - \lambda_{\rm eff}\, \dilaton_*^4/4\right)}, ~\qquad \qquad \text{if}~\chi_* \lesssim T/g_\chi, \vspace{0.25cm} \\
                 c_* \frac{\pi^2}{2}~\underset{\dilaton_*}{\textrm{Min}} \frac{\dilaton^4_*}{-\left(b \,N^2\,T^4 - \lambda_{\rm eff}\, \dilaton_*^4/4\right)}, \qquad \qquad\quad \text{if}~\chi_* \gtrsim T/g_\chi .
                \end{array}
              \right.
              \label{eq:S4_2_cases}
\end{equation}
Minimization of the expression for $\chi_* \gtrsim T/g_\chi$ leads to $\chi_*$ = 0, which is not a viable bounce solution. 
\paragraph{Final result.}
The minimization of the expression for $\chi_* \lesssim T/g_\chi$ leads to
\begin{equation}
S_4  = c_* ~ \frac{2\pi^2}{\lambda_*} \label{eq:S4_thickwall_light_dilaton},
\end{equation}
with
\begin{align}
&\lambda_* \equiv \lambda_{\rm eff}(\dilaton_*),\label{eq:lambda_star_O4} \quad \quad \chi_{*} =\left(4(4+\gamma_\epsilon)\,b \, N^2 \, T^4(Z \, f)^{2+\gamma_\epsilon}/m_\sigma^2 \right)^{\frac{1}{4+\gamma_\epsilon}}=   \left(\dfrac{16\,b\,N^2}{\lambda_*}\right)^{\!1/4}~T+\mathcal{O}(\gamma_\epsilon^2).
\end{align}
\paragraph{Backward check .}
We backward check that the thermal corrections are negligible at the tunneling point
\begin{equation}
\frac{g_\chi \,\chi_*}{T} \simeq 13 \left(\frac{5}{N} \right)^{\!1/2}\left(\frac{10}{N_e} \right)^{\!1/4}\left(\frac{0.25}{m_{\sigma}/f} \right)^{\!1/2}Z^{1/2} \gg 1,
\end{equation}
hence confirming that the second line of Eq.~\eqref{eq:S4_2_cases} is the correct solution.

\subsection{Analytical expression for the nucleation temperature}
\label{sec:nucl_temp_ana}

\begin{figure}[t]
\centering
\raisebox{0cm}{\makebox{\includegraphics[height=0.5\textwidth, scale=1]{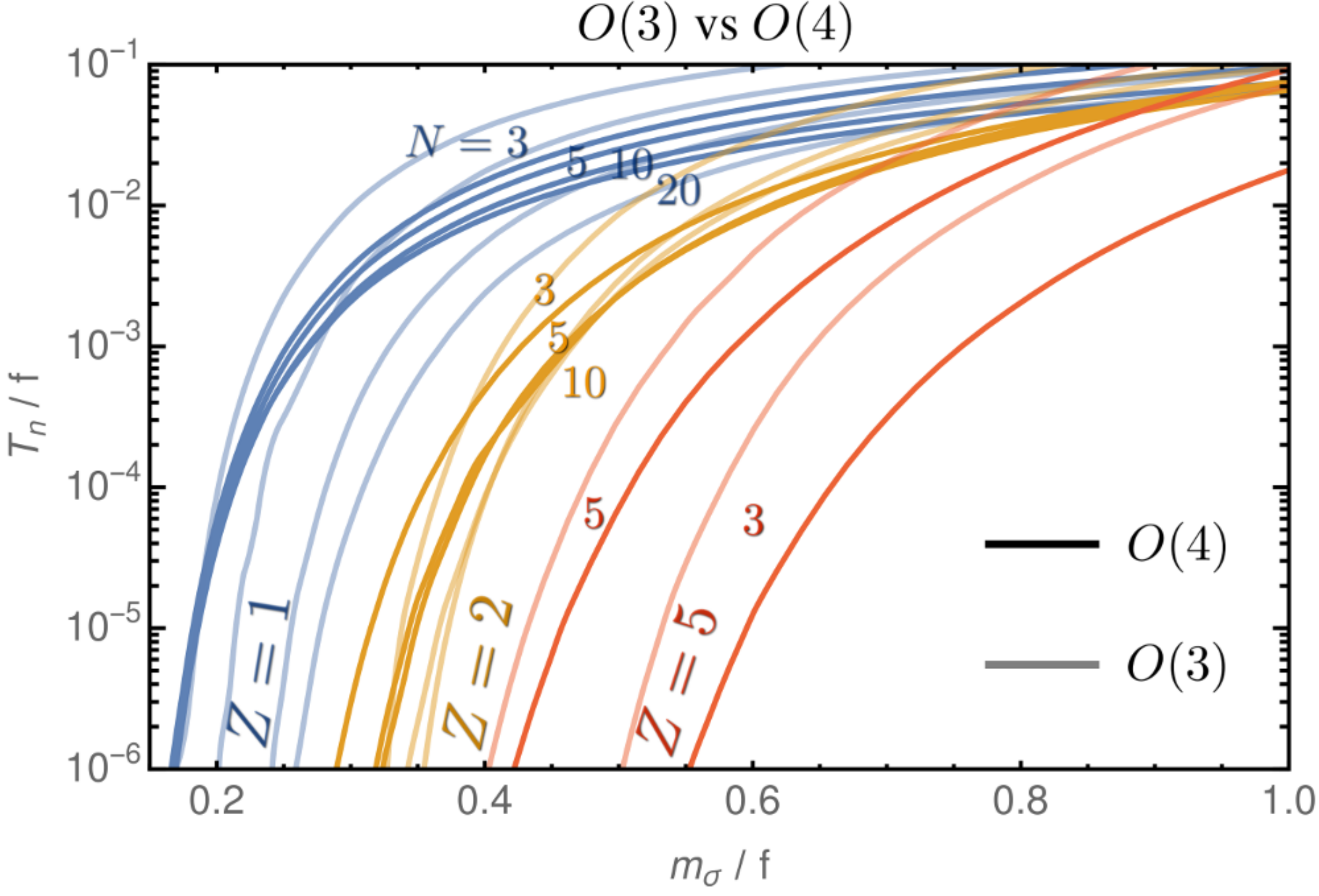}}} 
\caption{\it \small We compare the temperature of nucleation via $O(3)$-symmetric bounce to the temperature of nucleation via $O(4)$-symmetric bounce, computed numerically with an overshoot-undershoot method. We can see that the $O(4)$ bounce dominates over the $O(3)$ bounce for large supercooling $f/\Tnuc$, $Z= 1$ and $N \geq 3$. }
\label{fig:O4_vs_O3} 
\end{figure}

\begin{figure}[t]
\centering
\hspace{-1cm}
\raisebox{0cm}{\makebox{\includegraphics[width=0.50\textwidth, scale=1]{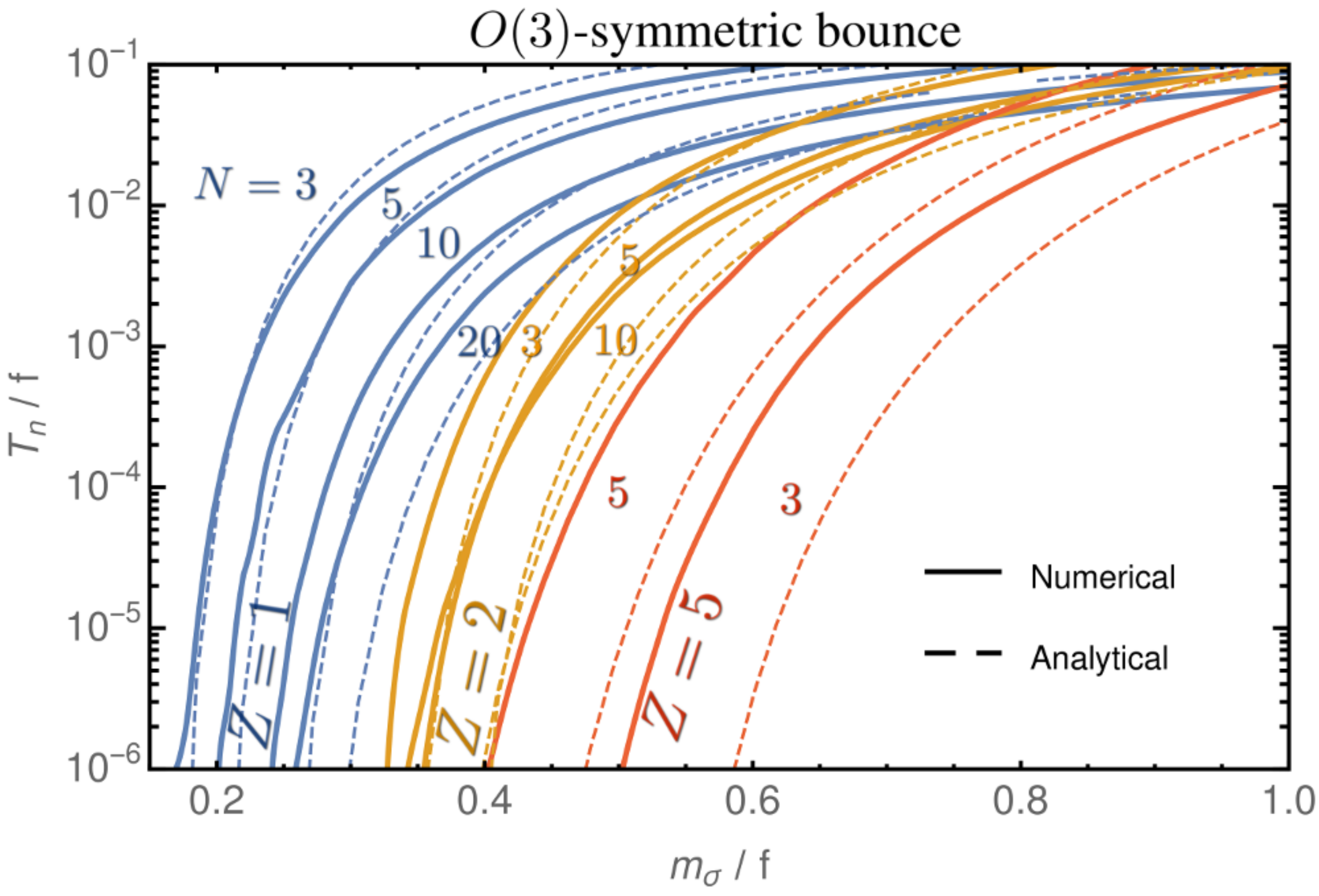}}} $\quad$
\raisebox{0cm}{\makebox{\includegraphics[width=0.50\textwidth, scale=1]{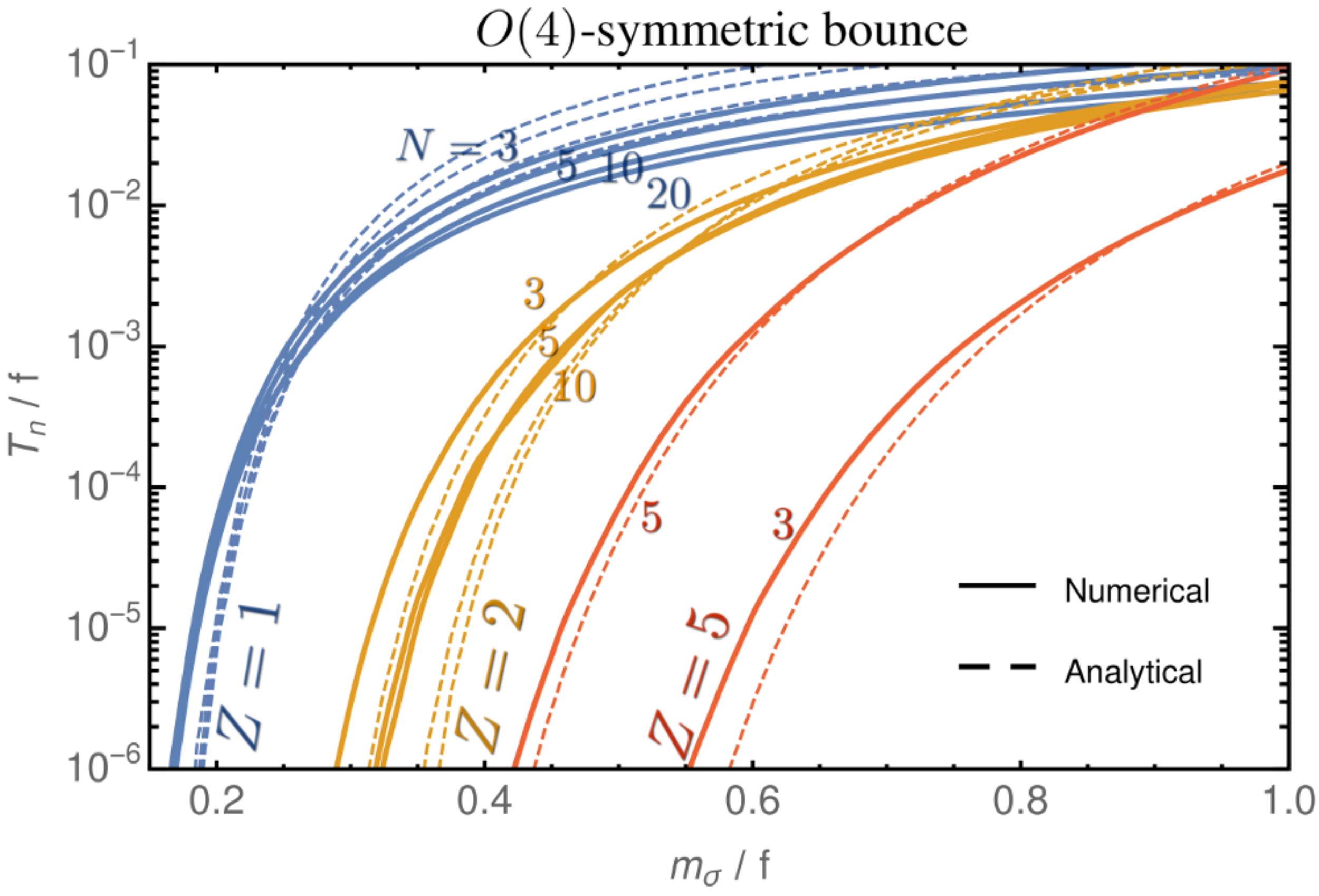}}}
\caption{\it \small We compare the nucleation temperature computed after numerically solving for the $O(3)$- or $O(4)$-bounce action with the overshoot-undershoot method, to the analytical approximations in the thick-wall limit presented in Sec.~\ref{sec:light_dilaton}. \textbf{Right:} The $O(4)$ analytical formula, in Eq.~\eqref{eq:S4_thickwall_light_dilaton}, is a quite good approximation. \textbf{Left:} The $O(3)$ case, in Eq.~\eqref{eq:S3_thickwall_light_dilaton}, while relatively good approximation for $Z=1$, worsens for larger $Z$. This was expected since it has been derived in the conformal limit $\gamma_\epsilon \to 0$.}
\label{fig:nucleation_temp_numVSana} 
\end{figure}

\subsubsection{$O(3)$ nucleation temperature.}
\paragraph{Instantaneous nucleation approximation.}
The nucleation happens when the tunneling rate in Eq.~\eqref{eq:tunneling_rate} becomes comparable to the Hubble expansion rate per Hubble volume, see App.~\ref{app:Bubble nucleation and percolation temperature} for a justification. For $O(3)$-dominance, the nucleation temperature is solution of 
\begin{equation}
\label{eq:nucleation_eq_S_3_light_dilaton}
\frac{S_{3}(T_{\rm nuc})}{T_{\rm nuc}} \simeq 4 \log{\frac{T_{\rm nuc}}{H(T_{\rm nuc})}} + \frac{3}{2} \log \frac{S_3/T_{\rm nuc}}{2\pi},
\end{equation}
where the Hubble parameter is defined by Eqs.~\eqref{eq:vacuum_energy} and \eqref{eq:free_energy_CFT},
\begin{equation}
H^2(T) = H^2_{\Lambda}+H^2_{\rm rad} = Z^2\frac{m_{\sigma}^2\,f^2/16}{3 M_{\rm pl}^2}+\frac{\pi^2 g_* T^4}{90M_{\rm pl}^2}, \quad g_* = 106.75 +\frac{45 N^2}{4} .
\end{equation}

\paragraph{$O(3)$ nucleation temperature.}
An analytical expression of the nucleation temperature can be obtained in the conformal limit $m_\sigma \to 0$ by inverting the thick-wall formula for the $S_3$ bounce action in Eq.~\eqref{eq:S3_thickwall_light_dilaton},
	\begin{equation}
	\label{eq:analyticalTnuc_O3_app}
	\left(\frac{T_{\rm nuc}}{T_c} \right)^4 \simeq \text{Exp}\left( -c_{*}^4\, \frac{1024\, \pi^4 }{ 3 (S_3^{\mathsmaller{\rm crit}}/T_{\rm nuc})^4 }b\,N^2 \frac{Z^2}{ y_{\sigma}^2} \right),
	\end{equation}
where $S_3^{\mathsmaller{\rm crit}}$ is solution of Eq.~\eqref{eq:nucleation_eq_S_3_light_dilaton}, where $y_\sigma \equiv m_\sigma/\dilatonvev$, and	where $T_c$ is the critical temperature, defined by Eq.~\eqref{eq:crit_temp_PT}, and which we recall here
\begin{equation}
\label{eq:crit_temp_PT_app}
T_c =  \left(  \frac{Z \,  \msigma \dilatonvev}{4\sqrt{b} N}  \right)^{1/2}.
\end{equation}
Note that from using Eq.~\eqref{eq:msigma} and $g_{\dilaton} = 4\pi/N$, we can rewrite Eq.~\eqref{eq:analyticalTnuc_O3_app} in terms of the anomalous dimension $\gamma_\epsilon$
	\begin{equation}
	\label{eq:analyticalTnuc_O(3)_N}
	\left(\frac{T_{\rm nuc}}{T_c} \right)^4 \simeq \text{Exp}\left( -c_{*}^4\, \frac{16 \,\pi^2\, b\, Z^4 \,N^4}{ 3\,c_{\dilaton}\,\left|\gamma_{\epsilon}\right|\,(S_3^{\mathsmaller{\rm crit}}/T_{\rm nuc})^4 } \right).
	\end{equation}

\subsubsection{$O(4)$ nucleation temperature.}
\paragraph{Instantaneous nucleation approximation.}
We now consider the case where tunneling occurs via $O(4)$-symmetric bounce. The nucleation temperature is solution of 
\begin{equation}
\label{eq:nucleation_eq_S_4_light_dilaton}
S_{4}(T_{\rm nuc}) \simeq 4 \log{\frac{R_c^{-1}}{H(T_{\rm nuc})}} + \frac{1}{2} \log \frac{S_4}{2\pi},
\end{equation}
where $R_c$ is the bubble radius which according to the thick-wall formula in Eq.~\eqref{S_4_thick_wall}, reads
\begin{equation}
R_c = \left(\frac{Z^3 f/m_{\rm \sigma}}{\sqrt{b}\,N\,\textrm{\rm log} \frac{T_c}{T}}\right)^{1/2}~T^{-1} \sim T^{-1}.
\end{equation}
\paragraph{Solution.}
From rewriting the $S_4$ bounce action in Eq.~\eqref{eq:S4_thickwall_light_dilaton} in the conformal limit $m_\sigma \to 0$ as 
\begin{equation}
S_4  = \frac{A_4}{\textrm{log}~T_c/T}, \qquad \textrm{with} \quad  A_4= c_*~\frac{\pi^2}{2}\frac{Z^4}{ \left|\gamma_{\epsilon}\right|\, c_\dilaton \,g_\dilaton^2} = c_*~2\pi^2 \frac{Z^2 f^2}{m_{\rm \sigma}^2},
\label{eq:S_4_light_dilaton_A_T_c}
\end{equation}
and from injecting it into Eq.~\eqref{eq:nucleation_eq_S_4_light_dilaton}, we obtain
\begin{equation}
\Tnuc = \sqrt{ H_{\Lambda}\,T_c} ~\left( \frac{2\pi}{S_4} \right)^{1/16}~\mathrm{Exp} \left( \frac{1}{2} \sqrt{-A_4 + \left( \log \frac{T_c}{H_{\Lambda}} + \frac{1}{8} \log \frac{S_4}{2\pi} \right)^2} \right).
\end{equation}
\paragraph{Universe stuck forever in false vacuum.}
Neglecting the $S_4/2\pi$ terms, we conclude that there is no nucleation solution when \cite{DelleRose:2019pgi}
\begin{equation}
A_4 \gtrsim  \log \frac{T_c}{H_{\Lambda}},
\label{eq:minimal_msigma_S4_light_dilaton}
\end{equation}
and that the minimal nucleation temperature is
\begin{equation}
T_{\rm nuc}^{\rm min} \simeq  \sqrt{ H_{\Lambda}\,T_c} \simeq 0.1\,f~Z^{3/4}\left(\frac{m_{\sigma}}{f}  \right)^{3/4}\left(\frac{f}{M_{\rm pl}}\right)^{1/2}.
\label{eq:minimal_nucleation_temp_S4_light_dilaton}
\end{equation}
That minimal temperature can be visualized as the end points of the lines in Fig.~\ref{fig:O4_vs_O3}. In the parameter space beyond those end points, we can rely on the confinement of an additional strong sector (possibly QCD) to boost the tunneling rate and trigger the nucleation \cite{vonHarling:2017yew, Baratella:2018pxi,Bloch:2019bvc,Fujikura:2019oyi}.
\paragraph{$O(4)$ nucleation temperature.}
Note that from inverting Eq.~\eqref{eq:S_4_light_dilaton_A_T_c}, we can write
	\begin{equation}
	\label{eq:analyticalTnuc_app}
	\left(\frac{T_{\rm nuc}}{T_c} \right)^4 \simeq \text{Exp}\left( -c_{*}\, \frac{2\, \pi^2}{ S_4^{\mathsmaller{\rm crit}} } \frac{Z^2}{ y_{\sigma}^2} \right),
	\end{equation}
where $S_4^{\mathsmaller{\rm crit}}$ is solution of Eq.~\eqref{eq:nucleation_eq_S_4_light_dilaton} and where $y_\sigma \equiv m_\sigma/\dilatonvev$.  The advantage of Eq.~(\ref{eq:analyticalTnuc_app}) is to make explicit that the nucleation temperature is exponentially suppressed for small dilaton mass $m_{\sigma}$, or equivalently for small anomalous coupling $|\gamma_{\epsilon}|$. Note also that from using Eq.~\eqref{eq:msigma} and $g_{\dilaton} = 4\pi/N$, we can rewrite Eq.~\eqref{eq:analyticalTnuc_app} as 
	\begin{equation}
	\label{eq:analyticalTnuc_N}
	\left(\frac{T_{\rm nuc}}{T_c} \right)^4 \simeq \text{Exp}\left( -c_{*}\, \frac{Z^4 N^2}{ 32\,c_{\dilaton}\,\left|\gamma_{\epsilon}\right|\,S_4^{\mathsmaller{\rm crit}} } \right),
	\end{equation}
and we recover that at fixed $\gamma_{\epsilon}$, the nucleation temperature scales like $\Tnuc/f \propto e^{-N^2}$, e.g.  \cite{Baratella:2018pxi}.

\subsubsection{Discussion.}
In contrast to the $O(3)$ thick-wall solution which can only be computed in the conformal limit, Eq.~\eqref{eq:lambda_star}, the $O(4)$ thick-wall solution can be computed for finite $\gamma_\epsilon$, Eq.~\eqref{eq:lambda_star_O4}. In Fig.~\ref{fig:nucleation_temp_numVSana}, we can see that the thick-wall formula for the $O(3)$ and $O(4)$ bounce compare quite well with the numerical solution. Only the $O(3)$ formula becomes wrong away from the conformal limit, e.g. for $Z \gtrsim 2$ or $N \gtrsim 20$.
We can compare the analytical formula for the $O(3)$- and $O(4)$-bounce actions in the $\gamma_\epsilon \to 0$ limit
\begin{equation}
\frac{S_4}{S_3/T} = 1.0 \left(\frac{10}{N_e}\right)^{\!1/4} \left(\frac{3}{N}\right)^{\!1/2} \left(\frac{0.2}{m_\sigma/f}\right)^{\!1/2} Z^{1/2}
\label{eq:S4_over_S3_dilaton}
\end{equation}
This analytical approximation predicts that the $O(4)$ bounce overtakes over the $O(3)$ bounce for $N \gtrsim 3$, $Z =1$ and large $N_e$. However, note the aforementioned inaccuracies for $O(3)$ for $Z \gtrsim 2$. Indeed, contrary to the prediction made in Eq.~\eqref{eq:S4_over_S3_dilaton}, the $O(3)$ bounce governs nucleation for $Z = 5$ in the numerical results of Fig.~\ref{fig:O4_vs_O3}.

\subsection{Precise percolation temperature}
\label{app:Bubble nucleation and percolation temperature}

\FloatBarrier

\begin{figure}[t]
\centering
\hspace{-1cm}
\raisebox{0cm}{\makebox{\includegraphics[width=0.7\textwidth, scale=1]{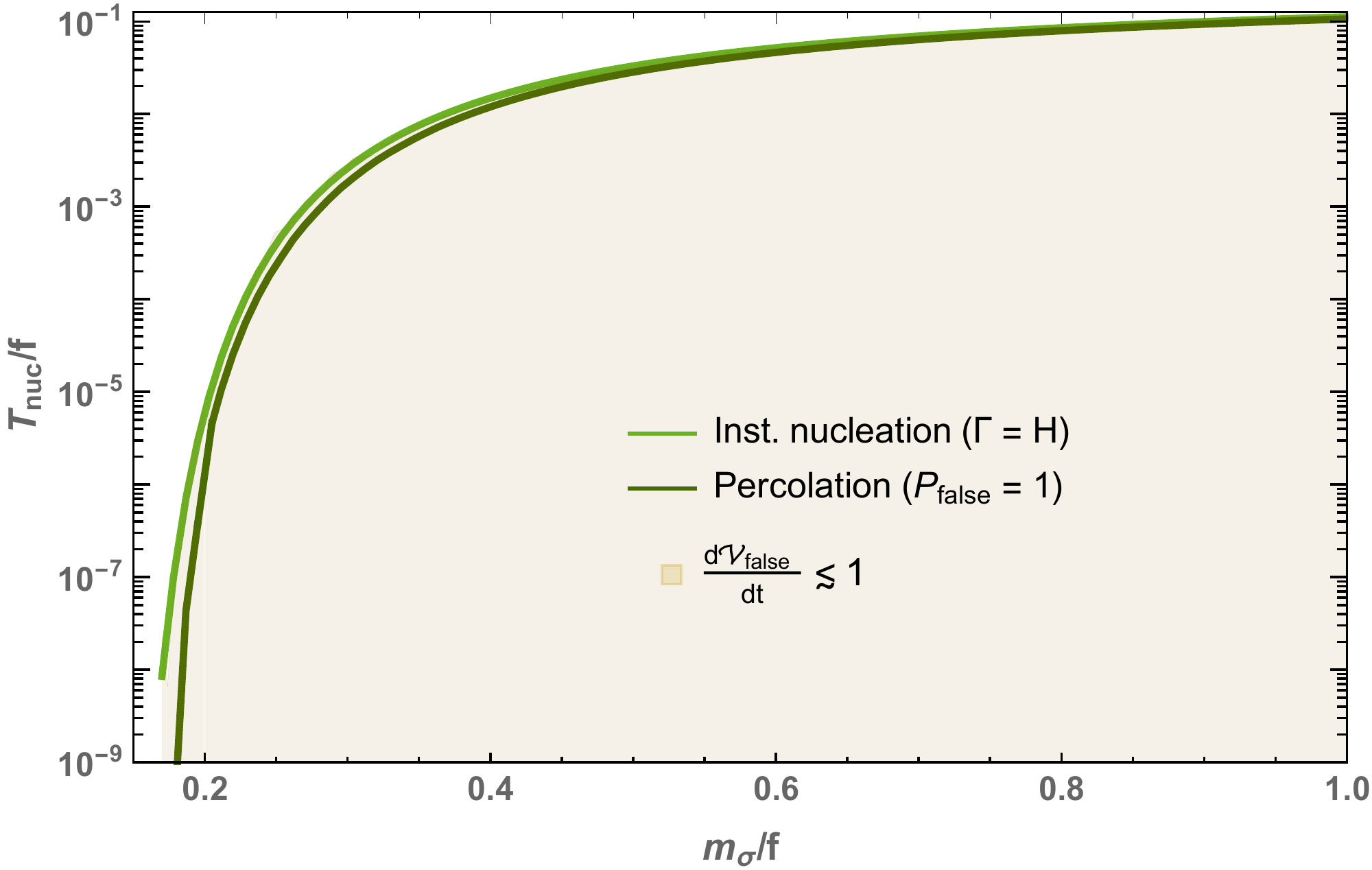}}}
\caption{\it \small We compare the nucleation temperature computed from the analytical $O(4)$ action in the thick-wall limit, using either the instantaneous approximation in Eq.~\eqref{eq:nucTempInstApprox}, or the refined percolation conditions in Eq.~\eqref{eq:percoTemp} and Eq.~\eqref{eq:percoTemp_PhysVolDec}.}
\label{fig:instant_nucleation_vs_percolation} 
\end{figure}

In this appendix we make a refined estimate of the temperature at which the phase transition completes, following the work of~\cite{Guth:1979bh,Guth:1981uk,Guth:1982pn,Enqvist:1991xw,Turner:1992tz}. 
\paragraph{Volume fraction converted into true vacuum.}
The number of bubbles formed per unit of time and per unit of comoving volume is given by
	\begin{equation}
	\Gamma(t') \, a(t')^3.
	\end{equation} 
The comoving volume of a bubble which has nucleated at $t'$ and expanded at the speed of light until $t$ is
	\begin{equation}
	\frac{4\pi}{3} \, \left( \int_{t'}^t \frac{d \tilde{t}}{a(\tilde{t})}  \right)^3.
	\end{equation}
We deduce the volume fraction converted into true vacuum at time $t$, including bubble overlapping\footnote{Note that we have neglected the initial size of the bubble just after it nucleates. Including it would give $ I(t) =  \int_{t_c}^t dt' \, \Gamma(t') \, \frac{4\pi}{3} \, \left( \int_{t'}^t \frac{a(t')}{a(\tilde{t})} \, d \tilde{t} + R(t') \right)^3 = \frac{4\pi}{3} \int_{T_{\rm nuc}}^{T_c} \frac{dT'}{T'} \, \frac{\Gamma(T') }{T'^3 \, H(T')} \, \left( \int_{T_{\rm nuc}}^{T'} \frac{d \tilde{T}}{H(\tilde{T})} + R(T') \, T'\right)^3$ where $R'$ is the bubble radius at nucleation. The correction is of order $R(T') \, H' \sim \frac{\text{TeV}}{\, \MPl}$ and is completely negligible. }
	\begin{equation}
	 I(t) =  \int_{t_c}^t dt' \, \Gamma(t') \, a(t')^3 \, \frac{4\pi}{3} \, \left( \int_{t'}^t \frac{d \tilde{t}}{a(\tilde{t})}  \right)^3.
	\end{equation}
We can also interpret $I(t)$ as the average number of bubbles inside which, a given point of space is contained. It can be larger than $1$ because it includes overlapping. Hence, one defines $P_{\rm false}(t) \equiv e^{-I(t)}$, where multiple counting has been subtracted by the exponentiation.\footnote{The probability $P_{\rm false}(t)$ that a given point remains in the false vacuum is the $N \to \infty$ limit of the product $\prod_{n=1}^N \left(1-dt\frac{dI(t_n)}{dt}\right)$ where $dt = \frac{t-t_c}{N}$ and $1-dt\frac{dI(t_n))}{dt}$ is the probability of remaining in the false vacuum between $t_n = t_c+(t-t_c)\frac{n}{N}$ and $t_{n+1}$.} This is the probability that a given point remains in the false vacuum at time $t$. We also define the physical volume remaining in the false vacuum per unit of comoving volume $\mathcal{V}_{\mathsmaller{\rm false}}(t) \equiv a(t)^3 \, P_{\rm false}(t)$. 
\paragraph{Percolation time.}
We consider the percolation to occur when 
	\begin{equation}
	P_{\rm false}(t) \lesssim 1/e \, \implies \, I(t) \gtrsim 1 \quad \text{ and} \quad \frac{\mathcal{V}_{\mathsmaller{\rm false}}(t)}{dt} \lesssim 0.
	\end{equation}
The second condition is necessary to prevent the possibility that the probability that a given point has been converted to the true vacuum tends to $1$, while the physical volume of false vacuum goes to infinity~\cite{Turner:1992tz}.
\paragraph{Percolation temperature.}
Finally, upon using the adiabatic relations $H \,dt = -dT/T$ and $d(T \, a)=0$, the nucleation temperature $T_{\rm nuc}$ is obtained from
	\begin{equation}
	\label{eq:percoTemp}
	\frac{4\pi}{3} \int_{T_{\rm nuc}}^{T_c} \frac{dT'}{T'} \, \frac{\Gamma(T') }{T'^3 \, H(T')} \, \left( \int_{T_{\rm nuc}}^{T'} \frac{d \tilde{T}}{H(\tilde{T})} \right)^3 \simeq 1,
	\end{equation}
after checking that 
	\begin{equation}
	\label{eq:percoTemp_PhysVolDec}
	3+T_{\rm nuc}\frac{dI(T)}{dT}\Big|_{T_{\rm nuc}} < 0.
	\end{equation}
\paragraph{Instantaneous nucleation approximation is enough.}
As shown in Fig.~\ref{fig:instant_nucleation_vs_percolation}, for the class of  logarithmic potential which we consider in this paper, the difference between the percolation temperature derived from Eq.~\eqref{eq:percoTemp} and Eq.~\eqref{eq:percoTemp_PhysVolDec}, and the instantaneous nucleation temperature derived from $\Gamma \simeq H^4$ is tiny. This justifies the use of the latter prescription in Sec.~\ref{sec:nucleation-temperature}.

\FloatBarrier

\resumetocwriting

\section{More plots}
\label{app:more_pheno_plots}
\stoptocwriting

\subsection{Non-supercooled dilaton-mediated composite DM}
\label{app:Standard dilaton-mediated DM}

\begin{figure}[h!]
\centering
\vspace{-1.5cm}
\raisebox{0cm}{\makebox{\includegraphics[height=0.45\textwidth, scale=1]{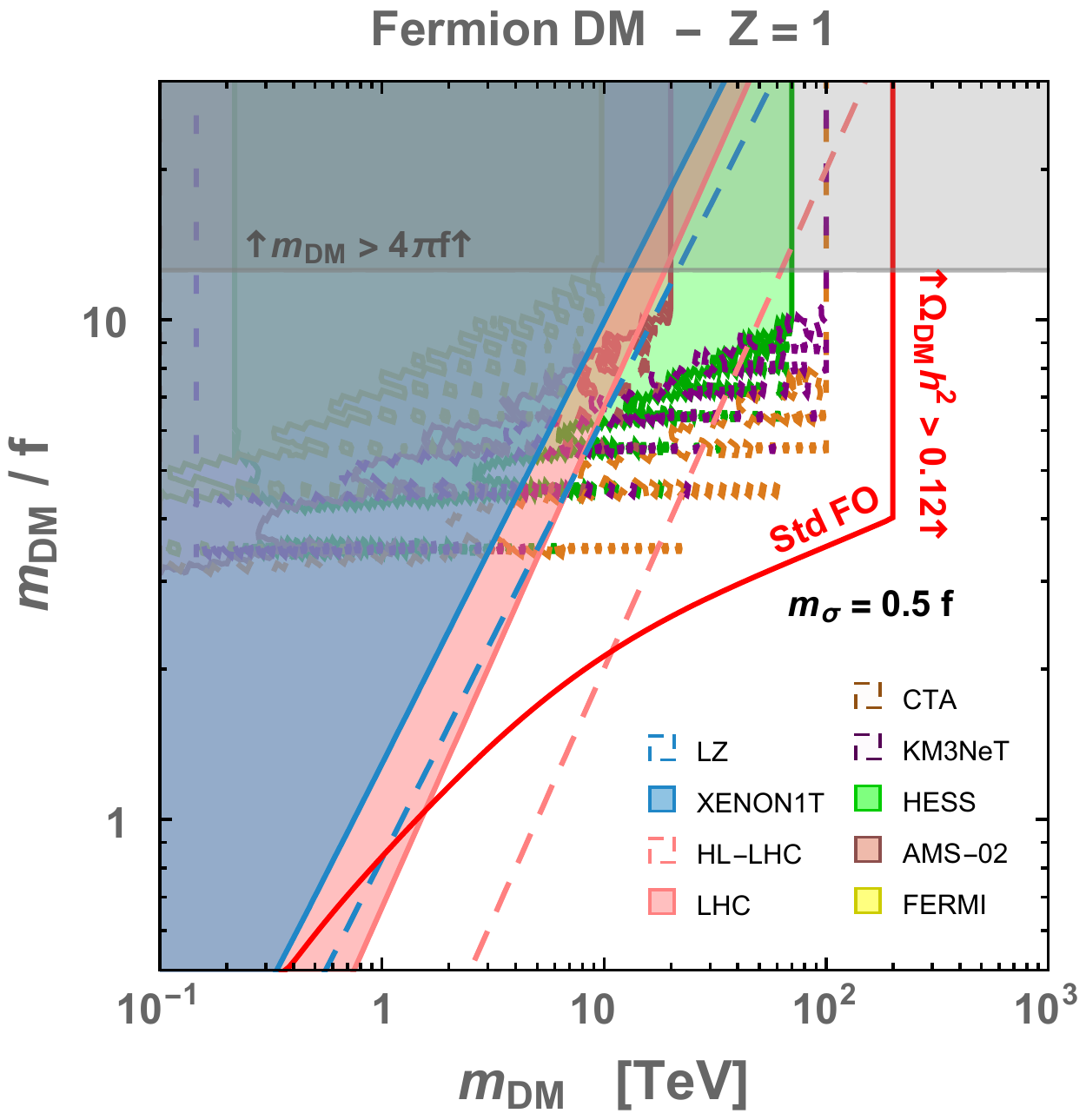}}}
\raisebox{0cm}{\makebox{\includegraphics[height=0.45\textwidth, scale=1]{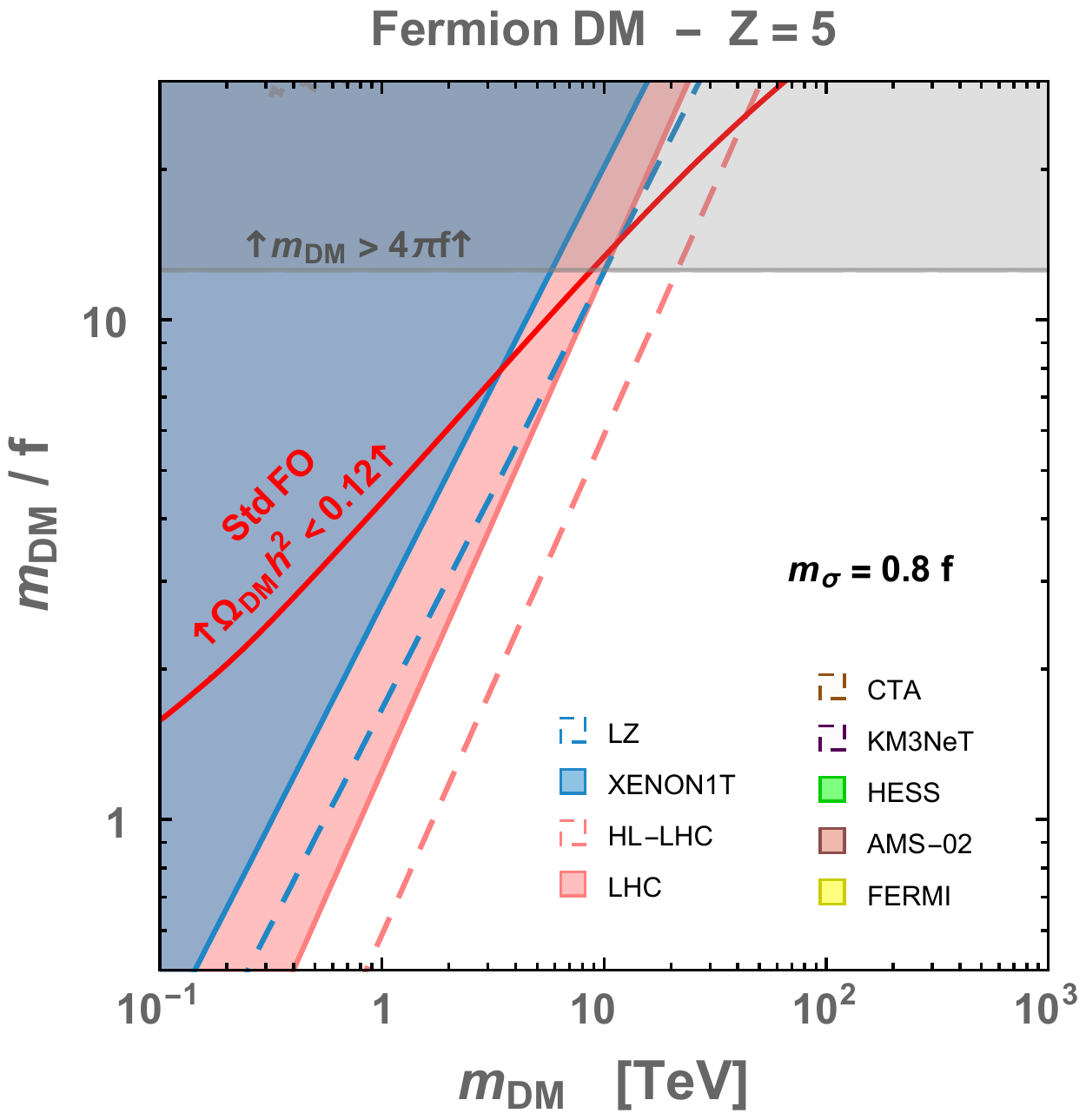}}}
\raisebox{0cm}{\makebox{\includegraphics[height=0.45\textwidth, scale=1]{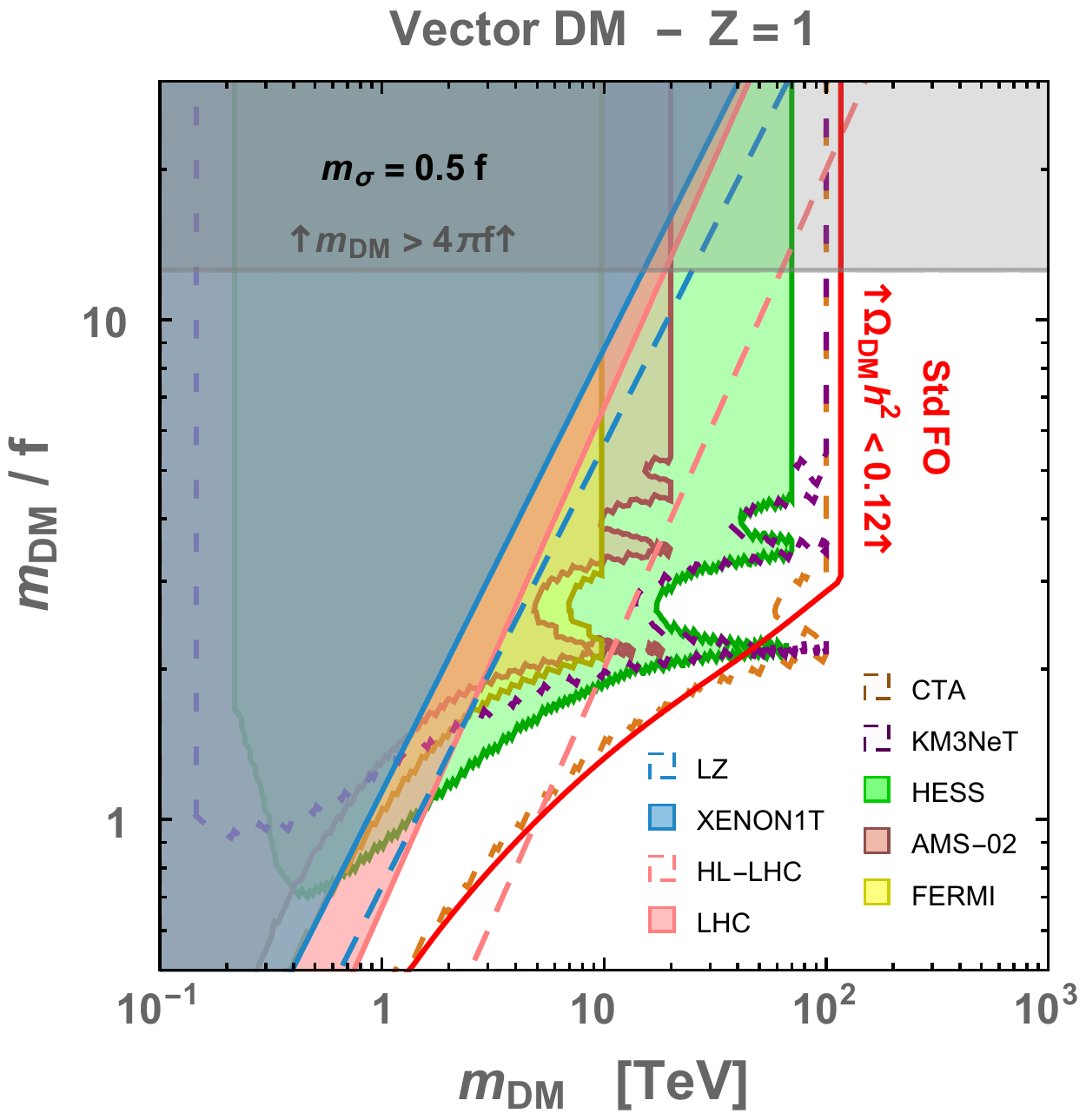}}}
\raisebox{0cm}{\makebox{\includegraphics[height=0.45\textwidth, scale=1]{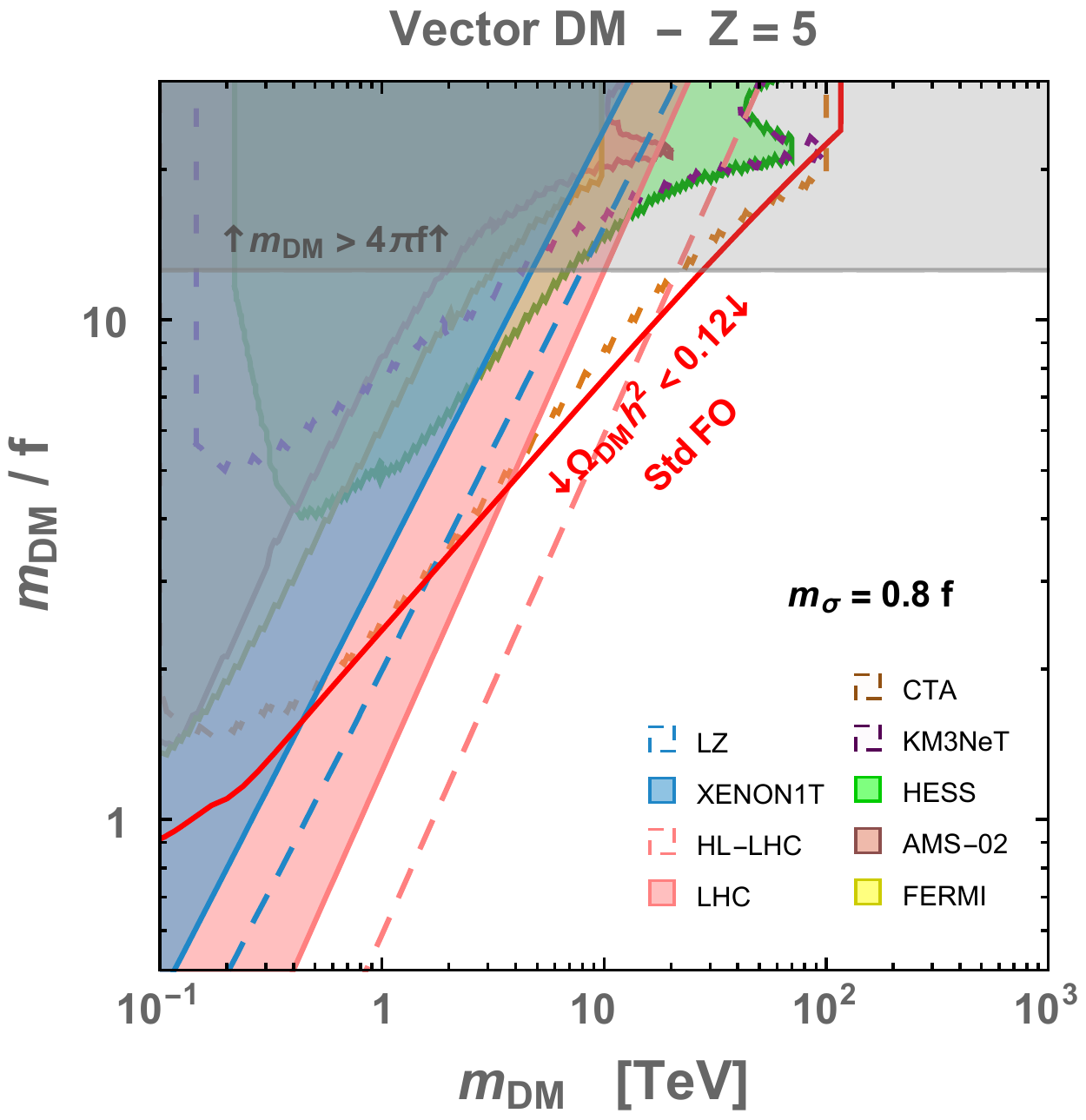}}}
\raisebox{0cm}{\makebox{\includegraphics[height=0.45\textwidth, scale=1]{Figures/dilaton_DM_scalar_Z1_stdFO}}}
\raisebox{0cm}{\makebox{\includegraphics[height=0.45\textwidth, scale=1]{Figures/dilaton_DM_scalar_Z5_stdFO}}}
\caption{\it \small  \textbf{Dilaton-mediated DM in the standard freeze-out case.} We display DM as a fermion (\textbf{top}), vector (\textbf{middle}) and scalar (\textbf{bottom}) resonance of the strong sector, cf. Sec.~\ref{sec:DM-Candidates}. 
Lines and shadings as in Fig.~\ref{fig:stdFO_pheno_msigma_p25a}. In the fermionic case, the DM annihilation cross-section is p-wave, cf. Eq.~\eqref{eq:DMannfactors}, so that the indirect-detection constraints are suppressed compared to the scalar and vector cases.}
\label{fig:stdFO_pheno_app} 
\thisfloatpagestyle{empty}
\end{figure}

We provide the phenomenological constraints, coming from direct detection, indirect detection and colliders on dilaton-mediated DM, in absence of supercooling.
Fig.~\ref{fig:stdFO_pheno_app} extends the scalar DM scenario of Fig.~\ref{fig:stdFO_pheno_msigma_p25a} in Sec.~\ref{sec:DM in Composite Higgs}, to the fermionic and vector DM cases. If DM is a fermion, the DM annihilation is p-wave and the ID signal is too weak to lead to any constraints. If DM is a vector, a scalar, or a pNGB, the ID signal is still too weak to be probed by currents ID experiments, however future experiments could probe region up to $100$~TeV, cf. Figs.~\ref{fig:stdFO_pheno_msigma_p25a} and~\ref{fig:stdFO_pheno_app}.

\begin{figure}[t]
\centering
\vspace{-1.5cm}
\raisebox{0cm}{\makebox{\includegraphics[height=0.5\textwidth, scale=1]{Figures/dilaton_DM_pNGB_Z1_top}}}
\raisebox{0cm}{\makebox{\includegraphics[height=0.5\textwidth, scale=1]{Figures/dilaton_DM_pNGB_Z5_top}}}\\
\vspace{0.5 cm}
\raisebox{0cm}{\makebox{\includegraphics[height=0.5\textwidth, scale=1]{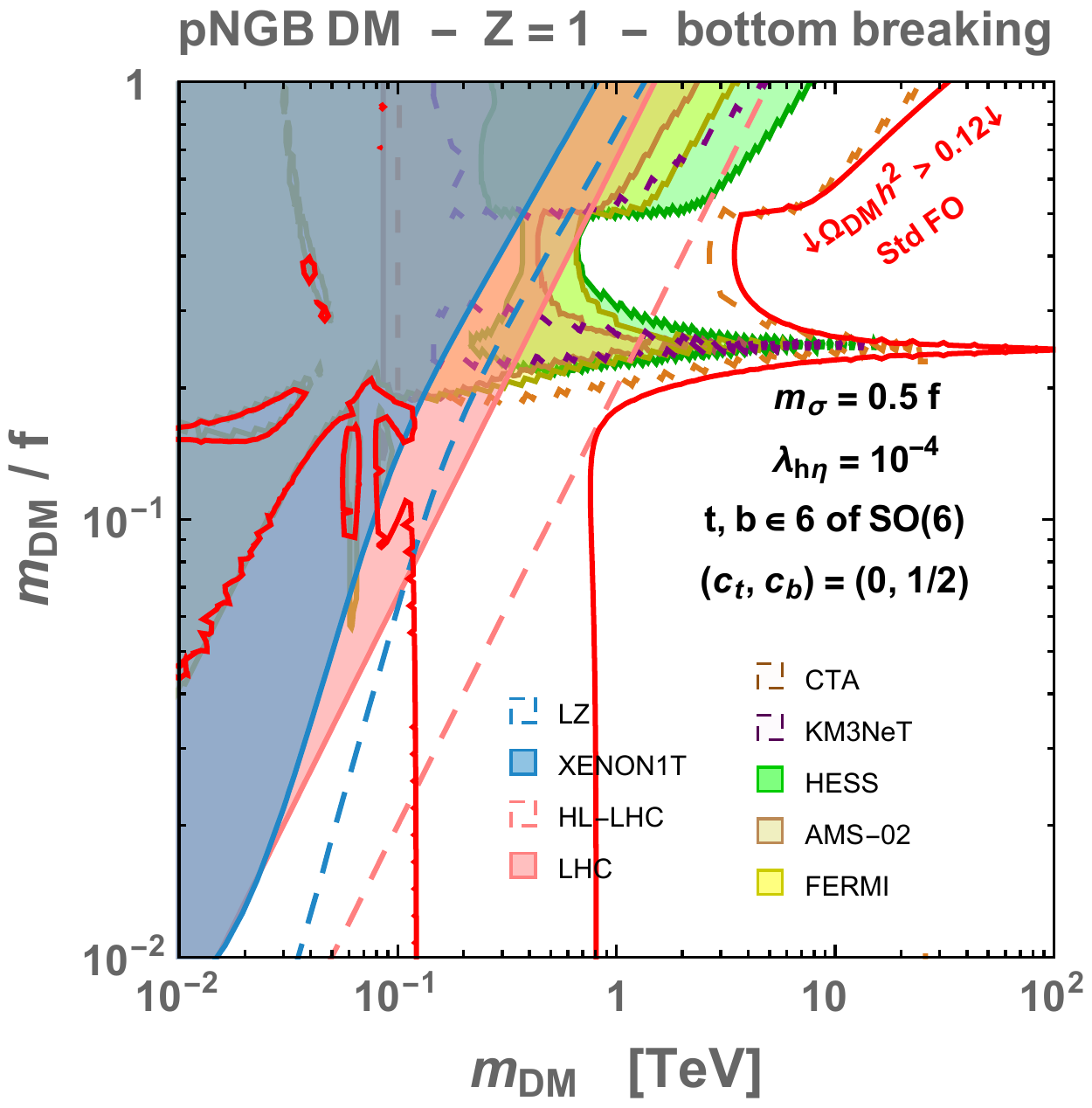}}}
\raisebox{0cm}{\makebox{\includegraphics[height=0.5\textwidth, scale=1]{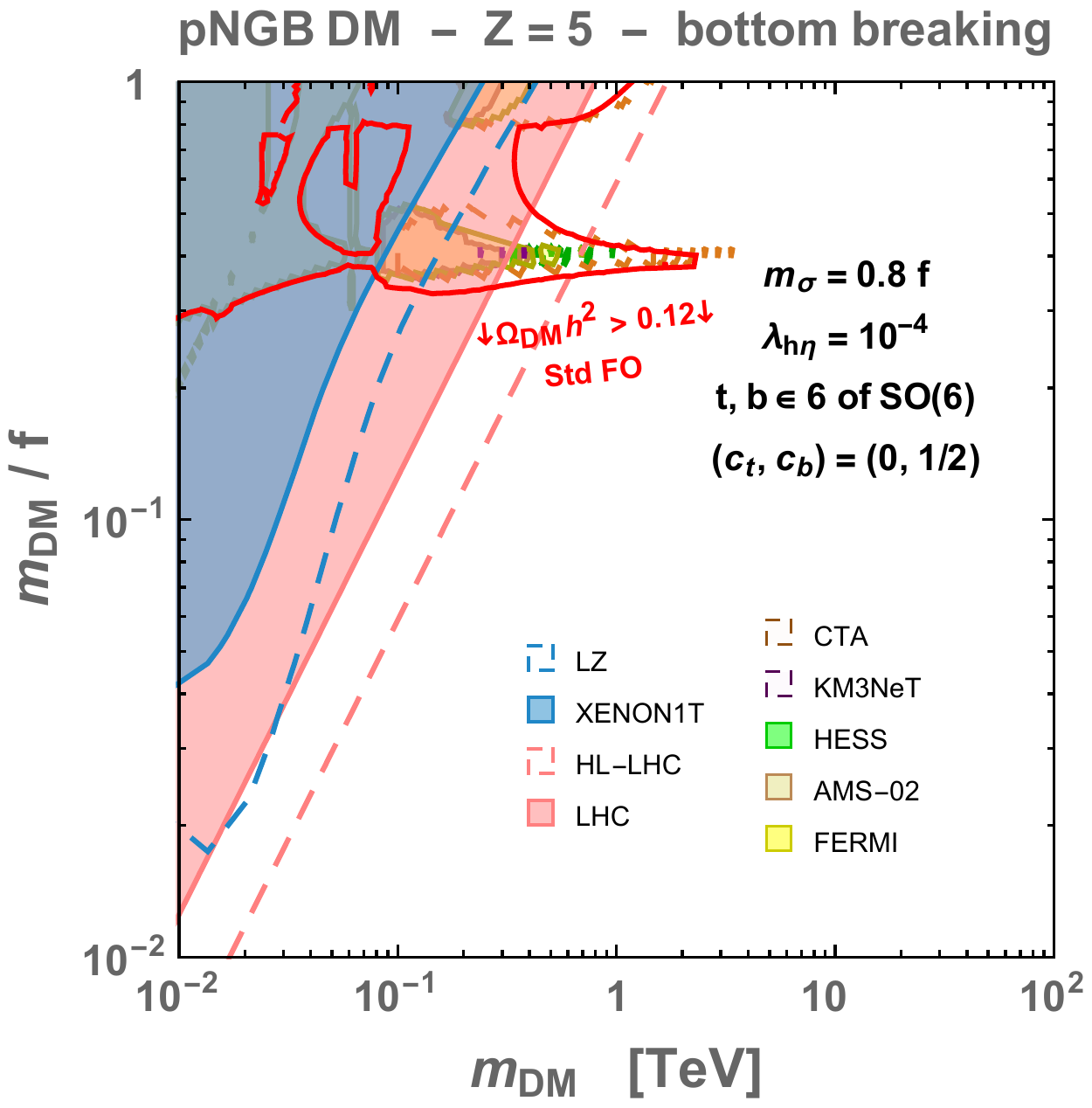}}}
\caption{\it \small \textbf{Dilaton-mediated O(6)/O(5) pNGB DM in the standard freeze-out case}.
Lines and shadings as in Fig.~\ref{fig:stdFO_pheno_msigma_p25a}.
All the rest as in Fig.~\ref{fig:stdFO_pheno_msigma_p25}, where here we additionally display the case where the shift symmetry is broken by the bottom quark, such that direct detection is dominated by dilaton exchange, as opposed to Higgs exchange in the case of top-breaking, see Eq.~\eqref{eq:DD_xsec_pNGB}.}
\label{fig:stdFO_pNGB_pheno_app} 
\thisfloatpagestyle{empty}
\end{figure}

Additionally, Fig.~\ref{fig:stdFO_pNGB_pheno_app} extends the pNGB scenario of Fig.~\ref{fig:stdFO_pheno_msigma_p25} to the case in which the pNGB mass receives its dominant contributions from the bottom quark instead of the top quark, such that the Higgs mixing $\lambda_{h\eta}$ is reduced.

\begin{figure}[ph!]
\centering
\vspace{-1.5cm}
\raisebox{0cm}{\makebox{\includegraphics[height=0.5\textwidth, scale=1]{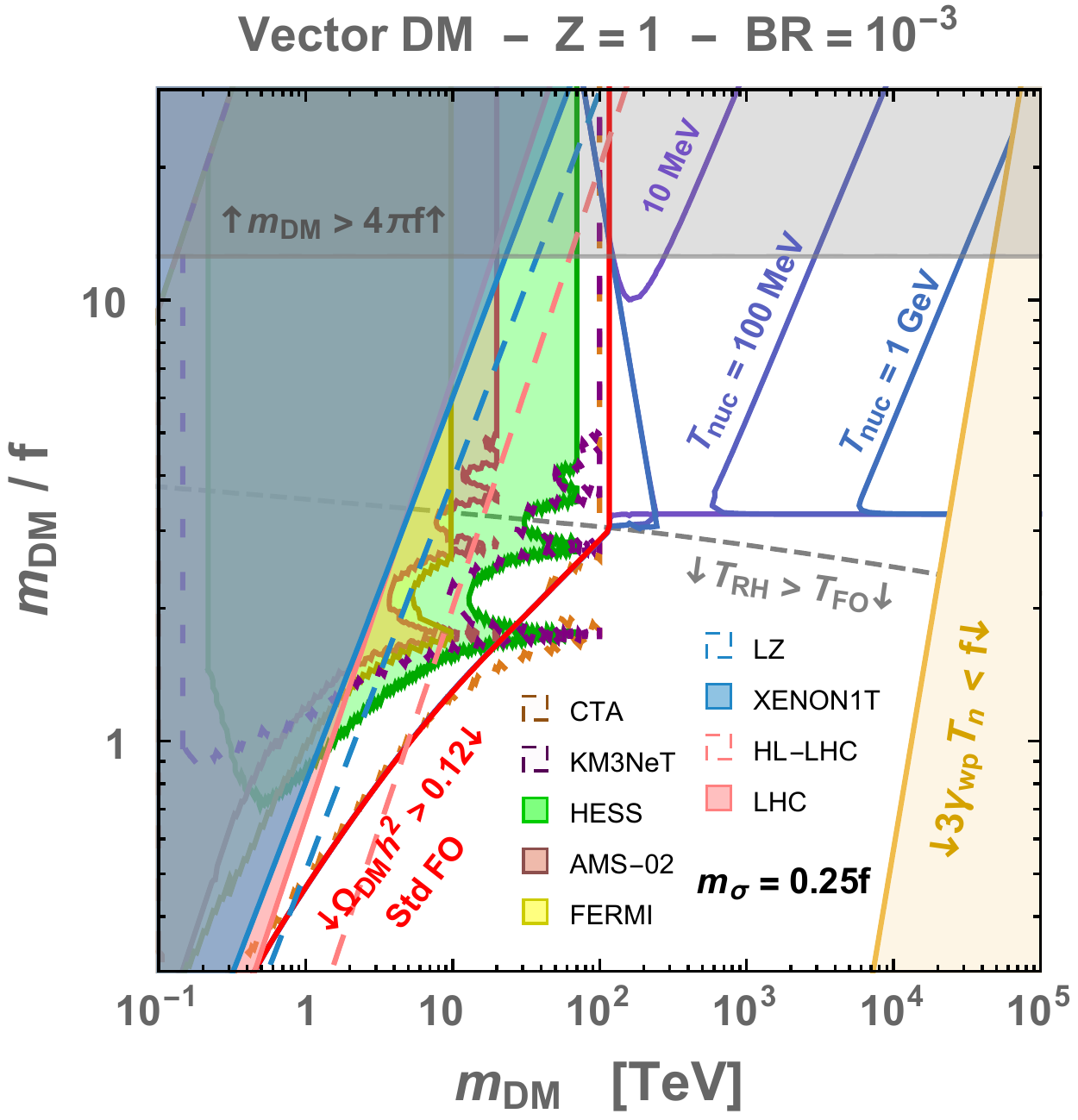}}}
\raisebox{0cm}{\makebox{\includegraphics[height=0.5\textwidth, scale=1]{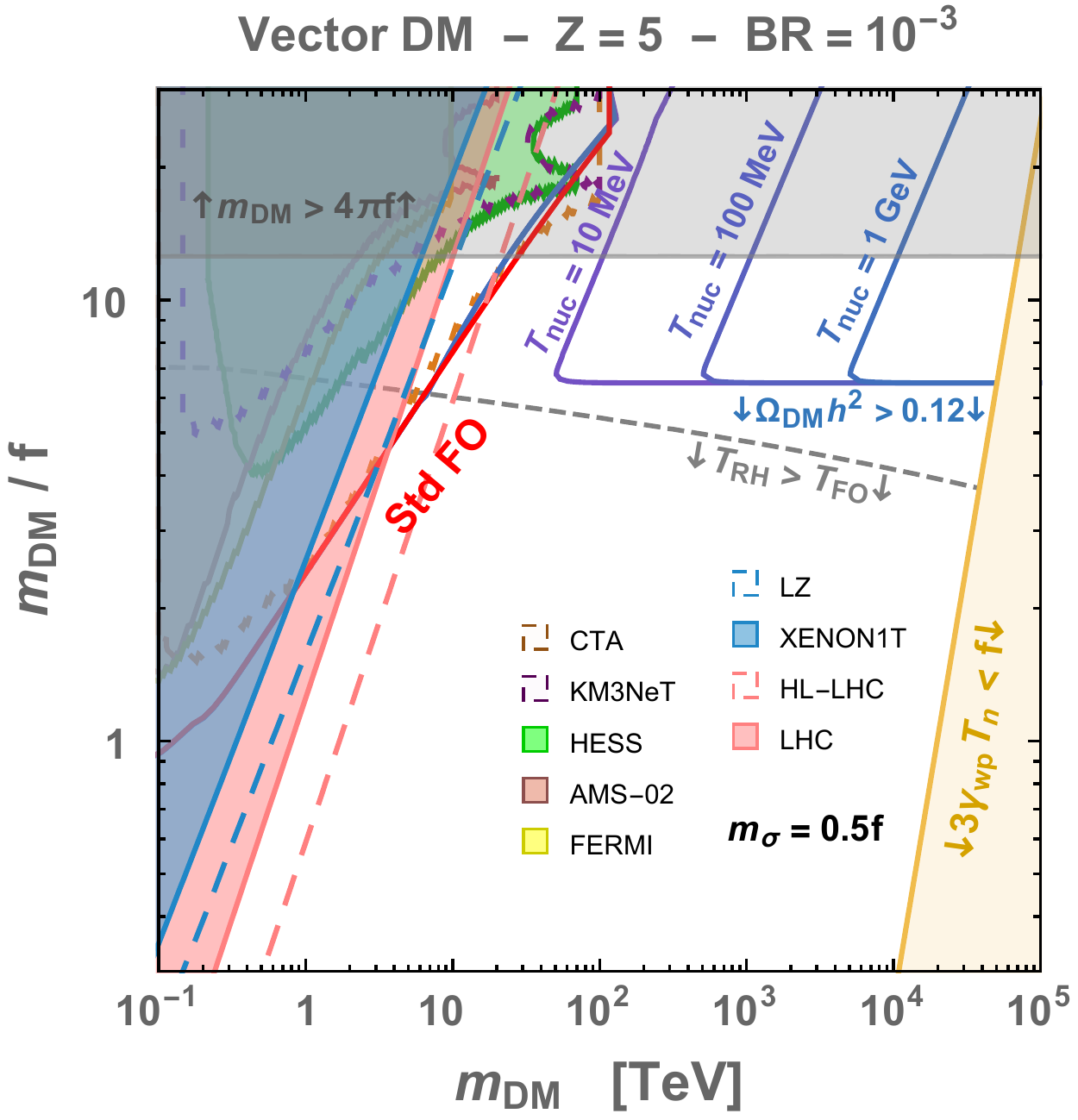}}}\\
\vspace{0.5 cm}
\raisebox{0cm}{\makebox{\includegraphics[height=0.5\textwidth, scale=1]{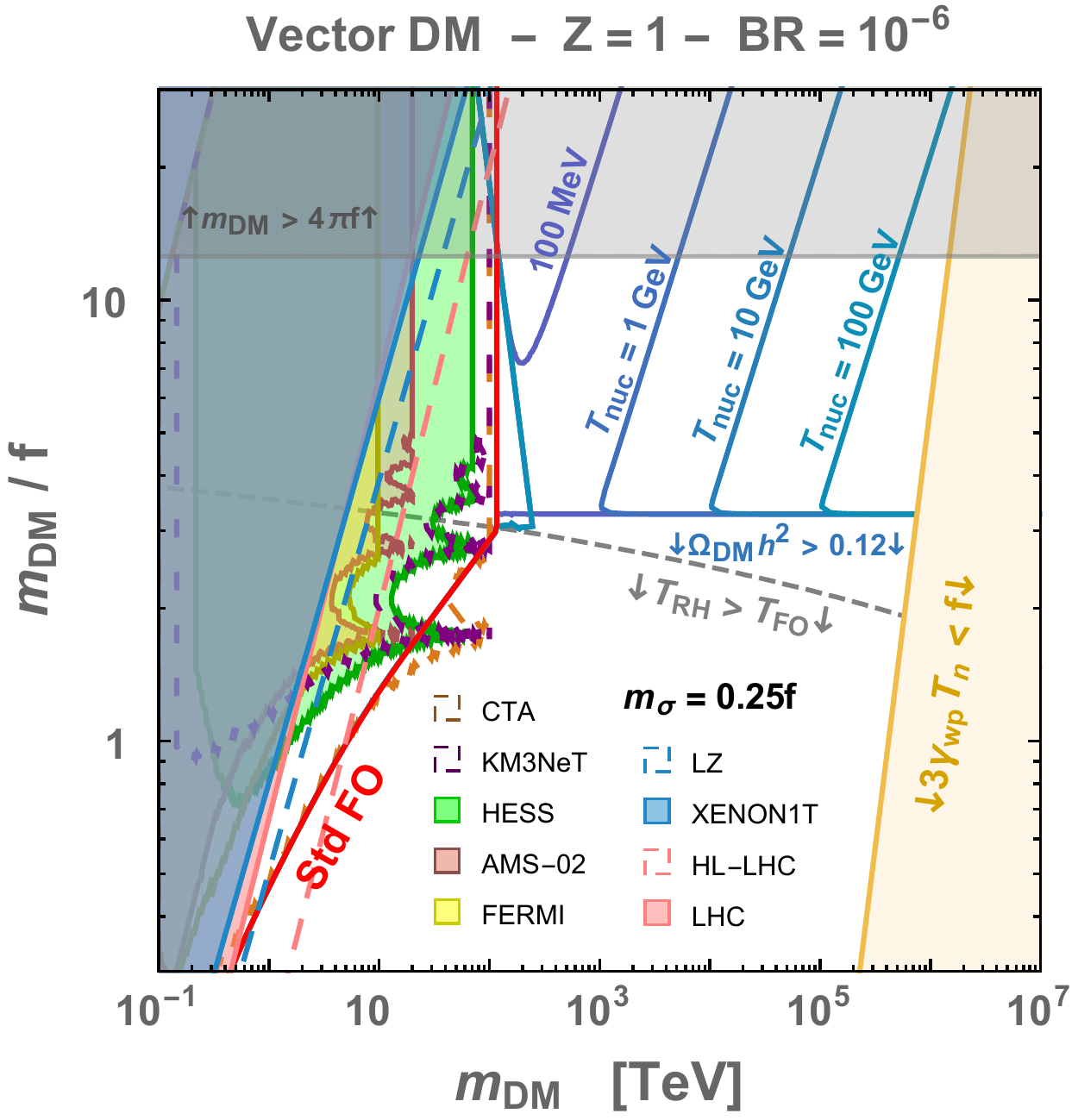}}}
\raisebox{0cm}{\makebox{\includegraphics[height=0.5\textwidth, scale=1]{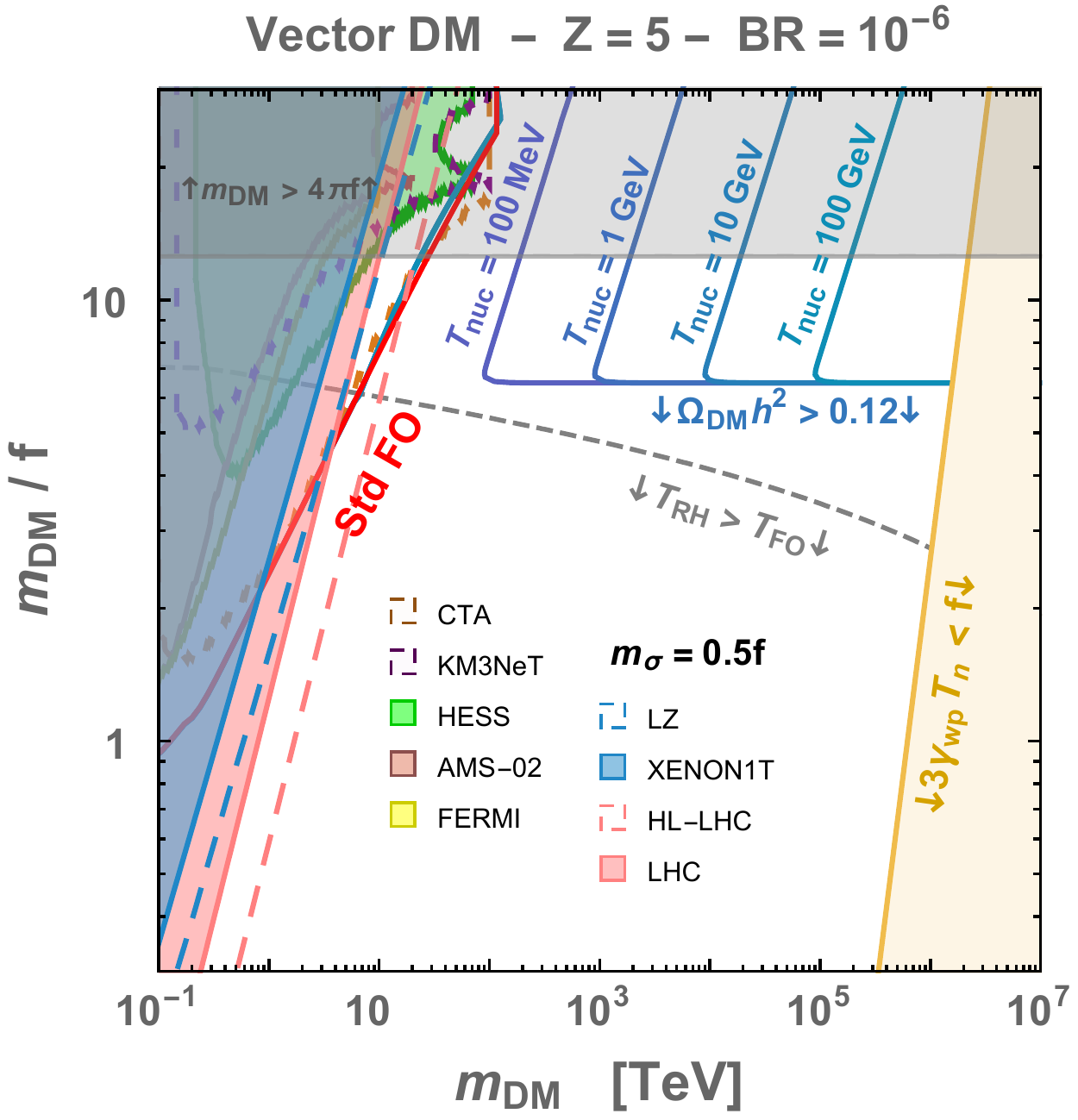}}}
\caption{\it \small \textbf{Supercooled dilaton-mediated vector DM.} Lines and shadings as in Fig.~\ref{fig:DM_SC_abund}.}

\label{fig:DM_SC_abund_vector} 
\thisfloatpagestyle{empty}
\end{figure}

\begin{figure}[ph!]
\centering
\vspace{-1.5cm}
\raisebox{0cm}{\makebox{\includegraphics[height=0.5\textwidth, scale=1]{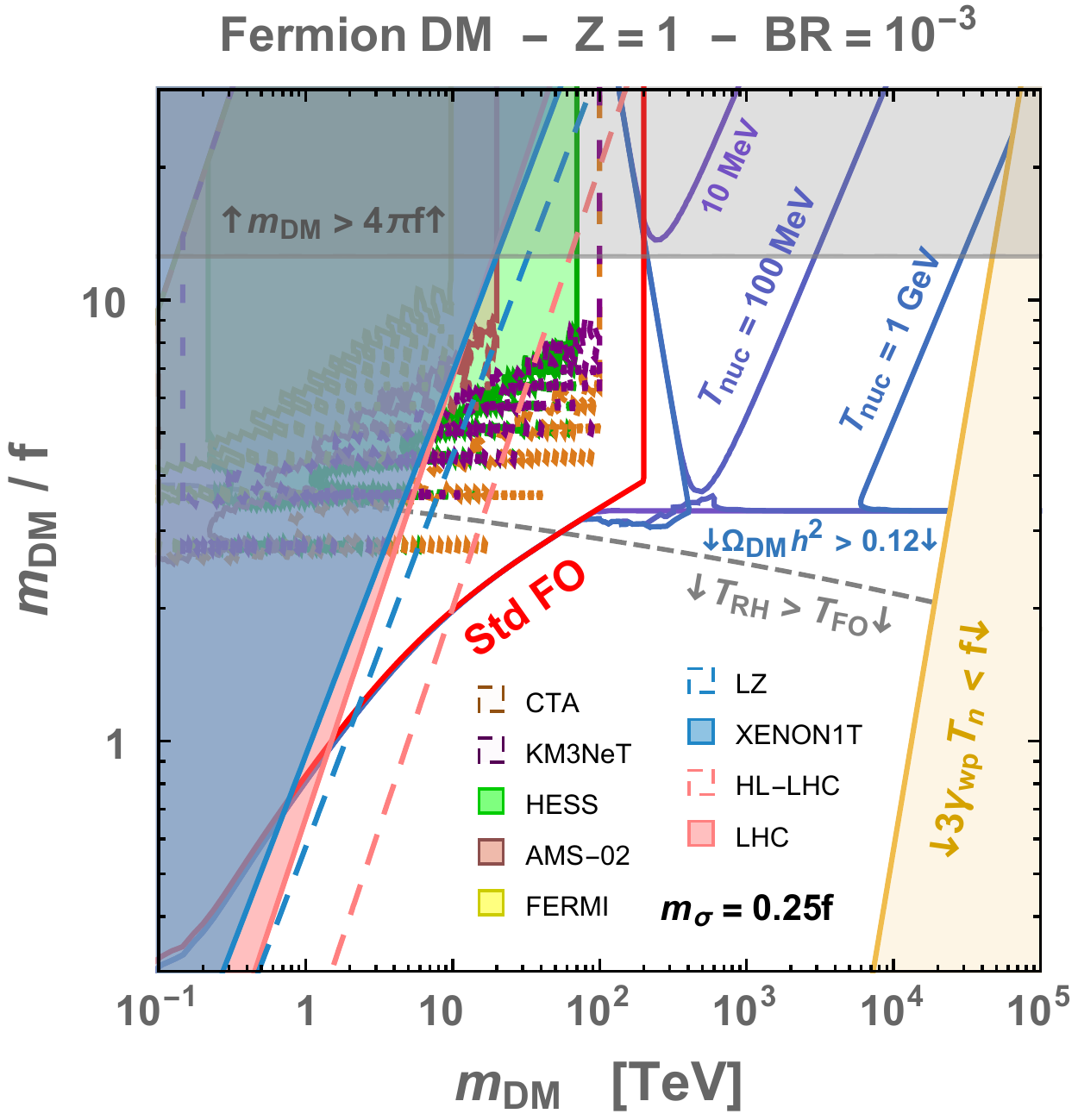}}}
\raisebox{0cm}{\makebox{\includegraphics[height=0.5\textwidth, scale=1]{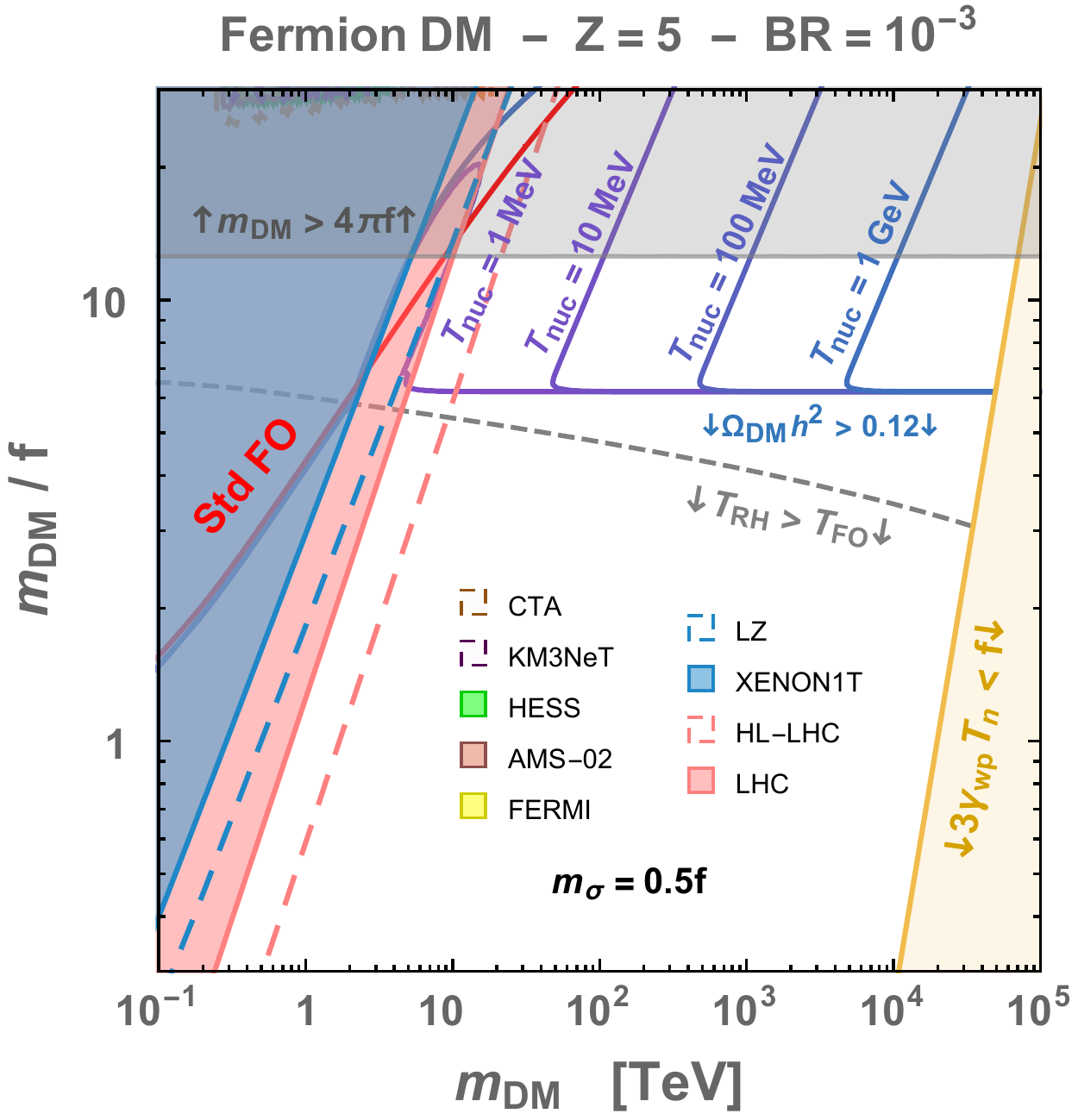}}}\\
\vspace{0.5 cm}
\raisebox{0cm}{\makebox{\includegraphics[height=0.5\textwidth, scale=1]{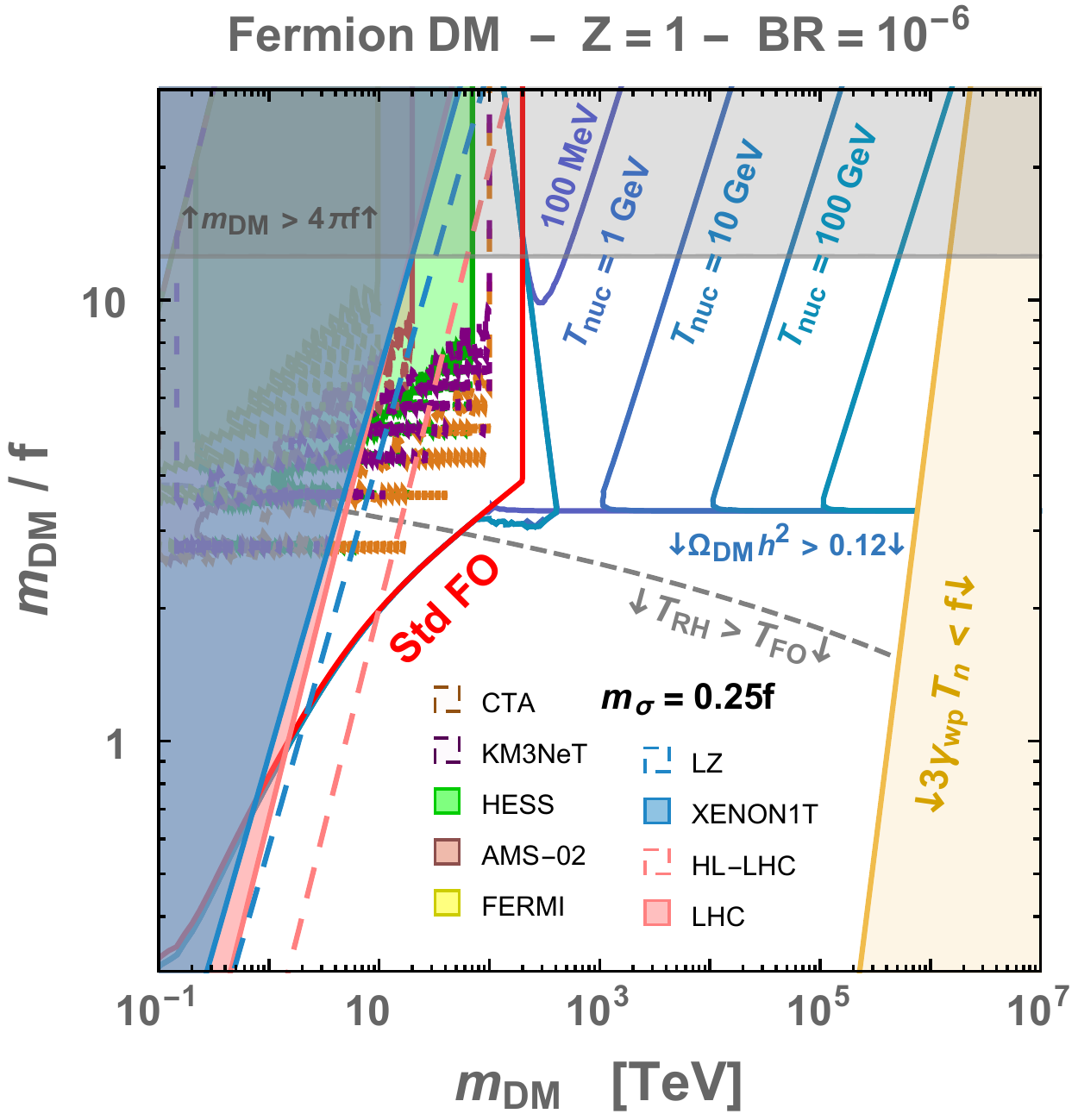}}}
\raisebox{0cm}{\makebox{\includegraphics[height=0.5\textwidth, scale=1]{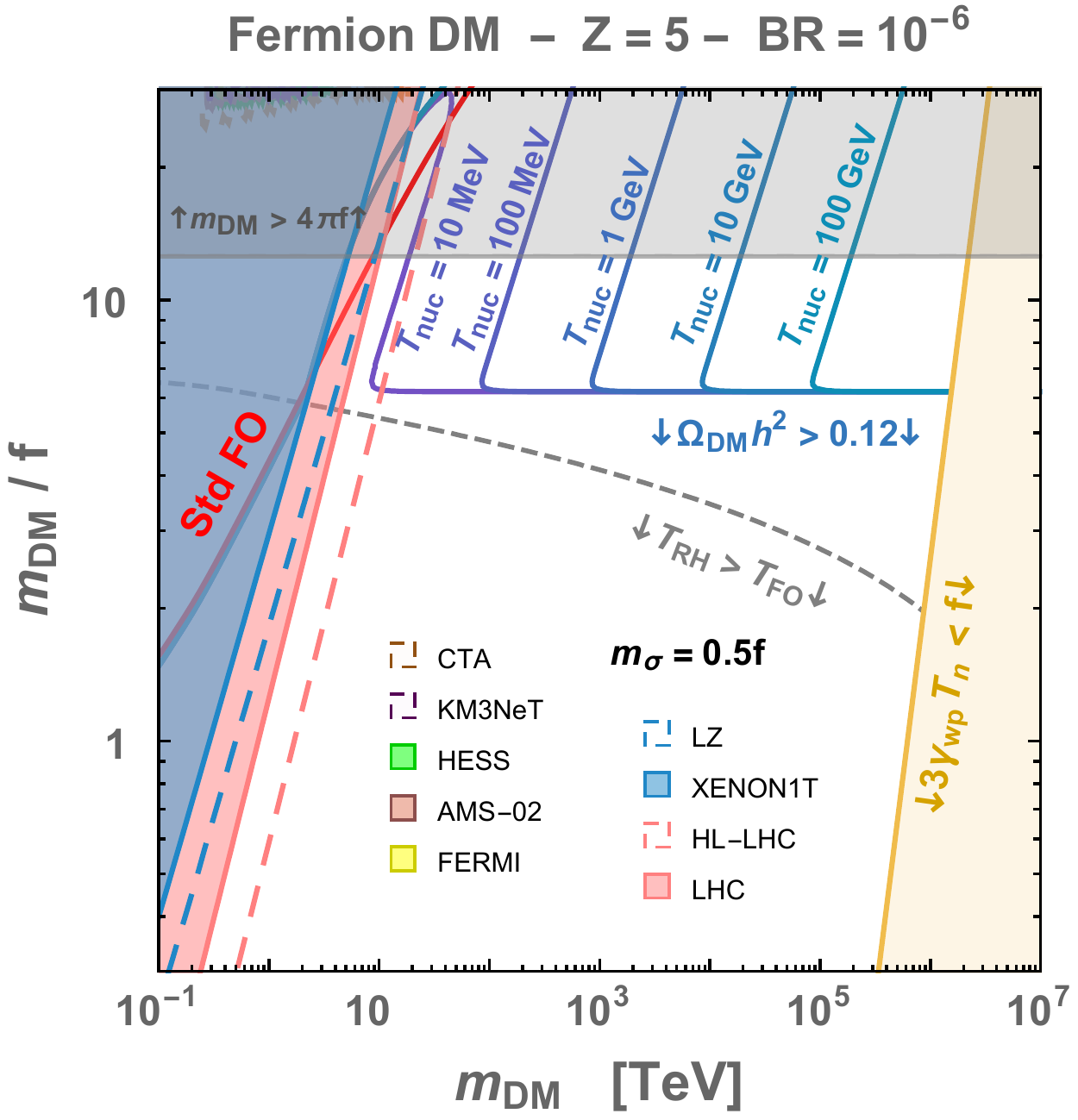}}}
\caption{\it \small \textbf{Supercooled dilaton-mediated fermionic DM.}
Lines and shadings as in Fig.~\ref{fig:DM_SC_abund}.
The indirect-detection constraints are suppressed compared to the scalar and vector cases, because the annihilation cross-section of fermion DM is p-wave, cf. Eq.~\eqref{eq:DMannfactors}.}
\label{fig:DM_SC_abund_fermion} 
\thisfloatpagestyle{empty}
\end{figure}
\subsection{Supercooled dilaton-mediated composite DM}
\label{app:Supercool dilaton-mediated DM}

We show the new parameter space which is opened when introducing a period of supercooling preceding the confinement. They complete Fig.~\ref{fig:DM_SC_abund} in Sec.~\ref{sec:SupercoolDM}.
We show the cases where DM is a vector in  Fig.~\ref{fig:DM_SC_abund_vector} and a fermion in  Fig.~\ref{fig:DM_SC_abund_fermion}. We vary the string-to-DM branching ratio from $10^{-3}$ to $10^{-6}$ and the dilaton normalization strength from $Z=1$ to $Z=5$.  We can see that the constraints coming from direct detection, indirect detection and colliders are far from reaching the new opened regions. However, they can be probed by GW, see Fig.~\ref{fig:summary}.

\FloatBarrier

\resumetocwriting

\section{DM non-relativistic potential}
\label{app:NR_potential}
\stoptocwriting

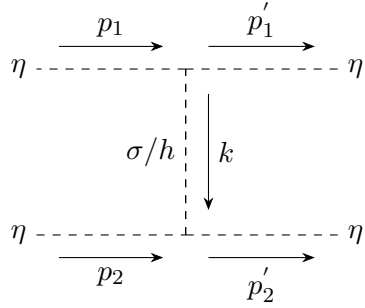
\begin{figure}[ph!]
\centering

{\begin{tikzpicture}
\begin{feynman}
   \vertex (a1) {\(\eta\)};
    \vertex[right=2.2cm of a1] (a2);
    \vertex[right=2cm of a2] (a3){\(\eta\)};
    \vertex[right=2cm of a3] (a4);

    \vertex[below=2.2cm of a1] (b1){\(\eta\)};
    \vertex[right=2.2cm of b1] (b2);
    \vertex[right=2cm of b2] (b3){\(\eta\)};

    \diagram*{
       {[edges=scalar, rmomentum'=\(p_1\)]
        (b2) -- [scalar,  rmomentum'=\(k\),edge label =\(\sigma/h\)] (a2),
      },
      (a2) -- [scalar,  rmomentum'=\(p_1\)] (a1),
      (a3) -- [scalar,  rmomentum'=\(p^{'}_1\)] (a2),
     
      (b1) -- [scalar, momentum'=\(p_2\)] (b2),
      (b2) -- [scalar, momentum'=\(p^{'}_2\)] (b3),

    };

\end{feynman}
\end{tikzpicture}
}
\caption{\it \small Feynman diagram of the 4-point function amplitude $\mathcal{W}(\vec{k}) $, either dilaton $\sigma$ or Higgs $h$ mediated, for pNGB DM $\eta$. Generalization to scalar, fermion and vector resonance DM is straightforward.}
\label{fig:4-pt_fct_amplitude} 
\thisfloatpagestyle{empty}
\end{figure}

\paragraph{Non-relativistic potential.}
The non-relativistic potential between the DM wave-functions can be computed as the Fourier transform
	\begin{equation}
	V(r) = -\frac{1}{4i\MDM^2} \int \frac{d\vec{k}^3}{(2\pi)^3} \mathcal{W}(\vec{k}) \, e^{-i\vec{k}\vec{r}} \label{eq:potential_def_FT}
	\end{equation}
of the four-point function amplitude $\mathcal{W}(\vec{k})$ with one-boson mediator exchange~\cite{Petraki:2015hla}, where $\vec{k}$ is the momentum of the exchanged boson. See an example in Fig.~\ref{fig:4-pt_fct_amplitude}.

\subsection{Dilaton-mediated channel}
\paragraph{Non-relativistic and instantaneous approximation.} 
In what follows, we work in the non-relativistic and instantaneous approximation limit \cite{Petraki:2015hla}
\begin{equation}
v_{\mathsmaller{\rm rel}},\, \alpha \lesssim 1 \quad \implies \quad k_0 \simeq \frac{\vec{k}^2}{2\mu}  \lesssim |\vec{k}| \quad \textrm{with}  ~|\vec{k}| \simeq \left( \mu v_{\mathsmaller{\rm rel}},\, \, \mu \alpha\right) ~ \textrm{and} ~ \mu \equiv \MDM/2. \label{eq:instantaneous_approx}
\end{equation}

\paragraph{4-point function amplitude.} 
For dilaton-mediated DM, scalar resonance, scalar pNGB, fermion resonance and vector resonance have the same 4-point amplitude
\begin{equation}
\mathcal{W}(\vec{k}) \simeq \frac{4 \MDM^4}{(Zf)^2}\frac{i}{\vec{k}^2 +m_\sigma^2} . \label{eq:wk_dilaton}
\end{equation}
We defined $k$ as the momentum running in the dilaton propagator and $p_1$, $p_2$ as the momenta of the incoming DM particles, see Fig.~\ref{fig:4-pt_fct_amplitude}.
To help the reader, we provide the detailed computations for scalar resonance DM, cf. Eq.~\eqref{eq:scalar_DM_dilaton}
\begin{align}
\mathcal{W}_{\rm scalar}(\vec{k}) &=\quad \frac{-2i\MDM^2}{(Zf)}\frac{i}{k^2 -m_\sigma^2} \frac{-2i\MDM^2}{(Zf)} \notag\\
&\overset{\mathclap{v_{\mathsmaller{\rm rel}}, \, \alpha ~\lesssim ~1}}{\simeq} \hspace*{1.5em}\frac{4 \MDM^4}{(Zf)^2}\frac{i}{\vec{k}^2 +m_\sigma^2} ,  \label{eq:wk_dilaton_scalar}
\end{align}
for scalar pNGB DM, cf. Eq.~\eqref{eq:derivative_coupling_dilaton_pNGB}
\begin{align}
\mathcal{W}_{\rm pNGB,\,\sigma}(\vec{k}) &= \frac{2i\left(p_1.(p_1-k) - 2 \MDM^2\right) }{Zf}\frac{i}{k^2 -m_\sigma^2} \frac{2i\left(p_2.(p_2+k) - 2 \MDM^2\right) }{Zf}\notag\\
&= -\frac{4 \MDM^4+8 \MDM^2 k(p_1-p_2) - 4(k p_1)(k p_2)}{(Zf)^2}\frac{i}{k^2 -m_\sigma^2} \notag \\
&\overset{\mathclap{v_{\mathsmaller{\rm rel}}, \, \alpha ~\lesssim ~1}}{\simeq} \hspace*{1.5em}\frac{4 \MDM^4}{(Zf)^2}\frac{i}{\vec{k}^2 +m_\sigma^2} , \label{eq:wk_dilaton_pNGB}
\end{align}
for fermion resonance DM, cf. Eq.~\eqref{eq:fermion_DM_dilaton}
\begin{align}
\mathcal{W}_{\rm fermion}(\vec{k}) &=(-1)\frac{-i\MDM}{Zf}\bar{u}(p_1^{'}) u(p_1)\frac{i}{k^2 -m_\sigma^2}  \frac{-i\MDM}{Zf}\bar{v}(p_2) v(p_2^{'}) \label{eq:minus_sign_fermion}\\
&\overset{\mathclap{v_{\mathsmaller{\rm rel}}, \, \alpha ~\lesssim ~1}}{\simeq} \hspace*{1.5em}(-1) \frac{\MDM^2}{(Zf)^2} \bar{u}(p_1)u(p_1)\frac{i}{\vec{k}^2 +m_\sigma^2} \frac{\MDM}{Zf}\bar{v}(p_2) v(p_2) \notag \\
&=\frac{4 \MDM^4}{(Zf)^2}\frac{i}{\vec{k}^2 +m_\sigma^2} , \label{eq:wk_dilaton_fermion}
\end{align}
and for vector resonance DM, cf. Eq.~\eqref{eq:vector_DM_dilaton}
\begin{align}
\mathcal{W}_{\rm vector}(\vec{k}) &=\quad \frac{-2i\MDM^2}{Zf}\epsilon^\mu(p_1)\epsilon^{*}_\mu(p_1^{'}) \frac{i}{k^2 -m_\sigma^2} \frac{-2i\MDM^2}{Zf} \epsilon^{*}_\alpha(p_2)\epsilon^\alpha(p_2^{'})\notag\\
&\overset{\mathclap{v_{\mathsmaller{\rm rel}}, \, \alpha ~\lesssim ~1}}{\simeq}\hspace*{1.5em}\frac{2\MDM^2}{Zf}\epsilon^\mu(p_1)\epsilon^{*}_\mu(p_1) \frac{i}{\vec{k}^2 +m_\sigma^2} \frac{2\MDM^2}{Zf} \epsilon^{*}_\alpha(p_2)\epsilon^\alpha(p_2)\notag\\
&= \frac{4\MDM^2}{(Zf)^2}\frac{i}{\vec{k}^2 +m_\sigma^2}. \label{eq:wk_dilaton_vector}
\end{align}
For the fermion and vector cases, we have used the normalization rules \cite{Peskin:1995ev}
\begin{align}
&\bar{u}^r(p)u^s(p)=2m \delta^{rs}, \qquad \bar{u}^r(p)u^s(p)=-2m \delta^{rs}, \\
&\epsilon_\mu^{*}(p) \epsilon^\mu(p) = -1.
\end{align}
The minus sign in Eq.~\eqref{eq:minus_sign_fermion} arises from fermion permutations \cite[Sec.~4.7]{Peskin:1995ev}.

\paragraph{Yukawa potential.} 
After injecting Eq.~\eqref{eq:wk_dilaton} into Eq.~\eqref{eq:potential_def_FT}, we get the Yukawa potential
\begin{align}
V_{\rm \sigma}(r) &= -\frac{2\pi}{(2\pi)^3 4i \MDM^2}  \frac{4\MDM^4}{(Zf)^2} \int_0^\infty dk \, k^2 \,  \int_0^\pi d\theta \, \sin{\theta}\,\frac{i}{k^2 +m_\sigma^2}  \, e^{-ikr\cos{\theta}}\notag\\
&= -\frac{1}{(2\pi)^2 4i \MDM^2}  \frac{4\MDM^4}{(Zf)^2} 2 \int_0^\infty dk \, k^2 \, \frac{i}{k^2 +m_\sigma^2}  \, \frac{\sin{kr}}{r}\notag\\
&= -\frac{1}{(2\pi)^2 4 \MDM^2}  \frac{4\MDM^4}{(Zf)^2} \, \pi  \frac{ e^{-m_{\dilaton} r} }{ r }\notag\\
&=-\alpha_{\sigma} \frac{ e^{-m_{\dilaton} r} }{ r },\label{eq:NR_pot_dilaton}
\end{align}
with
	\begin{equation}
	\alpha_{\sigma} = \frac{1}{4\pi}\frac{ \MDM^2 }{ (Z f)^2 }.
	\label{eq:alpha_def_app}
	\end{equation}
Our result for the pNGB case, Eq.~\eqref{eq:wk_dilaton_pNGB}, differs from \cite{Kim:2016jbz} where the authors find a value of $\alpha_{\sigma}$ which is $9$ times larger.

\begin{figure}[t]
\centering
\vspace{-1.5cm}
\raisebox{0cm}{\makebox{\includegraphics[width=0.6\textwidth, scale=1]{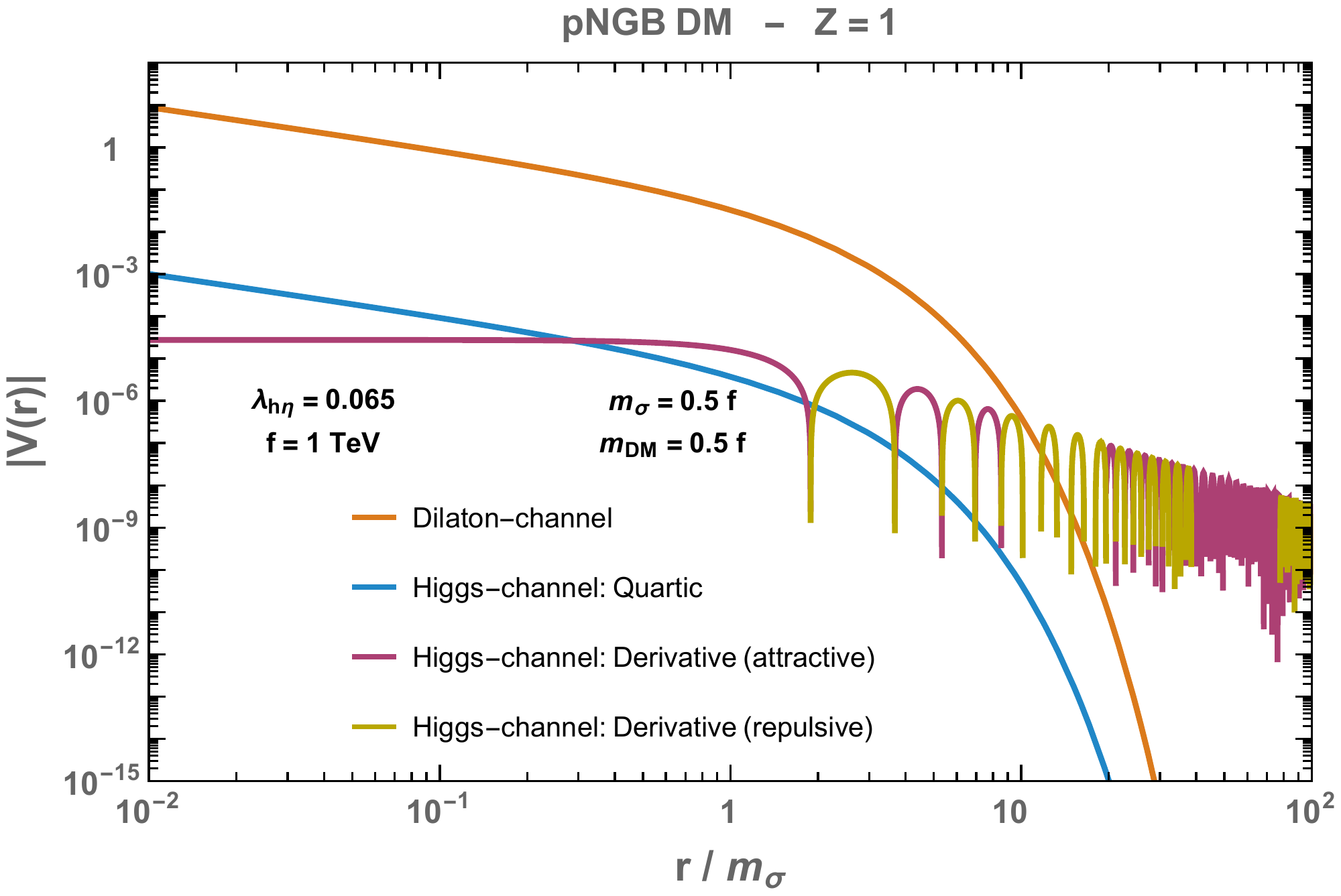}}}
\caption{\it \small Non-relativistic potentials between 2 pNGN DM particles assuming either the dilaton-mediated channel or the Higgs-mediated channel. In the later case, we consider either the quartic interaction or the derivative one, cf. Eq.~\eqref{eq:quartic_derivative}. }
\label{fig:NR_potential} 
\end{figure}
	
\subsection{Higgs-mediated channel}
\paragraph{Couplings.} 
Due to the coset structure of composite Higgs models, pNGB DM has also interactions with the Higgs boson.
The lagrangian in Eq.~\eqref{eq:pNGBDM_Higgs}, valid for the coset $SO(6)/SO(5)$, leads after SSB, $H = \left(0,\,(h + v_H)/\sqrt{2}\right)$, to the interaction terms
\begin{equation}
\mathcal{L} \supset \frac{\lambda_{h\eta} v_H}{2} h \,\eta^2 + \frac{v_H}{4(Zf)^2} (\partial_\mu h \partial^\mu \eta) \eta. \label{eq:quartic_derivative}
\end{equation}
\paragraph{4-point function.} 
They give the following respective Higgs-mediated four-point function amplitudes
\begin{equation}
\mathcal{W}_{\rm h,  \,quartic}(k) \simeq \lambda_{h\eta} v_H \frac{i}{|\vec{k}|^2 +m_\sigma^2}, \qquad \mathcal{W}_{\rm h, \, derivative}(k) \simeq \frac{v_H}{4(Zf)^4} \frac{i|\vec{k}|^4}{|\vec{k}|^2 +m_\sigma^2}. \label{eq:4-pt_fct_higgs}
\end{equation}
\paragraph{Higgs-mediated Yukawa potential.} 
From introducing the first term of Eq.~\eqref{eq:4-pt_fct_higgs} into Eq.~\eqref{eq:potential_def_FT}, we get a Yukawa potential
\begin{equation}
V_{\rm h,\, quartic}(r) =-\alpha_{\rm h,\, quartic} \frac{ e^{-m_{\dilaton} r} }{ r },\qquad \alpha_{\rm h,\, quartic} = \frac{1}{16\pi}\frac{\lambda_{h\eta}^2 v_{\rm H}^2 }{  \MDM^2 }. \label{eq:NR_potential_higgs_quartic}
\end{equation}
The injection of the second term of Eq.~\eqref{eq:4-pt_fct_higgs} into Eq.~\eqref{eq:potential_def_FT} leads to a divergent integral which must be cut-off at the EFT scale $|\vec{k}| \lesssim f$. We obtain
\begin{align}
V_{\rm h,\, derivative}(r) &\simeq \frac{v_H}{64\pi^2 \MDM^2 (Zf)^4} \frac{2 f r\left(6 +(m_{\sigma}^2 - f^2)r^2\right) \cos{f r} - 2 \left( 6 + (m_{\sigma}^2 - 3 f^2)r^2\right) \sin{f r} }{r^5} \notag\\
&\simeq \frac{v_H}{64\pi^2 \MDM^2 (Zf)^4} \times \left\{
                \begin{array}{ll}
                 2 f(m_{\sigma}^2 - f^2)\dfrac{\cos{f r}}{r^2}, \qquad r \gtrsim f^{-1} , \\ \\
               -\frac{2}{3} f^3 m_{\sigma}^2, \qquad \qquad ~ \qquad r \lesssim f^{-1} .
                \end{array}
              \right.\label{eq:NR_potential_higgs_derivative}
\end{align}
\paragraph{Higgs- and dilaton-mediated Yukawa potential.} 
The non-relativistic potential between pNGB DM particles is the sum of Eq.~\eqref{eq:NR_pot_dilaton}, Eq.~\eqref{eq:NR_potential_higgs_quartic} and Eq.~\eqref{eq:NR_potential_higgs_derivative}
\begin{equation}
V_{\rm pNGB}(r) = - \alpha_{\sigma} \frac{e^{-m_{\sigma} r}}{r} - \alpha_{\rm h, \,quartic} \frac{e^{-m_{\sigma} r}}{r} + V_{\rm h,\, derivative}(r) .
\end{equation}
We compare the three pNGB DM potentials in Fig.~\ref{fig:NR_potential}.  

In contrast to the other two, $V_{\rm h,\, derivative}(r)$ is not a Yukawa potential. Instead, at large distance $r \gtrsim f^{-1}$, the potential oscillates around attractive and repulsive behavior with a $1/r^2$ envelop, and at small distance $r \lesssim f^{-1}$, it flattens. 
The impact of such a potential on the annihilation cross-section of pNGB is beyond the scope of this work.
In any case, in the scenaro which we study in this work and for the chosen values of the parameters,
\begin{align}
&V_{\sigma}(r)  \gtrsim V_{\rm h,\, quartic}(r)  \quad \implies \quad \lambda_{h \eta} \lesssim \frac{6\MDM^2}{v_H (Z f)}, \\
&  V_{\sigma}(f^{-1})  \gtrsim V_{\rm h,\, derivative}(f^{-1}) \quad \implies \quad  \MDM^2 \gtrsim 6\sqrt{6\pi} Z m_{\sigma} v_H,
\end{align}
the full potential is always dominated by the dilaton-mediated contribution shown in orange in Fig.~\ref{fig:NR_potential}.
 \newpage

\begin{figure}[ph!]
\centering
\vspace{-1.5cm}
\raisebox{0cm}{\makebox{\includegraphics[height=0.45\textwidth, scale=1]{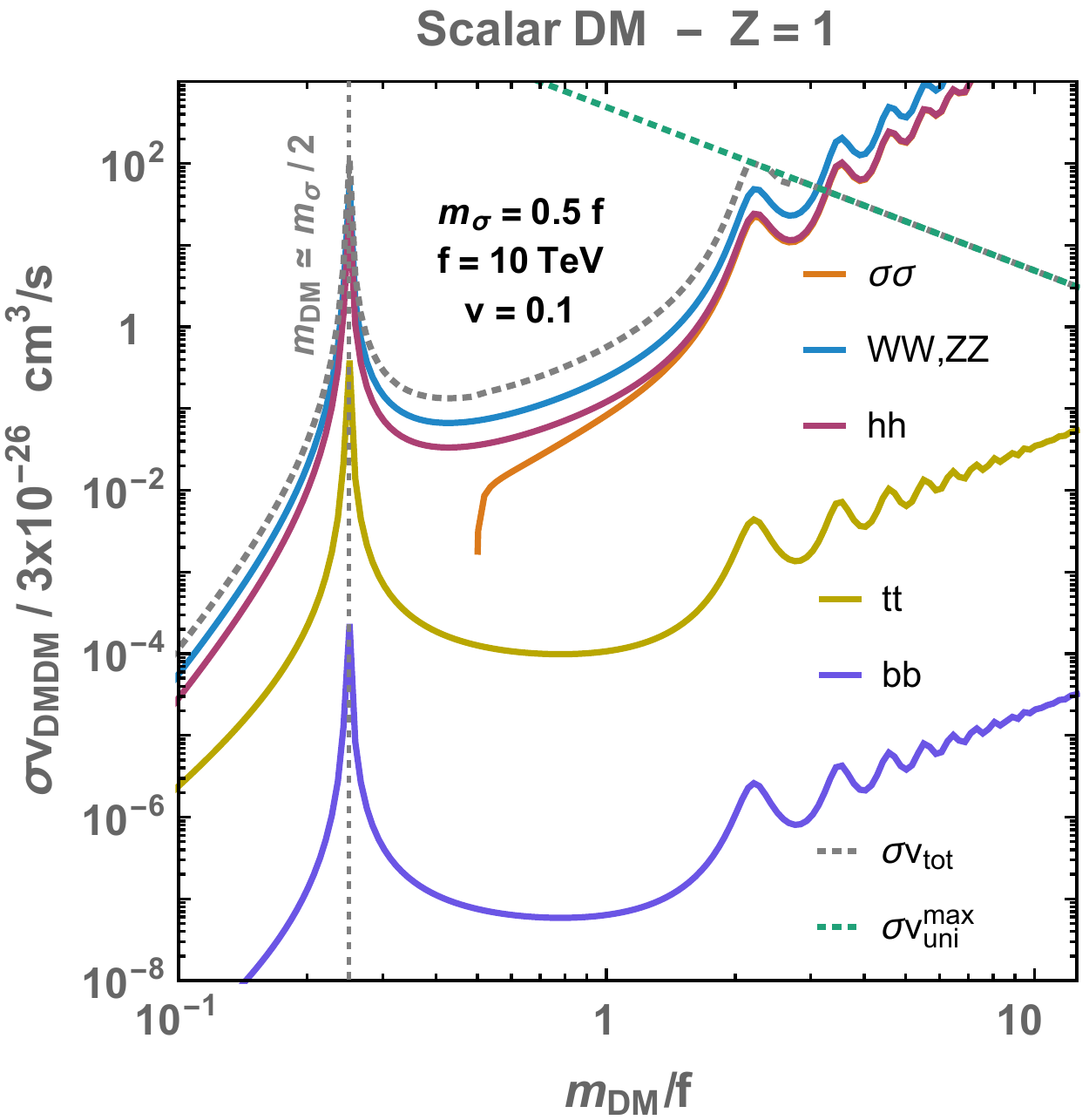}}}
\raisebox{0cm}{\makebox{\includegraphics[height=0.45\textwidth, scale=1]{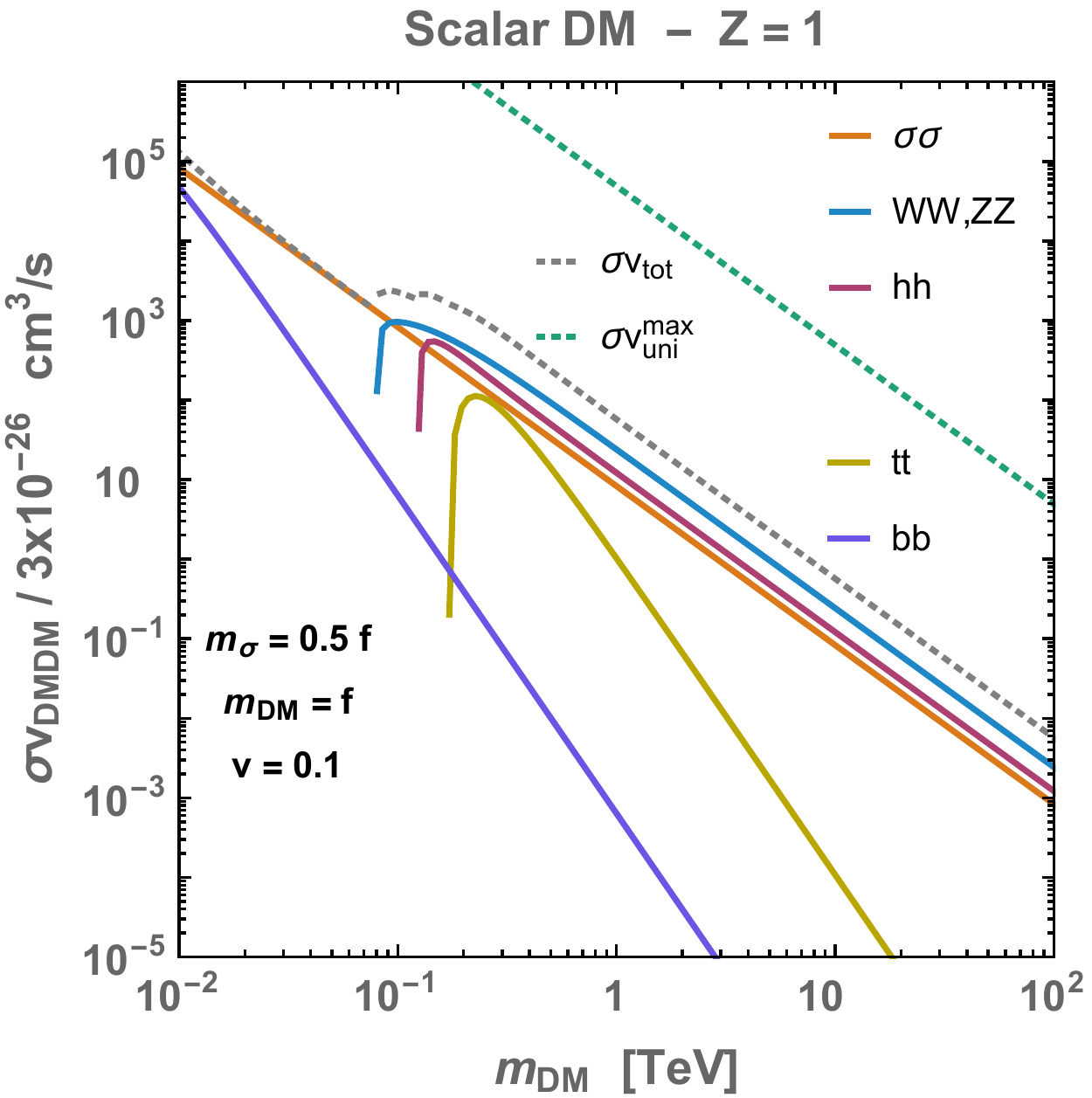}}}\\
\raisebox{0cm}{\makebox{\includegraphics[height=0.45\textwidth, scale=1]{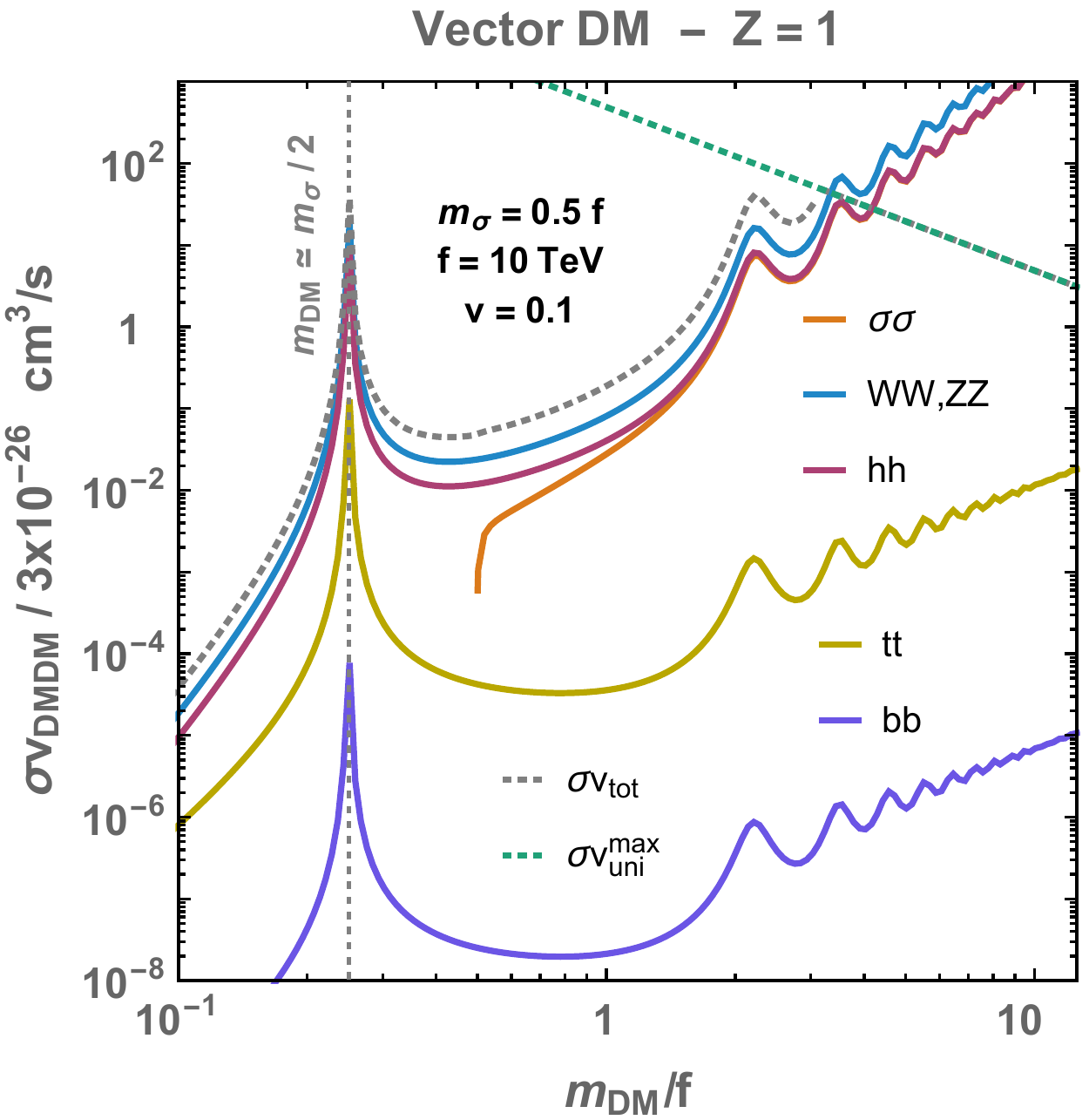}}}
\raisebox{0cm}{\makebox{\includegraphics[height=0.45\textwidth, scale=1]{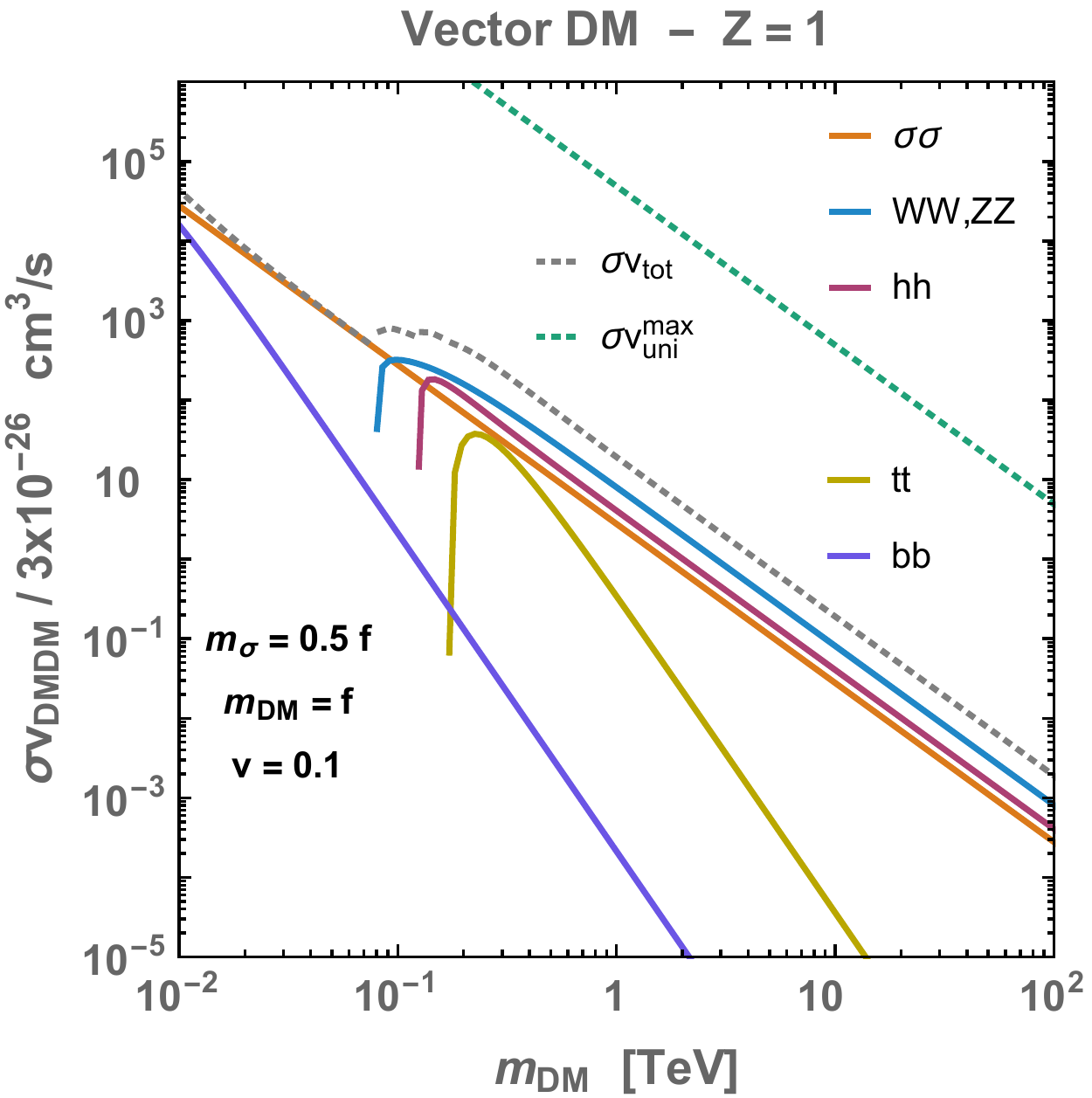}}}\\
\raisebox{0cm}{\makebox{\includegraphics[height=0.45\textwidth, scale=1]{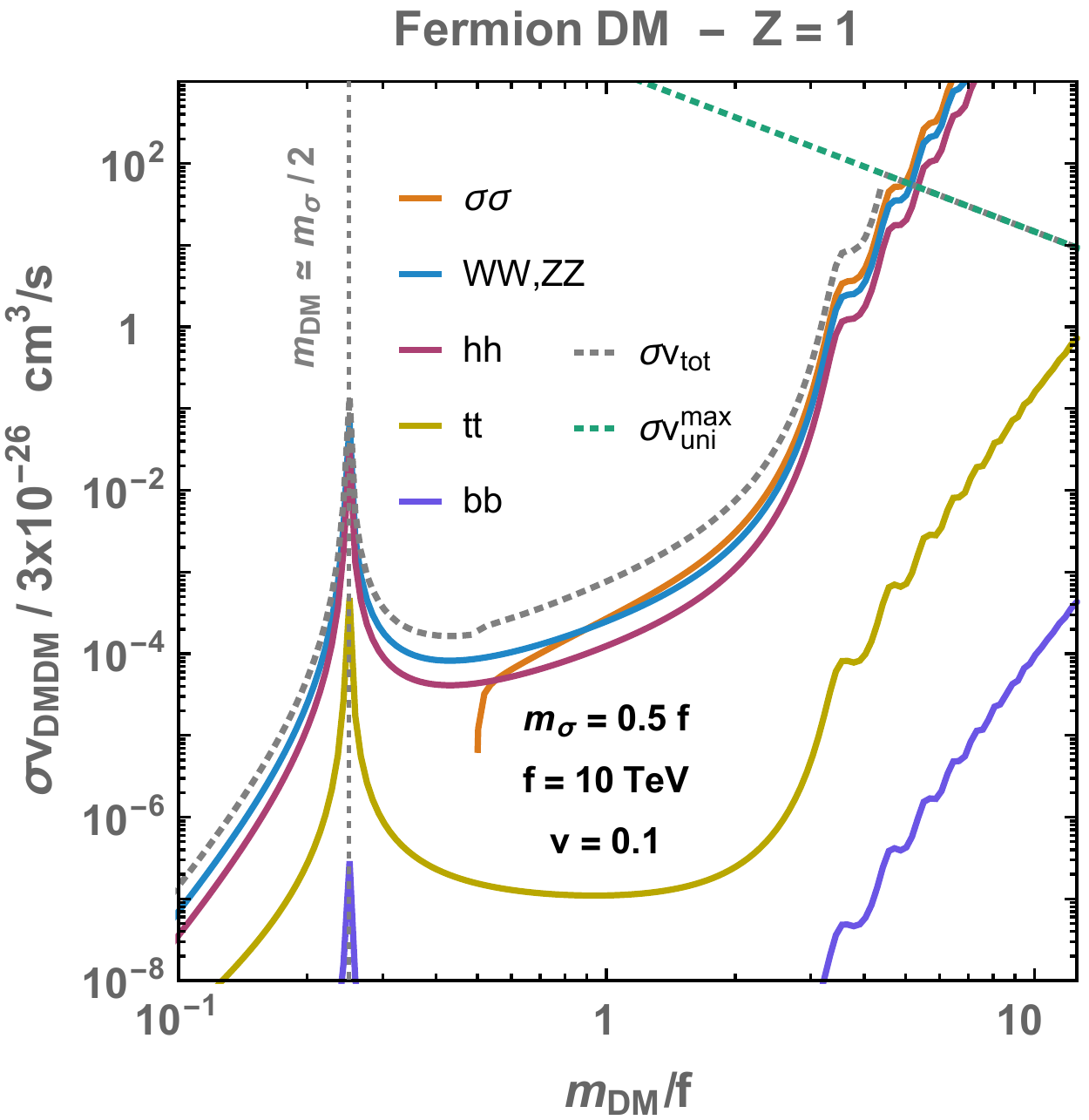}}}
\raisebox{0cm}{\makebox{\includegraphics[height=0.45\textwidth, scale=1]{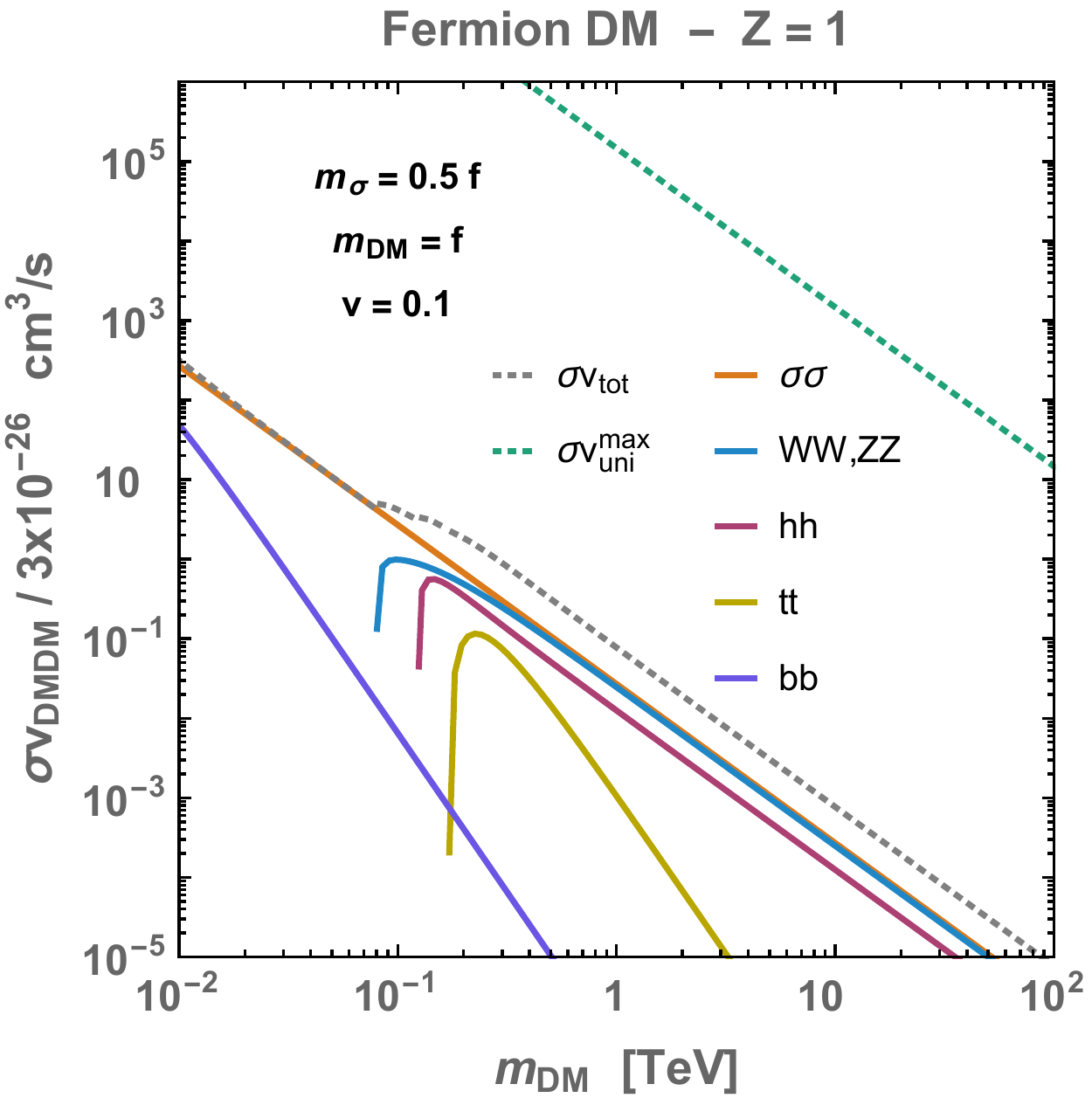}}}\\
\caption{\it \small \textbf{DM annihilation cross-section} in the scenarios where DM is a \textbf{scalar} (\textbf{top}), a \textbf{vector} (\textbf{middle}) or a \textbf{fermionic} (\textbf{bottom}) resonance of the confining sector. On the \textbf{left} panels we fix the confining scale $f = 10~\rm TeV$ while on the \textbf{right} panels, we fix it equal to the DM mass $f = \MDM$. For all plots, we set the dilaton mass to $m_{\sigma} = 0.5~f$ and the relative velocity to $v_{\mathsmaller{\rm rel}} = 0.1$. The wiggles are consequences of Sommerfeld effects, cf.~\ref{sec:unitary-bound}.}
\label{fig:sigmav_scalar_vector_fermion} 
\thisfloatpagestyle{empty}
\end{figure}

\begin{figure}[ph!]
\centering
\vspace{-1.5cm}
\raisebox{0cm}{\makebox{\includegraphics[height=0.5\textwidth, scale=1]{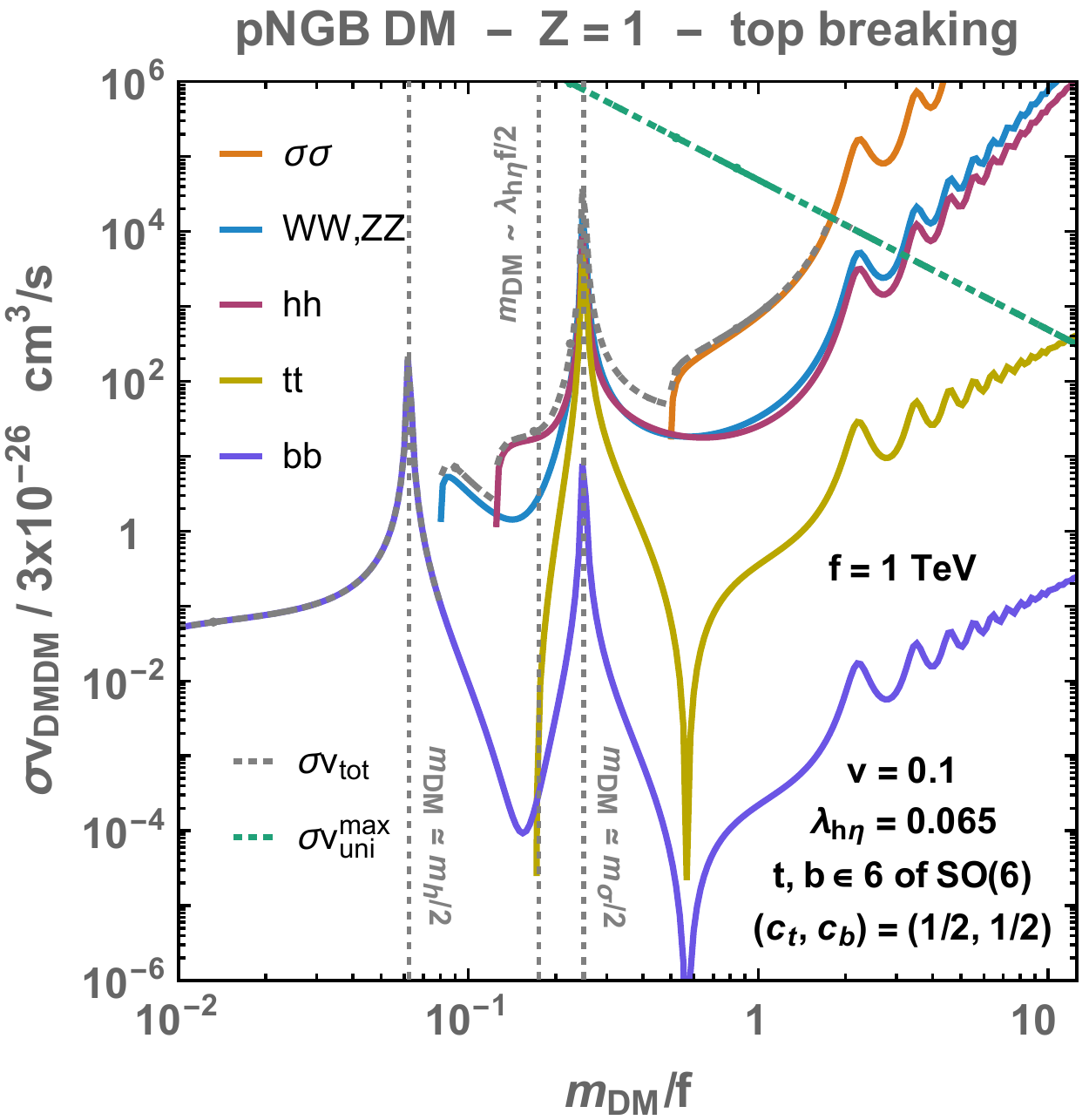}}}
\raisebox{0cm}{\makebox{\includegraphics[height=0.5\textwidth, scale=1]{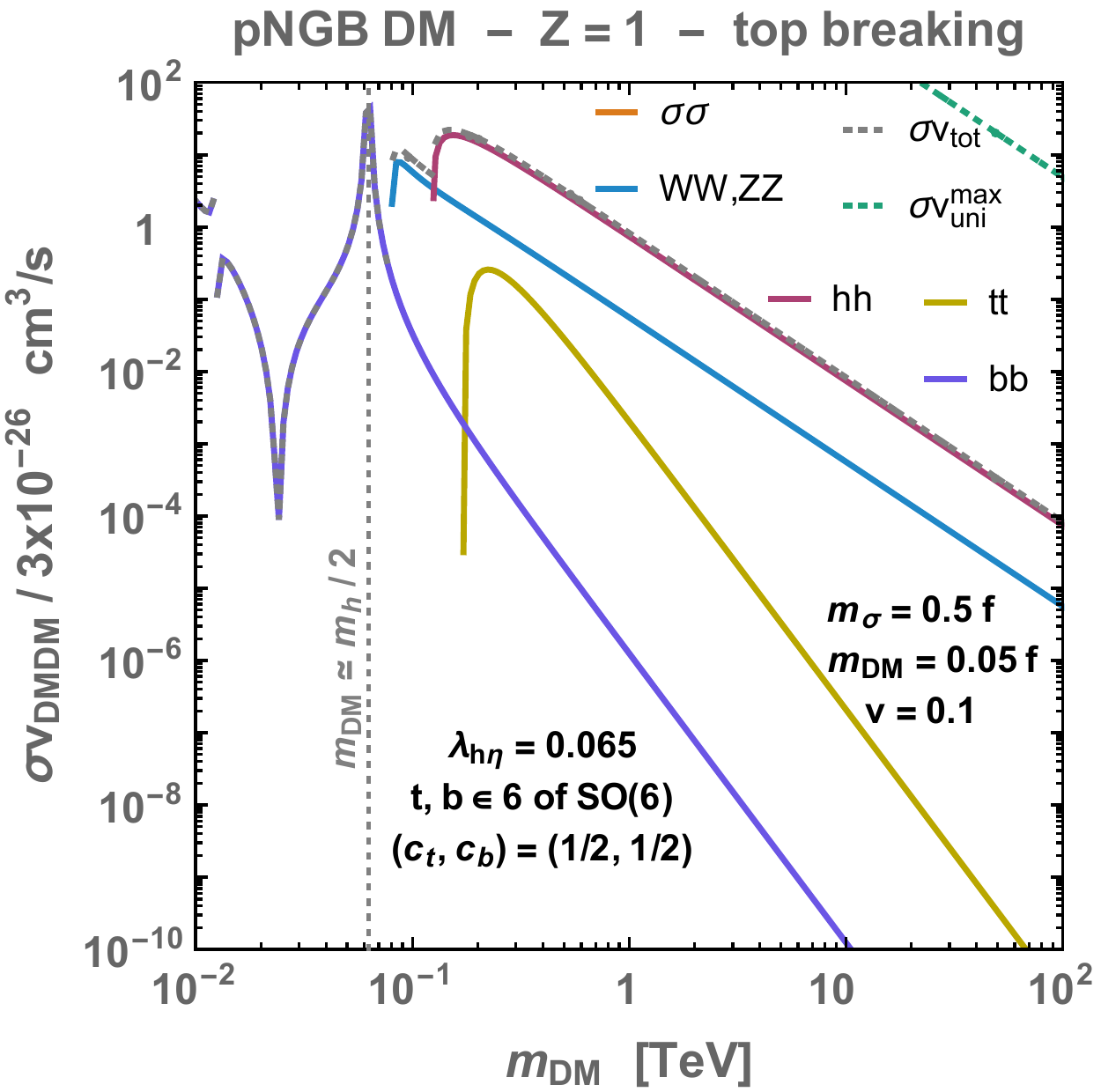}}}\\
\raisebox{0cm}{\makebox{\includegraphics[height=0.5\textwidth, scale=1]{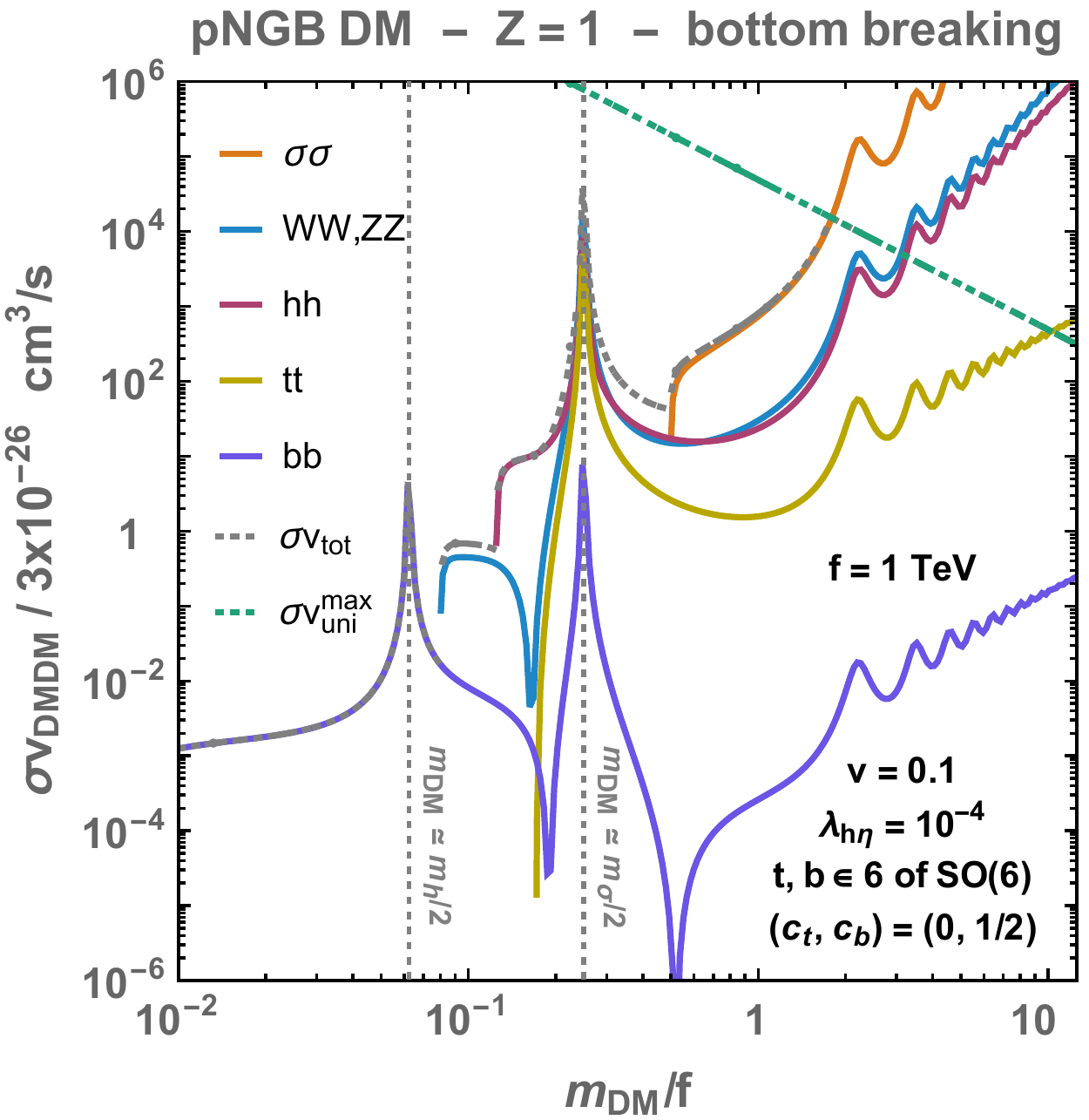}}}
\raisebox{0cm}{\makebox{\includegraphics[height=0.5\textwidth, scale=1]{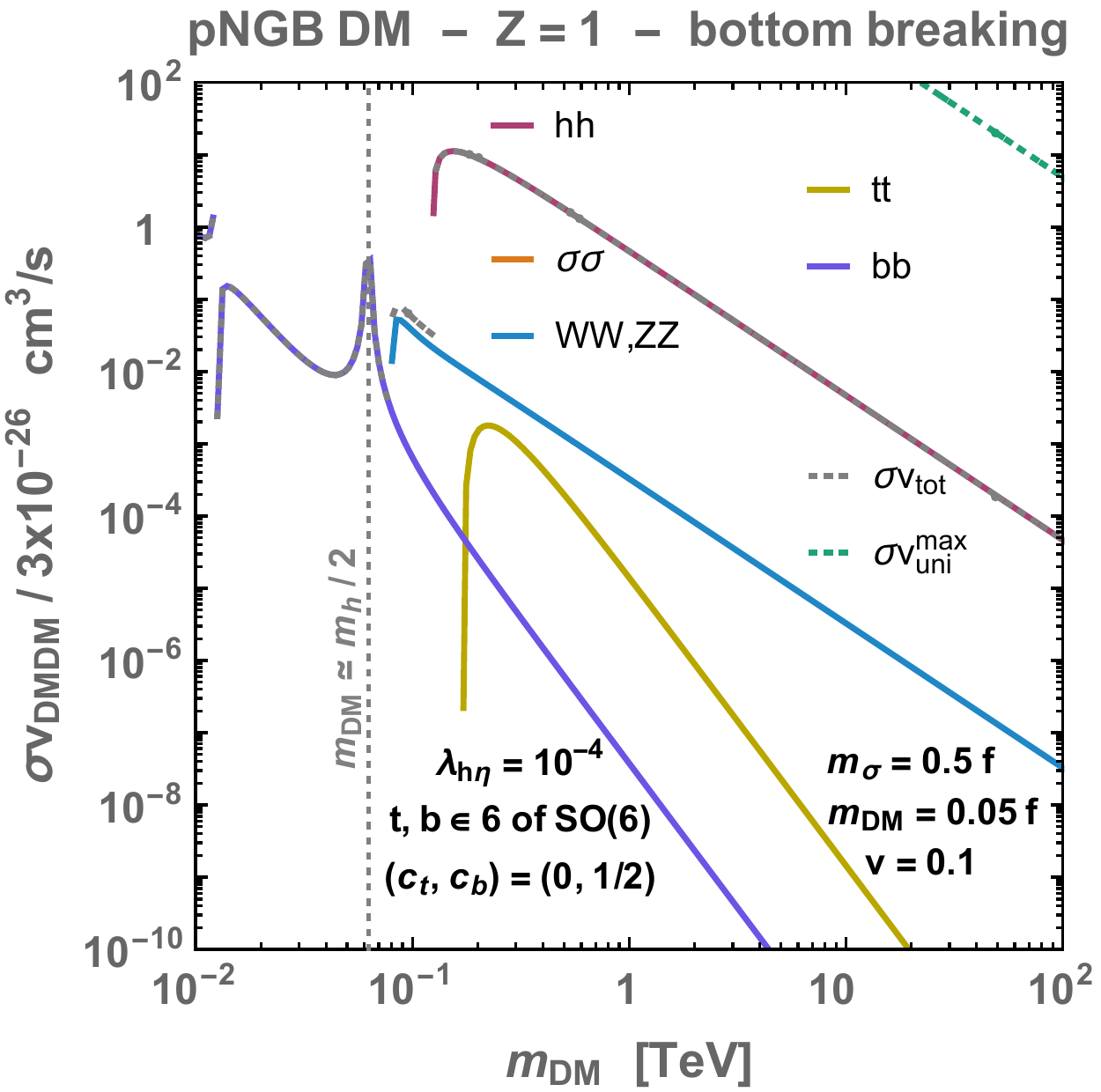}}}
\caption{\it \small \textbf{DM annihilation cross-section} in the scenario where DM is a \textbf{O(6)/O(5) pNGB}. We consider the pNGB shift symmetry to be explicitly broken either by the top interactions (\textbf{top}) or by the bottom ones (\textbf{bottom}). On the \textbf{left} panels, we fix the confining scale to $f = 1~\rm TeV$ while on the \textbf{right} panels, we fix it to $f = 20~\MDM$. In all plots, we set the dilaton mass to $m_{\sigma} = 0.5~f$ and the relative velocity to $v_{\mathsmaller{\rm rel}} = 0.1$. The wiggles are consequences of Sommerfeld effects, cf.~\ref{sec:unitary-bound}.}
\label{fig:sigmav_pNGB} 
\thisfloatpagestyle{empty}
\end{figure}

\resumetocwriting
\section{DM annihilation cross-sections}
\label{app:ann_cross-section}
\stoptocwriting

\paragraph{Thermally averaged annihilation cross-section.}
In order to compute the thermal effects on the DM abundance after the supercooling stage we need to compute the thermally averaged DM annihilation cross-section. Considering the process $A_{1} + A_{2} \longrightarrow A_{3} + A_{4}$ with $m_{1}=m_{2}=m_{i}$ and $m_{3}=m_{4}=m_{f}$, the reference \cite{Gondolo:1990dk} gives
\begin{equation}
\left< \sigma v_{\mathsmaller{\rm rel}} \right> = \frac{\int \sigma \, v_{\mathsmaller{\rm rel}} \, e^{-E_1/T}e^{-E_2/T} d^3 p_1 \, d^3 p_2}{\int  e^{-E_1/T}e^{-E_2/T} d^3 p_1 \, d^3 p_2} 
=  \frac{1}{8 m_{i}^4 T K_2^2\left(\frac{m_{i}}{T}\right)}\int_{4m_{i}^2}^{\infty}  (S-4m_{i}^2) \sqrt{S} K_{1}\left(\frac{\sqrt{S}}{T}\right) \, \sigma \, dS,
\end{equation}
where the annihilation cross-section $\sigma$ reads \cite{Peskin:1995ev}
\begin{equation}
\label{eq:cross_section_formula}
\sigma = \frac{1}{16\pi S} \sqrt{\frac{S-4m_{f}^2}{S-4m_{i}^2}} \int \frac{d\Omega}{4\pi} \, \frac{1}{N_{I}!}\frac{1}{\sigma_1 \sigma_2}\sum_{\sigma_1,\,\sigma_2\,\sigma_3\,\sigma_4} |M|^2,
\end{equation}
where $N_{I}$ is the number of identical particles in the final state and $\sigma_1$, $\sigma_2$ are the number of internal degrees of freedom of the initial states . We compute the transition amplitude $M$ with Feyncalc \cite{Shtabovenko:2016sxi}. 
\paragraph{Integration over the scattering angle.}
Integration over the scattering angle in the center of mass frame $d\Omega = 2\pi \sin{\theta} d\theta$ can be performed after replacing the Mandelstram variable $t = (p_1-p_3)^2$ by 
\begin{equation}
t = m_1^2 + m_3^2 - \frac{(S+m_1^2-m_2^2)(S+m_3^2-m_4^2)}{2S} + \frac{1}{2S}\lambda^{1/2}(S,\,m_1^2,\,m_2^2)\lambda^{1/2}(S,\,m_3^2,\,m_4^2) \, \cos{\theta},
\end{equation}
with $S=(p_1+p_2)^2$ and $\lambda(a,b,c)$ being the so-called triangle function \cite{Kallen:1964lxa}
\begin{equation}
\lambda(a,\,b,\,c) = a^2 + b^2 + c^2 - 2ab - 2ac - 2bc.
\end{equation}
\paragraph{Non-relativistic limit.}
Note that the non-relativistic limit, which is sufficient for computing the indirect detection constraints, is recovered by setting the Mandelstram variables $S$ to
\begin{align}
S=4m_{i}^2\left(1+\frac{v_{\mathsmaller{\rm rel}}^2}{4}\right),
\end{align}
where $v_{\mathsmaller{\rm rel}}$ is the relative velocity between the incoming particles in their center of mass frame.

\paragraph{Sommerfeld enhancement.}
As discussed in Sec.~\ref{par:sommerfeld_enhancement}, non-perturbative Sommerfeld enhancement of the interaction probability is included after multiplying the perturbative cross-section $ \sigma_{\rm pert}$ in Eq.~\eqref{eq:cross_section_formula} by the factor $S_{H}^{(l)}$ in Eq.~\eqref{eq:Sommerfeld_factor_Cassel}
\begin{equation}
\sigma_{\rm non-pert} = \sigma_{\rm pert}\times S_{H}^{(l)}.
\end{equation}

\paragraph{Goldstone equivalence theorem.}
The cross-sections for DM annihilation into Higgs pairs, $\sigma (\DM\DM \rightarrow hh)$, are different from \cite{Blum:2014jca} due to the coupling of the dilaton to the Higgs kinetic term in Eq.~\eqref{eq:eff_lag_dilaton}, needed in order to respect the Goldstone equivalence theorem
\begin{equation}
\sigma (\DM\DM \rightarrow hh) \underset{ S \gg m_Z}{\sim} \sigma (\DM\DM \rightarrow ZZ).
\end{equation} 
We find the cross-section for vector DM annihilation into dilaton pair $\approx 4$ times smaller than \cite{Blum:2014jca}, in the large $m_{\DM}$ limit.
\paragraph{Formulae.}
 In the next subsections, we give the analytical formulae of the DM annihilation cross-sections for scalar, vector and fermion DM and we plot them in Fig~\ref{fig:sigmav_scalar_vector_fermion}. Because of their length, we chose to not report the analytical formulae of the pNGB DM annihilation cross-sections in the present article, but instead we simply show their plot in Fig.~\ref{fig:sigmav_pNGB}.

\subsection{Fermion}
We compute
\begin{align}
\sigma v_{\mathsmaller{\rm rel}}(\DM\DM \rightarrow hh) =& ~ v_{\mathsmaller{\rm rel}}\frac{m_{\DM}^2 (S+2m_h^2)^2 \sqrt{(S-4 m_h^2)(S-4m_{\DM}^2})}{64 \pi (Zf)^4 S\left[(S-m_{\sigma}^2)^2 +m_{\sigma}^2 \Gamma_{\sigma}^2\right]} \\
\underset{ v_{\mathsmaller{\rm rel}}\ll1}{\Longrightarrow} & ~  v_{\mathsmaller{\rm rel}}^2 \frac{m_{\DM}\sqrt{m_{\DM}^2- m_h^2}(2m_{\DM}^2+ m_h^2)^2}{32\pi (Zf)^4 \left[(4m_{\DM}^2-m_{\sigma}^2)^2+m_{\sigma}^2 \Gamma_{\sigma}^2 \right]} \\
\underset{m_{\DM} \gg m_{\rm other}}{\Longrightarrow} & ~v_{\mathsmaller{\rm rel}}^2\frac{m_{\DM}^2}{128\pi (Zf)^4}.
\end{align}
\begin{align}
\sigma v_{\mathsmaller{\rm rel}}(\DM\DM \rightarrow ff)  =& ~ v_{\mathsmaller{\rm rel}} N_{c}\frac{m_{f}^2m_{\DM}^2(S-4 m_f^2)^{3/2}\sqrt{S-4m_{\DM}^2}}{16 \pi (Zf)^4 S  \left[(S-m_{\sigma}^2)^2 +m_{\sigma}^2 \Gamma_{\sigma}^2 \right]} \\
\underset{ v_{\mathsmaller{\rm rel}}\ll1}{\Longrightarrow} & ~ v_{\mathsmaller{\rm rel}}^2 N_{c}\frac{m_{f}^2m_{\DM} (m_{\DM}^2- m_f^2)^{3/2}}{8\pi (Zf)^4   \left[(4m_{\DM}^2-m_{\sigma}^2)^2+m_{\sigma}^2 \Gamma_{\sigma}^2 \right]} \\
\underset{m_{\DM} \gg m_{\rm other}}{\Longrightarrow} & ~v_{\mathsmaller{\rm rel}}^2N_c\frac{m_{f}^2}{128\pi (Zf)^4}.
\end{align}
\begin{align}
\sigma v_{\mathsmaller{\rm rel}}(\DM\DM \rightarrow WW)  =& ~2\,\sigma v_{\mathsmaller{\rm rel}}(\DM\DM \rightarrow ZZ)|_{m_{W} \leftrightarrow m_{Z}}\\
=& \, v_{\mathsmaller{\rm rel}} \frac{m_{\DM}^2\sqrt{S-4 m_W^2}\sqrt{S-4 m_{\DM}^2}}{32 \pi (Zf)^4 S} \frac{12m_W^4 -4m_W^4S+S^2}{ (S-m_{\sigma}^2)^2 + m_{\sigma}^2 \Gamma_{\sigma}^2 } \\
\underset{ v_{\mathsmaller{\rm rel}}\ll 1 }{\Longrightarrow} & ~  v_{\mathsmaller{\rm rel}}^2\frac{m_{\DM}\sqrt{m_{\DM}^2-m_W^2}}{16 \pi (Zf)^4 } \frac{3m_W^4 + 4m_{\DM}^4-4m_{\DM}^2 m_W^2}{ (4m_{\DM}^2-m_{\sigma}^2)^2 +m_{\sigma}^2 \Gamma_{\sigma}^2 } \\
\underset{m_{\DM} \gg m_{\rm other}}{\Longrightarrow} & ~v_{\mathsmaller{\rm rel}}^2\frac{m_{\DM}^2}{64\pi (Zf)^4}.
\end{align}

\begin{align}
\sigma v_{\mathsmaller{\rm rel}}(\DM\DM \rightarrow \sigma\sigma)  =& ~ v_{\mathsmaller{\rm rel}}\, m_{\DM}^2 \Bigg[ (S - 2 m_\sigma^2 ) \sqrt{(S-4m_\sigma^2)(S-4m_{\DM}^2)}\Big[m_\sigma^8(25 S - 66 m_{\DM}^2) \notag\\
&+ 2m_\sigma^6 m_{\DM}^2 (-3 \Gamma_\sigma^2 + 136 m_{\DM}^2 - 64 S) + m_\sigma^4(-32m_{\DM}^6 + 32 m_{\DM}^4(\Gamma_\sigma^2 + S) + 19 m_{\DM}^2 S^2) \notag\\
&+ 4 m_\sigma^2 m_{\DM}^4(16 m_{\DM}^2 S - \Gamma_\sigma^2(8m_{\DM}^2 + S))- 4m_{\DM}^4 S^2(8 m_{\DM}^2 + S)\Big]  \notag\\
&- 4 m_{\DM}^2(m_\sigma^4 - 4 m_\sigma^2 m_{\DM}^2 + m_{\DM}^2 S)\Big[(S - m_\sigma^2)[34 m_\sigma^6 + 2 m_\sigma^4 (5 S - 72 m_{\DM}^2)  \notag\\
&+ m_\sigma^2(32 m_{\DM}^4 + 48 m_{\DM}^2 S - 15 S^2 ) + S(-32 m_{\DM}^4 + 16 m_{\DM}^2 S + S^2)]  \notag\\
&- \Gamma_\sigma^2 m_\sigma^2 ( 6m_\sigma^4 - 4 m_\sigma^2(4m_{\DM}^2 + S) - 32 m_{\DM}^4 + 16m_{\DM}^2 S + S^2)\Big]\times \notag\\
&\textrm{ArcCoth}\left( \frac{2m_\sigma^2 - S}{\sqrt{(S-4m_\sigma^2)(S-4m_{\DM}^2)}} \right)\Bigg]/ \Big[ 64\pi (Zf)^4 S (S-2m_\sigma^2) (S-4m_{\DM}^2) \times \notag \\
&( m_\sigma^4 - 4m_\sigma^2 m_{\DM}^2 + m_{\DM}^2 S)(\Gamma_\sigma^2 m_\sigma^2 + (S - m_\sigma^2)^2) \Big] \\
\underset{ v_{\mathsmaller{\rm rel}}\ll1}{\Longrightarrow} & ~ v_{\mathsmaller{\rm rel}}^2 m_{\DM} \sqrt{m_{\DM}^2 - m_\sigma^2}\Big( 75 m_\sigma^{12} - 680 m_\sigma^{10} m_{\DM}^2 + 2672 m_\sigma^8 m_{\DM}^4 \notag\\
&+ 32 m_\sigma^6 m_{\DM}^4(\Gamma_\sigma^2 - 187 m_{\DM}^2)+ 32 m_\sigma^4(255 m_{\DM}^8 - 4 \Gamma_\sigma^2 m_{\DM}^6)  \notag\\
&+ 16 m_\sigma^2 m_{\DM}^8(9 \Gamma_\sigma^2 - 400 m_{\DM}^2) + 2304 m_{\DM}^{12}\Big)/ \nonumber \\
&\Big[ 384 \pi (Z f)^4(m_\sigma^2 - 2 m_{\DM}^2)^4(m_\sigma^4 + m_\sigma^2(\Gamma_\sigma^2 - 8 m_{\DM}^2) + 16 m_{\DM}^4) \Big] \\
\underset{m_{\DM} \gg m_{\rm other}}{\Longrightarrow} & ~v_{\mathsmaller{\rm rel}}^2\frac{3m_{\DM}^2}{128\pi (Zf)^4}.
\end{align}

\subsection{Scalar}
We compute
\begin{align}
\sigma v_{\mathsmaller{\rm rel}}(\DM\DM \rightarrow hh) =& ~ v_{\mathsmaller{\rm rel}}\frac{m_{\DM}^4 \sqrt{S-4 m_h^2}(S+2m_h^2)^2}{8 \pi (Zf)^4 S\sqrt{S-4m_{\DM}^2}\left[(S-m_{\sigma}^2)^2 +m_{\sigma}^2 \Gamma_{\sigma}^2\right]} \\
\underset{ v_{\mathsmaller{\rm rel}}\ll1}{\Longrightarrow} & ~   \frac{m_{\DM} \sqrt{m_{\DM}^2- m_h^2}(2m_{\DM}^2+ m_h^2)^2}{4\pi (Zf)^4 \left[(4m_{\DM}^2-m_{\sigma}^2)^2+m_{\sigma}^2 \Gamma_{\sigma}^2 \right]} \\
\underset{m_{\DM} \gg m_{\rm other}}{\Longrightarrow} & ~\frac{m_{\DM}^2}{16\pi (Zf)^4}.
\end{align}
\begin{align}
\sigma v_{\mathsmaller{\rm rel}}(\DM\DM \rightarrow ff)  =& ~ v_{\mathsmaller{\rm rel}} N_{c}\frac{m_{f}^2m_{\DM}^4(S-4 m_f^2)^{3/2}}{2 \pi (Zf)^4 S \sqrt{S-4m_{\DM}^2} \left[(S-m_{\sigma}^2)^2 +m_{\sigma}^2 \Gamma_{\sigma}^2 \right]} \\
\underset{ v_{\mathsmaller{\rm rel}}\ll1}{\Longrightarrow} & ~  N_{c}\frac{m_{f}^2m_{\DM} (m_{\DM}^2- m_f^2)^{3/2}}{\pi (Zf)^4   \left[(4m_{\DM}^2-m_{\sigma}^2)^2+m_{\sigma}^2 \Gamma_{\sigma}^2 \right]} \\
\underset{m_{\DM} \gg m_{\rm other}}{\Longrightarrow} & ~N_c\frac{m_{f}^2}{16\pi (Zf)^4}.
\end{align}
\begin{align}
\sigma v_{\mathsmaller{\rm rel}}(\DM\DM \rightarrow WW)  =& ~2\,\sigma v_{\mathsmaller{\rm rel}}(\DM\DM \rightarrow ZZ)|_{m_{W} \leftrightarrow m_{Z}}\\
=& \, v_{\mathsmaller{\rm rel}} \frac{m_{\DM}^4\sqrt{S-4 m_W^2}}{4 \pi (Zf)^4 S\sqrt{S-4 m_{\DM}^2}} \frac{12m_W^4 -4m_W^4S+S^2}{ (S-m_{\sigma}^2)^2 +m_{\sigma}^2 \Gamma_{\sigma}^2 } \\
\underset{ v_{\mathsmaller{\rm rel}}\ll1}{\Longrightarrow} & ~  \frac{m_{\DM}\sqrt{m_{\DM}^2-m_W^2}}{2 \pi (Zf)^4 } \frac{3m_W^4 + 4m_{\DM}^4-4m_{\DM}^2 m_W^2}{ (4m_{\DM}^2-m_{\sigma}^2)^2 +m_{\sigma}^2 \Gamma_{\sigma}^2 } \\
\underset{m_{\DM} \gg m_{\rm other}}{\Longrightarrow} & ~\frac{m_{\DM}^2}{8\pi (Zf)^4}.
\end{align}
\begin{align}
\sigma v_{\mathsmaller{\rm rel}}(\DM\DM \rightarrow \sigma\sigma)  =& ~ v_{\mathsmaller{\rm rel}} \Bigg[m_{\DM}^4(8m_\sigma^4 - 6m_\sigma^2 S + S^2)(4m_{\DM}^2-S)\Big(16m_{\sigma}^8  \notag\\
&+ m_{\sigma}^6(\Gamma_\sigma^2 - 64 m_{\DM}^2 +8S) + m_\sigma^4(8 m_{\DM}^4 - 4 m_{\DM}^2 (\Gamma_\sigma^2 + 4S) + S^2)  \notag\\
&+ m_\sigma^2 m_{\DM}^2(8m_{\DM}^2( \Gamma_\sigma^2 - 2 S)  + S(\Gamma_\sigma^2 + 4 S)) + m_{\DM}^2 S^2 (8m_{\DM}^2 + S)\Big)   \notag\\
&- 16 m_{\DM}^6 \sqrt{(S-4m_\sigma^2)(S-4m_{\DM}^2)}(m_\sigma^4 - 4 m_\sigma^2 m_{\DM}^2 + m_{\DM}^2 S)\times \notag \\
& \Big(\Gamma_\sigma^2 m_\sigma^2 ( S - 2(m_\sigma^2+m_{\DM}^2)) + (m_\sigma^2  - S)(8 m_\sigma^4 - 2 m_\sigma^2(m_{\DM}^2 + S) + S(2m_{\DM}^2 - S))\Big) \notag \\
& \times \textrm{ArcCoth}\left( \frac{2m_\sigma^2 - S}{\sqrt{(S-4m_\sigma^2)(S-4m_{\DM}^2)}} \right)\Bigg]   / \Big[ 8\pi (Zf)^4 S (2m_\sigma^2 - S) \sqrt{S-4m_\sigma^2} \times \notag \\
&( S - 4 m_{\DM}^2)^{3/2} ( m_\sigma^4 - 4m_\sigma^2 m_{\DM}^2 + m_{\DM}^2 S)(\Gamma_\sigma^2 m_\sigma^2 + (m_\sigma^2 - S)^2) \Big] \\
\underset{ v_{\mathsmaller{\rm rel}}\ll1}{\Longrightarrow} & ~ \frac{m_{\DM}\sqrt{m_{\DM}^2-m_\sigma^2}\Big(\Gamma_\sigma^2 m_\sigma^2 (m_\sigma^2 + 2 m_{\DM}^2)^2 + 16 (m_\sigma^4 - 2m_\sigma^2 m_{\DM}^2 + 2 m_{\DM}^4)^2\Big)}{16\pi(Z f)^4 (m_\sigma^2 - 2m_{\DM}^2)^2(\Gamma_\sigma^2 m_\sigma^2 + (m_\sigma^2 - 4 m_{\DM}^2)^2)}\\
\underset{m_{\DM} \gg m_{\rm other}}{\Longrightarrow} & ~\frac{m_{\DM}^2}{16\pi (Zf)^4}.
\end{align}

\subsection{Vector}
We compute
\begin{align}
\sigma v_{\mathsmaller{\rm rel}}(\DM\DM \rightarrow hh) =& \, v_{\mathsmaller{\rm rel}}\frac{\sqrt{S-4 m_h^2}(S+2m_h^2)^2}{288 \pi (Zf)^4  S\sqrt{S-4m_{\DM}^2}}\frac{12m_{\DM}^4 -4m_{\DM}^4S+S^2}{\left[(S-m_{\sigma}^2)^2 +m_{\sigma}^2 \Gamma_{\sigma}^2\right]} \\
\underset{ v_{\mathsmaller{\rm rel}}\ll1}{\Longrightarrow} & ~  \frac{m_{\DM} \sqrt{m_{\DM}^2- m_h^2}(2m_{\DM}^2+ m_h^2)^2}{12 \pi (Zf)^4 \left[(4m_{\DM}^2-m_{\sigma}^2)^2+m_{\sigma}^2 \Gamma_{\sigma}^2 \right]}, \\
\underset{m_{\DM} \gg m_{\rm other}}{\Longrightarrow} & ~\frac{m_{\DM}^2}{48\pi (Zf)^4}.
\end{align}
\begin{align}
\sigma v_{\mathsmaller{\rm rel}}(\DM\DM \rightarrow ff)  =& \, v_{\mathsmaller{\rm rel}} N_{c}\frac{m_{f}^2(S-4 m_f^2)^{3/2}}{72 \pi (Zf)^4 S \sqrt{S-4m_{\DM}^2}}\frac{12m_{\DM}^4 -4m_{\DM}^4S+S^2}{\left[(S-m_{\sigma}^2)^2 +m_{\sigma}^2 \Gamma_{\sigma}^2\right]} \\
\underset{ v_{\mathsmaller{\rm rel}}\ll1}{\Longrightarrow} & ~ N_{c}\frac{m_{f}^2m_{\DM} (m_{\DM}^2- m_f^2)^{3/2}}{3\pi (Zf)^4   \left[(4m_{\DM}^2-m_{\sigma}^2)^2+m_{\sigma}^2 \Gamma_{\sigma}^2 \right]} \\
\underset{m_{\DM} \gg m_{\rm other}}{\Longrightarrow} & ~N_c\frac{m_{f}^2}{48\pi (Zf)^4}.
\end{align}
\begin{align}
\sigma v_{\mathsmaller{\rm rel}}(\DM\DM \rightarrow WW)  =& ~2\,\sigma v_{\mathsmaller{\rm rel}}(\DM\DM \rightarrow ZZ)|_{m_{W} \leftrightarrow m_{Z}}\\
=& \, v_{\mathsmaller{\rm rel}} \frac{\sqrt{S-4 m_W^2}}{144 \pi (Zf)^4 S\sqrt{S-4 m_{\DM}^2}} \frac{\left(12m_{\DM}^4 -4m_{\DM}^4S+S^2\right)\left(12m_W^4 -4m_W^4S+S^2\right)}{ (S-m_{\sigma}^2)^2 +m_{\sigma}^2 \Gamma_{\sigma}^2 } \\
\underset{ v_{\mathsmaller{\rm rel}}\ll1}{\Longrightarrow} & ~ \frac{m_{\DM}\sqrt{m_{\DM}^2-m_W^2}}{6 \pi (Zf)^4 } \frac{3m_W^4 + 4m_{\DM}^4-4m_{\DM}^2 m_W^2}{ (4m_{\DM}^2-m_{\sigma}^2)^2 +m_{\sigma}^2 \Gamma_{\sigma}^2 }\\
 \underset{m_{\DM} \gg m_{\rm other}}{\Longrightarrow} & ~\frac{m_{\DM}^2}{24\pi (Zf)^4}.
\end{align}

\begin{align}
\sigma v_{\mathsmaller{\rm rel}}(\DM\DM \rightarrow \sigma\sigma)  =& ~ v_{\mathsmaller{\rm rel}} (12m_{\DM}^4-4m_{\DM}^2 S + S^2) \Bigg[16 m_{\DM}^2 (m_{\sigma}^4 - 4m_{\sigma}^2 m_{\DM}^2 + m_{\DM}^2 S) \times \notag \\
&\Big[\Gamma_{\sigma}^2 m_{\sigma}^2 (S - 2(m_{\sigma}^2 + m_{\DM}^2)) + (m_{\sigma}^2 - S)\Big(8m_{\sigma}^4 - 2 m_{\sigma}^2(m_{\DM}^2 + S) \notag \\
&+ S(2m_{\DM}^2 - S)\Big)\Big]\textrm{ArcCoth}\left( \frac{2m_\sigma^2 - S}{\sqrt{(S-4m_\sigma^2)(S-4m_{\DM}^2)}} \right) \notag \\
&- (2m_{\sigma}^2 - S) \sqrt{(S-4m_{\sigma}^2)(S-4m_{\DM}^2)}\Big[ 16 m_{\sigma}^8 + m_{\sigma}^6(\Gamma_\sigma^2 - 64 m_{\DM}^2 + 8 S) \notag \\
&+ m_{\sigma}^4(8m_{\DM}^4 - 4m_{\DM}^2(\Gamma_\sigma^2 + 4 S) + S^2) + m_{\sigma}^2 m_{\chi}^2(8m_{\chi}^2(\Gamma_{\sigma}^2 - 2 S) + S(\Gamma_{\sigma}^2 + 4S)) \notag \\
&+ m_{\chi}^2 S^2 (8m_{\chi}^2 + S)\Big] \Bigg] / \Bigg[288 \pi (Zf)^4 S (S-2m_{\sigma}^2)(S - 4 m_{\DM}^2)\times \notag \\
&(m_{\sigma}^4 - 4m_{\sigma}^2  m_{\DM}^2 + m_{\chi}^2 S)(\Gamma_{\sigma}^2 m_{\sigma}^2 + (S - m_{\sigma}^2)^2) \Bigg] \\
\underset{ v_{\mathsmaller{\rm rel}}\ll1}{\Longrightarrow} & ~ \frac{m_{\DM}\sqrt{m_{\DM}^2-m_\sigma^2}\Big(\Gamma_\sigma^2 m_\sigma^2 (m_\sigma^2 + 2 m_{\DM}^2)^2 + 16 (m_\sigma^4 - 2m_\sigma^2 m_{\DM}^2 + 2 m_{\DM}^4)^2\Big)}{48\pi(Z f)^4 (m_\sigma^2 - 2m_{\DM}^2)^2(\Gamma_\sigma^2 m_\sigma^2 + (m_\sigma^2 - 4 m_{\DM}^2)^2)}\\
\underset{m_{\DM} \gg m_{\rm other}}{\Longrightarrow} & ~\frac{m_{\DM}^2}{48\pi (Zf)^4}.
\end{align}

\resumetocwriting

\section{Direct-detection cross-sections}
\label{app:DD_cross-section}
\FloatBarrier
\stoptocwriting
We compute the DM-nucleon scattering cross-section for fermion, scalar and vector DM resonance. The constants $y_n$ and $c_n$ are defined in Eq.~\eqref{eq:y_n_def} and Eq.~\eqref{eq:cn_def}, respectively.

\subsection{Fermion}
\begin{align}
\sigma(\DM n \rightarrow \DM n) = \frac{y_n^2}{2\pi}\left(\frac{m_{n}m_{\DM}}{m_{n}+m_{\DM}}\right)^2 \left(\frac{m_{\DM}}{Zf}\right)^2 \frac{1}{ m_{\sigma}^4} .
\end{align}
\subsection{Vector}
\begin{align}
\sigma(\DM n \rightarrow \DM n) = \frac{3y_n^2}{2\pi}\left(\frac{m_{n}m_{\DM}}{m_{n}+m_{\DM}}\right)^2 \left(\frac{m_{\DM}}{Zf}\right)^2 \frac{1}{m_{\sigma}^4} .
\end{align}
\subsection{Scalar}
\begin{align}
\sigma(\DM n \rightarrow \DM n) = \frac{y_n^2}{\pi}\left(\frac{m_{n}m_{\DM}}{m_{n}+m_{\DM}}\right)^2 \left(\frac{m_{\DM}}{Zf}\right)^2 \frac{1}{ m_{\sigma}^4} .
\end{align}

\subsection{pNGB}
\begin{equation}
\sigma({\rm DMn \to DM n}) \simeq
\left\{
                \begin{array}{ll}
                \dfrac{y_n^2}{\pi}\left(\dfrac{m_{n}m_{\DM}}{m_{n}+m_{\DM}}\right)^2 \left(\dfrac{m_{\DM}}{Zf}\right)^2 \dfrac{1}{ m_{\sigma}^4}
                \quad \text{bottom-breaking}\,,\\ \\
               \dfrac{c_n^2}{\pi} \left(\dfrac{m_n}{m_n+\MDM} \right)^2
               \quad \text{top-breaking}\,,
                \end{array}
              \right.
              \end{equation}
\begin{equation}
\sigma({\rm DMn \to DM n}) \simeq \frac{y_n^2}{\pi}\left(\frac{m_{n}m_{\DM}}{m_{n}+m_{\DM}}\right)^2 \left(\frac{m_{\DM}}{Zf}\right)^2 \frac{1}{ m_{\sigma}^4} + \frac{c_n^2}{\pi} \left(\frac{m_n}{m_n+\MDM} \right)^2.
\end{equation}

\resumetocwriting
\section{Dilaton decay width}
\label{app:dilaton_decay_width}
\stoptocwriting
\FloatBarrier
We compute the dilaton decay width into SM. The decay width $\Gamma (\sigma \rightarrow hh) $ is different from \cite{Blum:2014jca} due to the coupling of the dilaton to the Higgs kinetic mixing in Eq.~\eqref{eq:eff_lag_dilaton}, needed in order to respect the Goldstone equivalence theorem
\begin{equation}
\Gamma (\sigma \rightarrow hh) \underset{ m_{\sigma} \gg m_Z}{\sim} \Gamma (\sigma \rightarrow ZZ).
\end{equation} 
\subsection{Into WW and ZZ}
\begin{align}
\Gamma (\sigma \rightarrow WW) & = ~ \frac{m_W^4}{4\pi m_\sigma (Zf)^2} \sqrt{1-4\frac{m_W^2}{m_\sigma^2}} \left( 2 + \frac{(m_\sigma^2-2m_W^2)^2}{4m_W^4}   \right) \\
& \underset{m_\sigma \gg m_W}{\Longrightarrow}  ~\frac{m_{\sigma}^3}{16\pi (Zf)^2}.
\end{align}

\begin{align}
\Gamma (\sigma \rightarrow ZZ) & = ~ \frac{m_Z^4}{8\pi m_\sigma (Zf)^2} \sqrt{1-4\frac{m_Z^2}{m_\sigma^2}} \left( 2 + \frac{(m_\sigma^2-2m_Z^2)^2}{4m_Z^4}   \right) \\
& \underset{m_\sigma \gg m_Z}{\Longrightarrow}  ~\frac{m_{\sigma}^3}{32\pi (Zf)^2}.
\end{align}

\subsection{Into Higgs}
\begin{align}
\Gamma (\sigma \rightarrow hh)  &= ~ \frac{m_\sigma^3}{32\pi (Zf)^2}\left(1-4\frac{m_h^2}{m_\sigma^2}\right)^{1/2}\left(1+2\frac{m_h^2}{m_\sigma^2}\right)^{2} \\
 &\underset{m_\sigma \gg m_h}{\Longrightarrow} ~ \frac{m_\sigma^3}{32\pi (Zf)^2}.
\end{align}

\subsection{Into fermions}
\begin{align}
\Gamma (\sigma \rightarrow ff)  = ~ N_c\frac{m_\sigma m_f^2}{8\pi (Zf)^2}\left(1-4\frac{m_f^2}{m_\sigma^2}\right)^{3/2} .
\end{align}

\subsection{Into pNGB DM}
\begin{align}
\Gamma (\sigma \rightarrow \eta\eta)  &= ~ \frac{m_\sigma^3}{32\pi (Zf)^2}\left(1-4\frac{\MDM^2}{m_\sigma^2}\right)^{1/2}\left(1+2\frac{\MDM^2}{m_\sigma^2}\right)^{2} \\
 &\underset{m_\sigma \gg \MDM}{\Longrightarrow}  ~ \frac{m_\sigma^3}{32\pi (Zf)^2}.
\end{align}

\resumetocwriting

\medskip
\small

\bibliographystyle{JHEP}
\bibliography{SupercoolingCompositeHiggs}
\end{document}